\documentclass{aa}

\usepackage{graphicx}
\usepackage{txfonts}

\begin{document}

\title{Near-infrared and optical observations of galactic warps: A
common, unexplained feature of most discs}

\author{A. Guijarro\inst{1,2}, R. F. Peletier\inst{3}, E. Battaner\inst{1}, J. Jim\'enez-Vicente\inst{1}, R. de Grijs\inst{4,5} \and E. Florido\inst{1}}

\institute{Dpto. F\'{\i}sica Te\'orica y del Cosmos, Universidad de Granada, Granada, Spain
           \and Centro Astron\'omico Hispano Alem\'an, Almer\'{\i}a, Spain
           \and Kapteyn Astronomical Institute, Groningen, The Netherlands 
           \and Kavli Institute for Astronomy and Astrophysics and Department of Astronomy, Peking University, Beijing, China
           \and Department of Physics \& Astronomy, The University of Sheffield, UK
           }

\offprints{Ana Guijarro,
           \email{aguijarr@ugr.es}}
		
\date{Received / Accepted}
\titlerunning{Near-infrared and optical observations of galactic warps}
\authorrunning{Guijarro et al.}

\abstract{ \underline{Context.} Warps occurring in galactic discs have
been studied extensively in HI and in the optical, but rarely in the
near-infrared (NIR) bands that trace the older stellar populations.\\
\underline{Aims.} We provide NIR data of nearby edge-on galaxies,
combined with optical observations, for direct comparison of the
properties of galactic warps as a function of wavelength, and calculate
warp curves for each galaxy and obtain the characteristic warp
parameters. We discuss these properties as possible constraints to the
different mechanisms that have been proposed for the development and
persistence of galactic warps.\\ 
\underline{Methods.} We observed 20 galaxies that were selected from a
statistically complete diameter-limited subsample of edge-on disc
galaxies. We used the Cerro Tololo Infrared Imager (CIRIM) at the CTIO
1.5m Ritchey-Chretien telescope to acquire the NIR data. We used the
1.54m Danish and 0.92m Dutch telescopes at the European Southern
Observatory's La Silla site for our optical observations.\\
\underline{Results.} Our results show that 13 of our 20 sample
galaxies are warped, with the warp more pronounced in the
optical than at NIR wavelengths. In the remaining seven galaxies, no
warp is apparent within the limitations of our automated detection
method. The transition between the unperturbed inner disc and the
outer, warped region is rather abrupt. S0 galaxies exhibit very small
or no warps. The magnetic model remains one of a number of interesting
formation scenarios.\\
 }

\maketitle

\section{Introduction}

Most spiral galaxies, including our own, exhibit warped discs. This
has long been known based on observations of both the extended
neutral-gas (HI) component (Sancisi 1976; Bosma 1981; Briggs 1990;
Garc\'{\i}a-Ruiz et al. 2002b) and optical starlight
(S\'anchez-Saavedra et al. 1990; Florido et al. 1991; Reshetnikov \&
Combes 1998; Ann \& Park 2006). The first detections of warped discs
came from 21 cm line observations of our own Galaxy (Burke 1957;
Kerr 1957; Burton 1988). Based on a sky survey covering the Northern
Hemisphere, S\'anchez-Saavedra et al. (1990) first reported the high
frequency of optical warps in disc galaxies, which they later
confirmed in S\'anchez-Saavedra et al. (2003) based on an extended
catalogue that also included the Southern Hemisphere. These studies
took into account that warps can only be detected when their
orientation (projection) is favourable (i.e., not in the line of
sight). A noteworthy and important result, established in the later
paper, is that they did not find any warped S0 galaxies. Here, we
therefore pay special attention to S0 galaxies. Reshetnikov et
al. (2002) find that warps were also common in the past, with even
greater magnitudes at $z \approx 1$.

Edge-on galaxies are usually chosen to study warps, although dynamical
warps in the HI distribution of galaxies of intermediate inclination have also been reported, e.g., in M83 (Rogstad et al. 1974), NGC 5033, NGC 5055, NGC2841, and NGC7331 (Bosma 1978), M31 (Brinks \& Burton 1984), and M33 (Corbelli et al. 1989). Garc\'{\i}a-Ruiz et al. (2002b)  analysed HI observations and find that 
all edge-on galaxies that have an extended HI disc (with respect to the
optical component) are warped. 

Ann \& Park (2006) find that 73\% of the
325 galaxies in their well-defined sample of almost perfectly edge-on
galaxies ($a/b > 9.5$, where $a/b$ is the major-to-minor axis ratio at
$\mu_{B} = 25$ mag arcsec$^{-2}$) exhibit optical warps. Recently, Reyl\'e et
al. (2009) have found
that a warp is present in the stellar, dust, and gas discs of the
Milky Way, which are all asymmetric and characterised by a similar
line of nodes. The Milky Way's HI gas warp is the strongest, followed
by the dust warp, while the stellar warp is significantly smaller (by
a factor of approximately two). They conclude that this comparison
shows that the different components react differently to the forces
responsible for the origin of the warp. The most important additional
information that we provide here (not yet been discussed
by other authors) is a multiwavelength study of galactic warps in one
NIR and three optical filters.

Considerable effort has gone into establishing the physical origin and
stability of warps. Numerous mechanisms have been proposed and
explored, but they have thus far not provided a definitive and
satisfactory physical explanation (see, e.g., the reviews by Binney
1992; Battaner et al. 1997; van der Kruit 2007). Kerr (1957), Hunter
\& Toomre (1969), Weinberg (1998), Weinberg \& Blitz (2006), and
others have suggested that warps can result from tidal interactions with a
satellite or neighbouring galaxy. This hypothesis can explain
warps neither in isolated galaxies (Sancisi 1976; Tubbs \& Sanders 1979;
Sparke 1984) nor in the Milky Way, if the latter was produced by
an interaction with the Large Magellanic Cloud (see Garc\'{\i}a-Ruiz
et al. 2002a). Hunter \& Toomre (1969) and Dekel \& Shlosman (1983)
proposed that the Galactic warp is a result of oscillations in the
disc, either triggered by an interaction with another galaxy in the
past or caused by the disc being embedded in a nonspherical halo of dark
matter. Sparke \& Casertano (1988) showed that some discrete
oscillation modes can survive. This theory is not without problems,
however; for instance, Binney (1991) showed that the halo should then
respond to its misalignment with the disc, thus destroying the warp in
a few orbital periods. Alternatively, a nonlinear coupling between a
spiral wave and two warp waves could create a warp (Masset \& Tagger
1997). The efficiency of this mechanism is too low in the stellar disc
except at its outer edge (near a Lindblad resonance), where the spiral
wave slows down and is coupled efficiently to the warp waves. At those
radii, the energy of the spiral arm can almost be completely converted
into transmitted and reflected warp waves, which can be observed as
corrugations in the disc.

Warping could thus be a natural response of the outer disc to a series
of stimulations, and it seems that the responsible mechanism does not
necessarily have to be the same in each galaxy. In any case,
observations of large edge-on galaxies could contribute to the
definitive determination of the dominant scenario. The intergalactic
medium could also play an important role by accretion of matter as
galaxies move along their paths (Kahn \& Woltjer 1959;
L\'opez-Corredoira et al. 2002; S\'anchez-Salcedo 2006). Battaner et
al. (1990) explained warps by intergalactic magnetic fields,
which would produce a direct distortion in the distribution of the
gas. The stellar system would also indirectly reach a warped
distribution because stars are formed from gas, but this would result
in a difference between young and old stars. Colour gradients in warps
can thus provide an important clue. Based on
this model, gas-poor S0 galaxies should not exhibit warps.

In this paper, we present new, deep NIR and optical observations of 20
edge-on galaxies. NIR wavelengths have rarely been used to date to
study warps. They are, however, essential for better tracing the stellar
mass distribution, because at infrared (IR) wavelengths ($1 < \lambda <
5 \mu{\rm{m}}$) dust extinction is minimised (the absorption, in
magnitudes, in the $K$ filter is 10\% of that in the $V$ band,
$A_K/A_V \sim 0.1$). NIR emission is also a better tracer of old
stellar populations, one of our major
objectives in this study. Therefore, these wavelengths provide essential
complementary information for studying stellar warps.

\section{Sample}

Our sample was selected by de Grijs (1997, 1998) from the Surface
Photometry Catalogue of the ESO-Uppsala Galaxies (ESO-LV; Lauberts \&
Valentijn 1989). This catalogue contains large numbers of galaxies
that have been selected and parameterised uniformly. The galaxies were
originally selected on ESO-Schmidt survey plates covering declinations
$\delta \le -17.5^\circ$, excluding the area within $15^\circ$ of the
Galactic Equator, at Galactic latitudes $|b| < 15^\circ$. We selected
20 candidate galaxies from a statistically complete diameter-limited
subsample of edge-on disc galaxies (de Grijs 1997, 1998), with the
following characteristics (see Table 1):
 
\begin{itemize}
\item $D_{25} (B) \geq 2.2'$, i.e., with optical diameters (at a
surface brightness level of $\mu_{B} = 25$ mag arcsec$^{-2}$) larger
than $2.2'$;
\item Isolated galaxies, i.e., without significant neighbours within a
distance greater than or equal to five times their optical
diameters. They should be classified as noninteracting and
undisturbed;
\item Inclination, $i > 87^\circ$;
\item Morphological type range from S0 to Sd, i.e., revised Hubble types
later than T $= -2$ (spiral and lenticular galaxies).
\end{itemize}

The inclinations were determined assuming an intrinsic flattening
$(b/a)_{0} = 0.11$ (de Grijs 1997, 1998), so that our sample contains
highly inclined galaxies in which warps should be easy to
detect. This could introduce a bias against early types or galaxies with prominent bulges that should be taken into account in the discussion of the warp phenomena. Using this approach, we try to avoid the common confusion
between spiral arms, corrugations, and warps.

\begin{table*}
\centering
\caption[ ]{Global galaxy parameters from the ESO-LV Catalogue (Lauberts \& Valentijn 1989).}
\begin{flushleft}
\begin{tabular}{|l|c|c|r|c|c|c|c|c|r|} \hline
  Galaxy &  RA(J2000.0)  &  Dec(J2000.0)   &  Hubble &  $a/b$       &  $D_{25} (B)$  &  
$B$ mag  &  $B-R$ (tot)  & Radial velocity &  PA (deg.)     \\
         &  (hh:mm:ss.s) & (dd:mm:ss)      &  type   & (axis ratio) &  (arcsec)    &  
(total)  &  (mag)        & (km s$^{-1}$)     &  (N$\rightarrow$E)    \\
\hline
\hline
  ESO026-G06  & 20:48:28.3 & $-$78:04:09 & 6.7$\pm$1 & 9.3 & 134.9 &  
15.48$\pm$0.30 & +1.37 & 2323$^{1)}$ & 76.0 \\
  ESO033-G22  & 05:31:41.8 & $-$73:44:58 & 6.8$\pm$1 & 12.5 & 131.8 &
15.56$\pm$0.19 & +0.86 & 3932$^{1)}$ & 170.0 \\
  ESO142-G24  & 19:35:42.3 & $-$57:31:10 & 6.6$\pm$1 & 8.4 & 237.1 &
14.03$\pm$0.24 & +1.30 & 2027$^{1)}$ &  6.0 \\
  ESO157-G18  & 04:17:54.4 & $-$55:56:04 & 6.5$\pm$1 & 6.6 & 186.2 &
13.90$\pm$0.48 & +0.96 & 1139 & 18.0 \\
  ESO201-G22  & 04:09:00.4 & $-$48:43:36 & 4.8$\pm$1 & 7.3 & 151.4 &
14.73$\pm$0.32 & $\cdots$ & 4014 & 59.0 \\
  ESO202-G35  & 04:32:15.6 & $-$49:40:31 & 3.4$\pm$1 & 6.2 & 184.1 &
13.38$\pm$0.31 & +1.06 & 1856 & 133.0 \\
  ESO235-G53  & 21:05:10.4 & $-$47:47:17 & 3.0$\pm$0 & 6.0 & 153.1 &
14.53$\pm$0.26 & +1.85 & 5110 & 49.0 \\
  ESO240-G11  & 23:37:49.3 & $-$47:43:34 & 4.7$\pm$1 & 10.0 & 331.1 &
13.05$\pm$0.21 & +1.35 & 2817 & 129.0 \\
  ESO288-G25  & 21:59:17.6 & $-$43:52:02 & 4.2$\pm$1 & 6.5 & 153.1 &
13.97$\pm$0.21 & +1.30 & 2484 & 53.0 \\
  ESO311-G12  & 07:47:34.2 & $-$41:27:07 & 0.1$\pm$1 & 6.4 & 251.2 &
12.83$\pm$0.28 & +1.69 & 848$^{1)}$ & 14.0 \\
  ESO340-G08  & 20:17:11.0 & $-$40:55:22 & 6.0$\pm$1 & 24.0 & 192.8 &
15.45$\pm$0.20 & +0.95 & 3004$^{1)}$ & 34.0 \\
  ESO340-G09  & 20:17:20.2 & $-$38:40:29 & 7.0$\pm$1 & 9.0 & 160.3 &
14.52$\pm$0.27 & $\cdots$ & 2546 & 98.0 \\
  ESO358-G26  & 03:35:30.8 & $-$34:26:49 & $-$1.6$\pm$2 & 4.2 & 153.1 &
12.90$\pm$0.40 & +1.19 & 1660 & 84.0 \\
  ESO358-G29  & 03:36:31.4 & $-$35:17:38 & $-$2.0$\pm$1 & 3.9 & 153.1 &
12.47$\pm$0.18 & +1.45 & 1740 & 139.2 \\
  ESO416-G25  & 02:48:41.3 & $-$31:32:09 & 3.2$\pm$1 & 6.0 & 141.3 &
14.68$\pm$0.25 & +1.38 & 4992 & 31.0 \\
  ESO460-G31  & 19:44:21.2 & $-$27:24:22 & 4.8$\pm$1 & 9.5 & 154.9 &
15.15$\pm$0.21 & $\cdots$ & 5352$^{1)}$ & 92.0 \\
  ESO487-G02  & 05:21:48.6 & $-$23:48:45 & 3.9$\pm$1 & 7.5 & 199.5 &
13.62$\pm$0.26 & +1.35 & 2088$^{1)}$ & 60.0 \\
  ESO531-G22  & 21:40:29.5 & $-$26:31:40 & 4.5$\pm$1 & 8.7 & 154.2 &
13.94$\pm$0.35 & $\cdots$ & 3639$^{1)}$ & 8.0 \\
  ESO555-G36  & 06:07:41.5 & $-$19:54:54 & 5.5$\pm$1 & 11.5 & 139.6 &
15.60$\pm$0.58 & +1.62 & $\cdots$ & 146.0 \\
  ESO564-G27  & 09:11:54.3 & $-$20:07:04 & 6.1$\pm$1 & 14.0 & 278.6 &
14.39$\pm$0.90 & +1.36 & 2178 & 168.0 \\
\hline
\end{tabular}
$^{1)}$ Heliocentric radial velocities from Mathewson et al. (1992).
\end{flushleft}
\end{table*}

\section{Observations and data reduction}

The sample was observed at

\begin{itemize}
\item \textbf{Near-infrared} wavelengths, with the \textit{Cerro
Tololo Infrared IMager} (CIRIM) at the 1.5 m Ritchey-Chretien
Telescope of the Cerro Tololo Inter-American Observatory (CTIO) on October 22- -27, 1998 (Prop0020). CIRIM
is equipped with a 256$\times$256-pixel HgCdTe NICMOS3 array (Rockwell
Science) with a pixel scale of $1.16 \arcsec$. We
observed our sample galaxies through the $K_{\rm{short}}$ filter
($\equiv K_{\rm s}$), characterised by an effective wavelength
$\lambda_{\rm eff} = 2.15 \mu$m (Wainscoat \& Cowie 1992). We opted
for this filter instead of the longer-wavelength $K$ filter
($\lambda_{\rm eff} = 2.20 \mu$m) because the sky brightness in
$K_{\rm{s}}$ is lower than in the $K$ band (for instance, at
UKIRT, Hawaii, the sky brightness is 13.5 and 13.0 mag arcsec$^{-2}$
in $K_{\rm{s}}$ and $K$, respectively\footnote{\tt
http://www.jach.hawaii.edu/UKIRT/instruments/uist/\newline imaging/imaging.html}) and $K_{\rm{short}}$ is almost as little affected by
dust as the $K$ band, $A_{K_{\rm s}} \sim 0.11$ mag airmass$^{-1}$; and

\item \textbf{Optical} wavelengths. We obtained our observations using
two telescopes of the European Southern Observatory (ESO) at La Silla,
Chile, including the 1.54 m Danish telescope (equipped with a
1081$\times$1040-pixel TEK CCD with $0.36 \arcsec$ pixel$^{-1}$) and
the 0.92 m Dutch telescope (equipped with a 512$\times$512-pixel TEK
CCD with $0.44 \arcsec$ pixel$^{-1}$), both in the standard Johnson
$B$ and $V$ and Thuan \& Gunn (1976) $i$ filters.
\end{itemize}

The observations and optical data-reduction techniques on which the
results presented in this paper are based were described in detail in
de Grijs (1998). Here, we focus on the reduction of the IR data.

\begin{figure*}
\centering
\includegraphics[width=17cm]{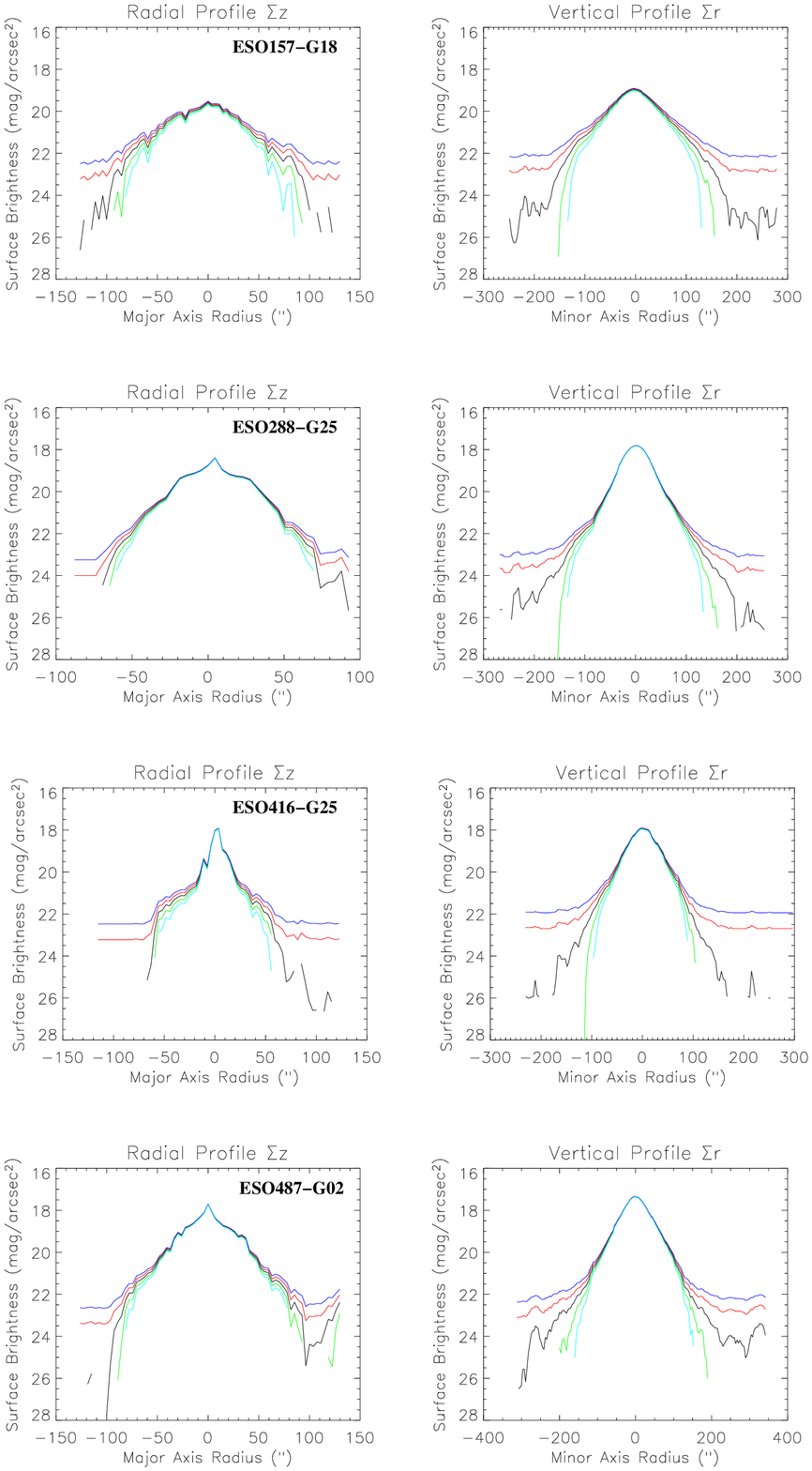}
\caption{$K_{\rm{s}}$-band luminosity profiles of four galaxies. The
lines in each panel represent, from top to bottom, $-2\sigma$ (blue),
$-1\sigma$ (red), profile after subtraction of our best estimate of
the sky background (black), $+1\sigma$ (green), and $+2\sigma$ (sky
blue). ($\sigma$ is the sky noise in the regions we used to determine
the sky-background levels.)}
\label{mag}
\end{figure*}

To study stellar warps in disc galaxies, we need to obtain reliable
data at the very low surface brightnesses in their outermost
regions. Our observational data reached surface brightnesses fainter
than 26 mag arcsec$^{-2}$ in the $B$ band. This is only 6\% of the sky
brightness ($\mu_{B,{\rm sky}}=23$ mag arcsec$^{-2}$), so that we had
to be particularly careful in measuring the sky brightness. The sky
background in NIR passbands is much higher than in the optical
($\mu_{K,{\rm sky}}= 13$ mag arcsec$^{-2}$) and can change
significantly on temporal and spatial scales the order of two
minutes and two arcmin, respectively. It was therefore necessary to
obtain as many sky frames as object frames to perform sky subtraction with
sufficiently high accuracy. We took sky and object frames alternately,
using the same exposure time for each individual frame (60 s) and
employing interframe shifts of $\sim2$ arcmin. An additional reason
we needed to obtain separate sky frames is that the IR
arrays used were not large enough to properly sample the sky on the
science frames. Incorrect background subtraction can cause the sky
level to be either underestimated (so that sky flux will contribute to
the surface brightness at large galactic radii) or overestimated
(resulting in a fictitious radial cut-off). Figure~\ref{mag}
illustrates the importance of careful background subtraction in the
$K_{\rm{s}}$ band, where the sky contribution is greater than in
optical passbands. The figure shows the major- and the minor-axis
(vertical, $z$) surface brightness profiles of four galaxies, binned
radially. The effects of oversubtraction can be seen clearly in the
minor-axis surface brightness profiles: they show artificial cut-offs, 
and the negative background values result in undefined surface
brightnesses at these $z$ heights.

We used a special software package for our NIR data reduction, customised to reduce and analyse very deep data of extended objects
observed in the IR, as a set of \textsc{iraf}\footnote{\textsc{iraf}
is distributed by the National Optical Astronomy Observatories, which
are operated by the Association of Universities for Research in
Astronomy, Inc., under cooperative agreement with the US National
Science Foundation.} tasks. The program removes spurious stars in sky
frames, subtracts cleaned, scaled sky frames from object frames taken
immediately prior to or after the relevant object frame, and
flatfields the differences. It uses a median sky frame to detect and
remove stars in the sky frames.

We observed several standard stars from the SAAO/ESO/ISO Faint
Standard Stars Catalogue (Carter \& Meadows 1995) to calibrate our
targets and used the Wainscoat \& Cowie (1992) correction to convert
$K_{\rm s}$ to $K$ magnitudes. These stars were observed at least
three times per night, at different airmasses, to calculate the
prevailing atmospheric extinction with sufficient accuracy. We
observed a total of 21 standard stars. These observations reached a
surface brightness of approximately 23 mag arcsec$^{-2}$ in $K_{\rm
s}$.

\section{Data analysis}

First, we aligned the images in all of the $B, V, I$, and $K_{\rm s}$
filters. We took special care to subtract foreground stars. This is
necessary and important because foreground stars projected close to a
galaxy can contaminate the galaxy's luminosity distribution and lead
to erroneous results. We used a master mask constructed by adding the
individual masks for each of the passbands. In each mask, the stars
were marked with circles of radius around twice the FWHM to avoid
contributions from any residual starlight. In some galaxies observed
near the Galactic plane, this meant that a large number of pixels had
to be masked.

To determine the warp curve (tracing the position of the intensity
maxima as a function of radius), we employed three different methods.
\begin{enumerate}
\item First, we calculated the intensity maxima using vertical cuts
(perpendicular to the galactic plane). This resulted in very noisy
profiles for the optical data and also sometimes for the NIR data,
because the presence of a dust lane frequently resulted in a
double-peaked luminosity distribution.
\item Second, we calculated the luminosity distribution's first
moment, with which we obtained reliable results (although it also
depended on the position of the dust lane). This method was applied
mainly at blue wavelengths to compensate for background stars.
\item Third, we fitted Gaussian profiles. We fitted each vertical
trace (parallel to the minor axis of the galaxy) in the selected zone
to a Gaussian function. We considered this the best approach, because
the contribution of the dust lane is less important in optical data
than when using the other two methods. Gaussians can be matched to
vertical profiles that contain a dip caused by a dust lane without
problems in almost perfectly edge-on galaxies.
\end{enumerate} 

\subsection{Warp curves from the Gaussian fit}

As a first step, we rotated our galaxies (to align their major axes
with the horizontal axis) using an iterative method until the position
angle (PA) had been determined to an accuracy of approximately 0.05
degrees. We used the PA from the ESO-LV Catalogue as a starting point;
we then iteratively fitted a straight line to the central region of
the galaxy. Our results are shown in Fig.~\ref{isofotas} as isophote
maps of our 20 galaxies.

\begin{figure*}
\resizebox{\hsize}{!}{\includegraphics{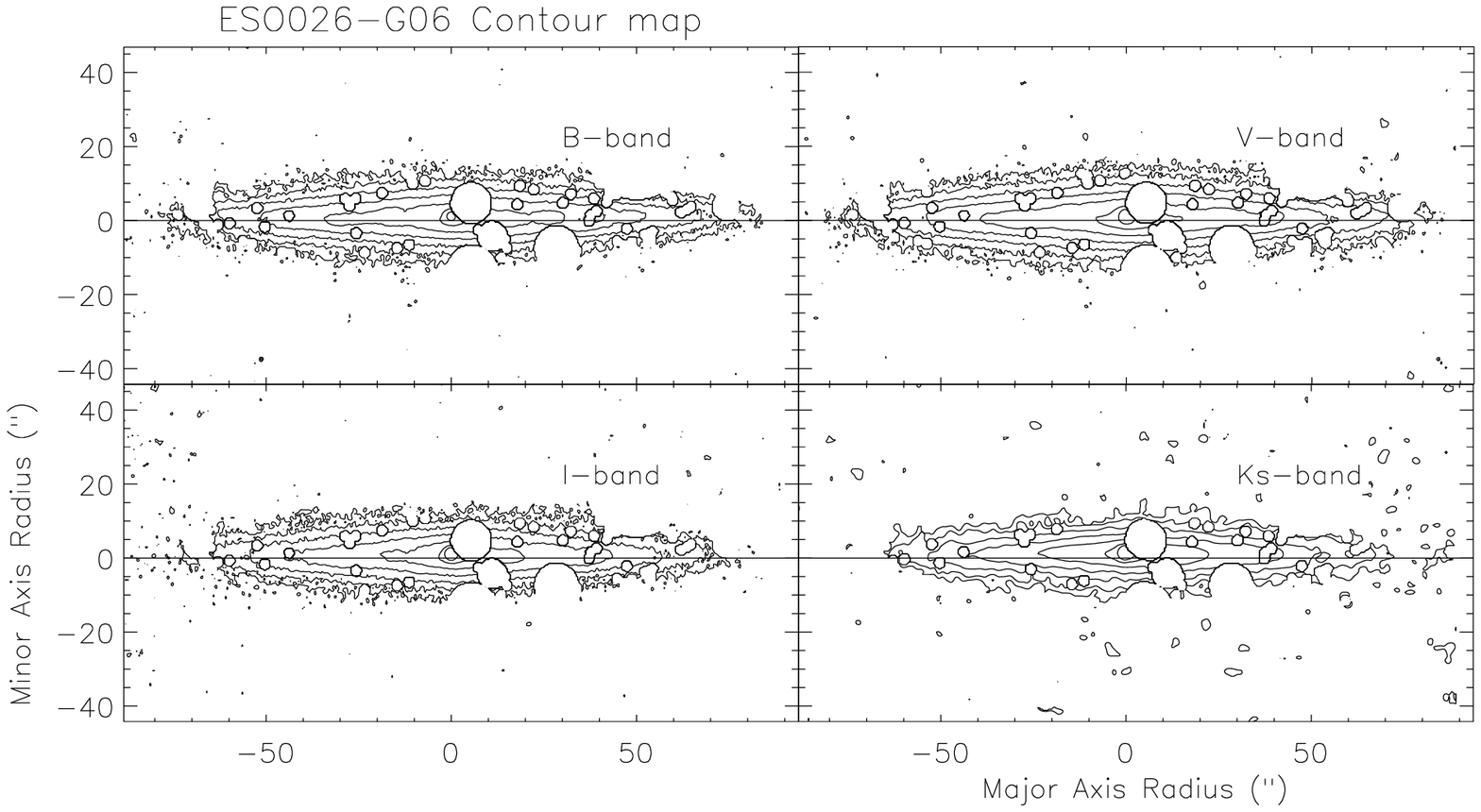}\includegraphics{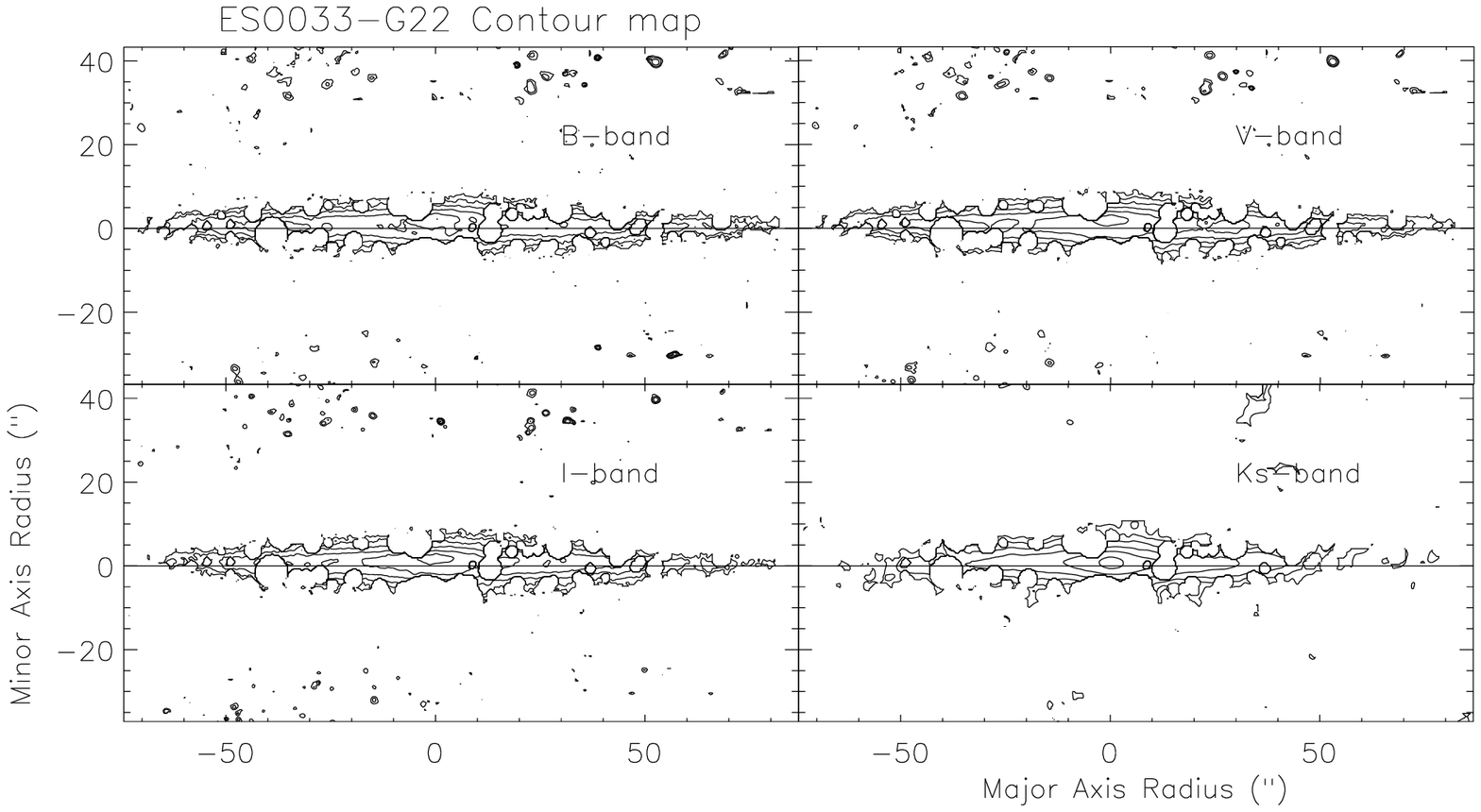}}
\resizebox{\hsize}{!}{\includegraphics{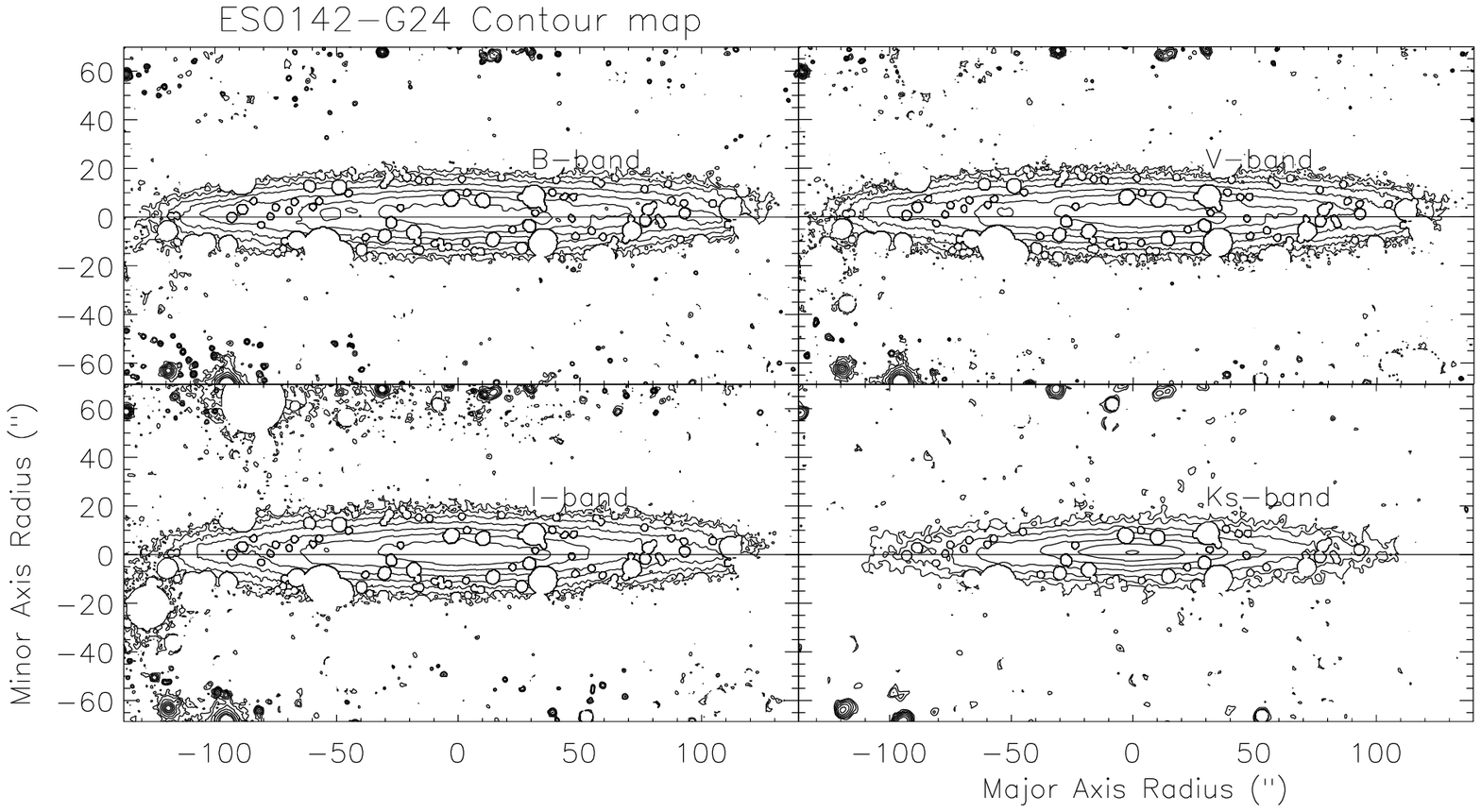}\includegraphics{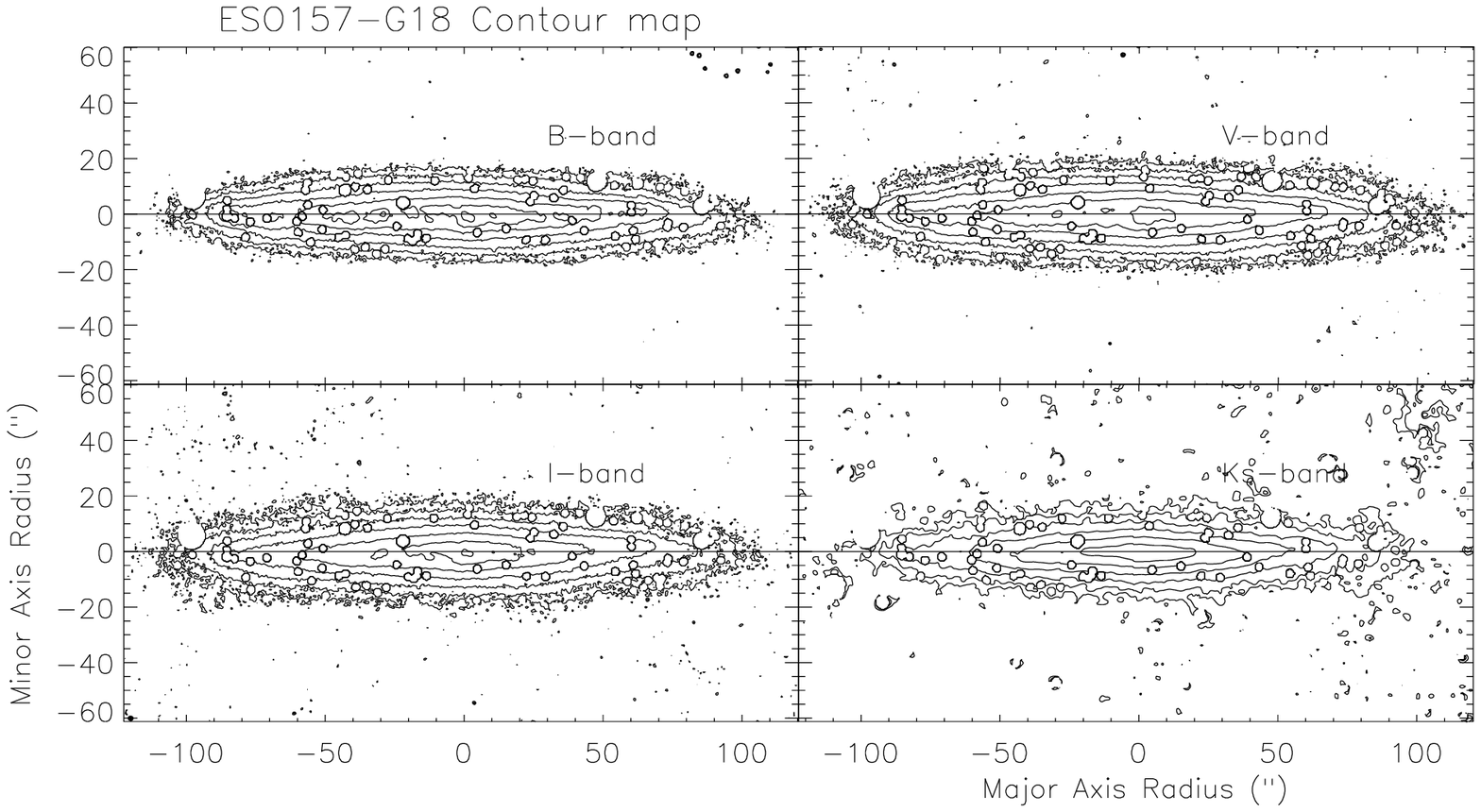}}
\resizebox{\hsize}{!}{\includegraphics{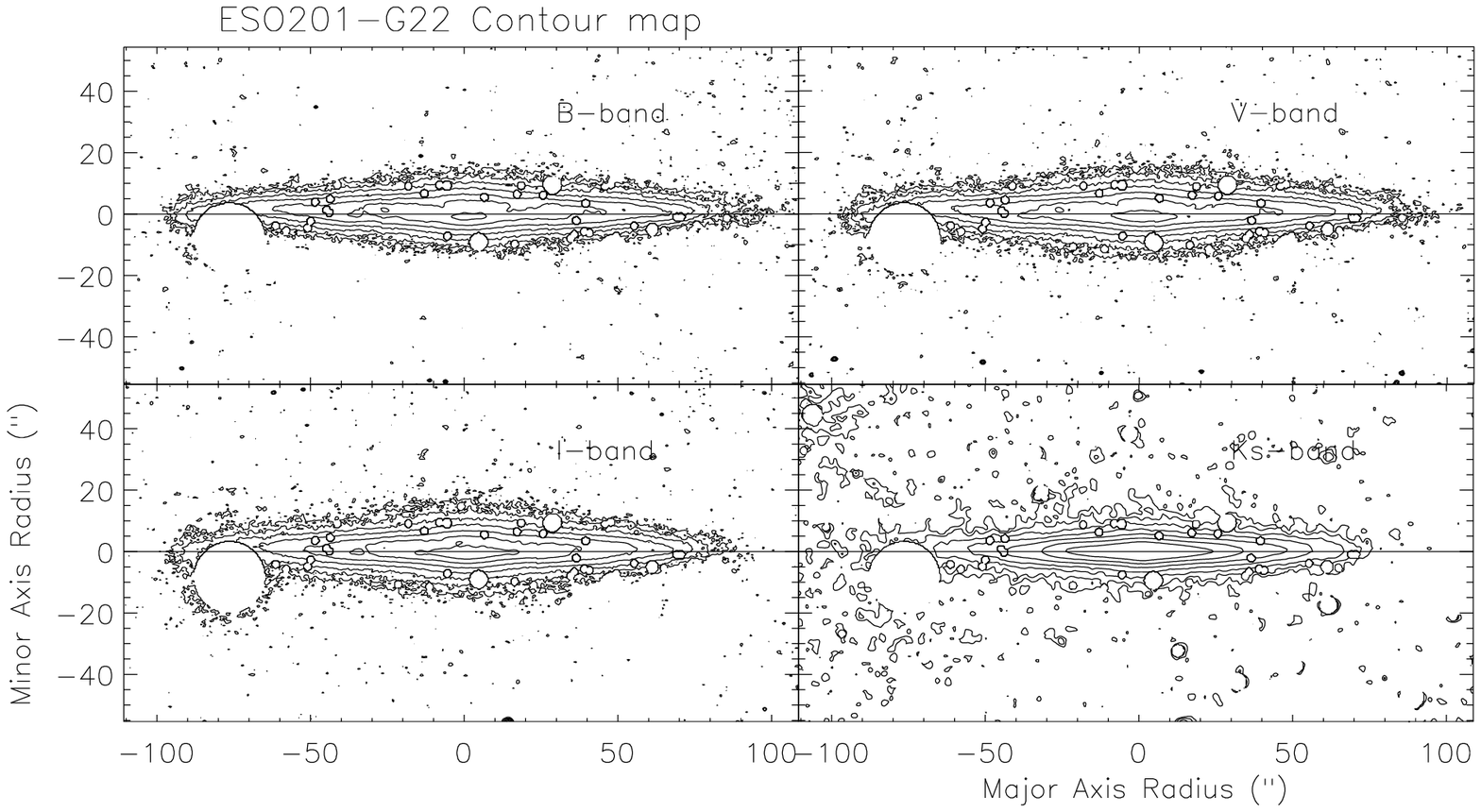}\includegraphics{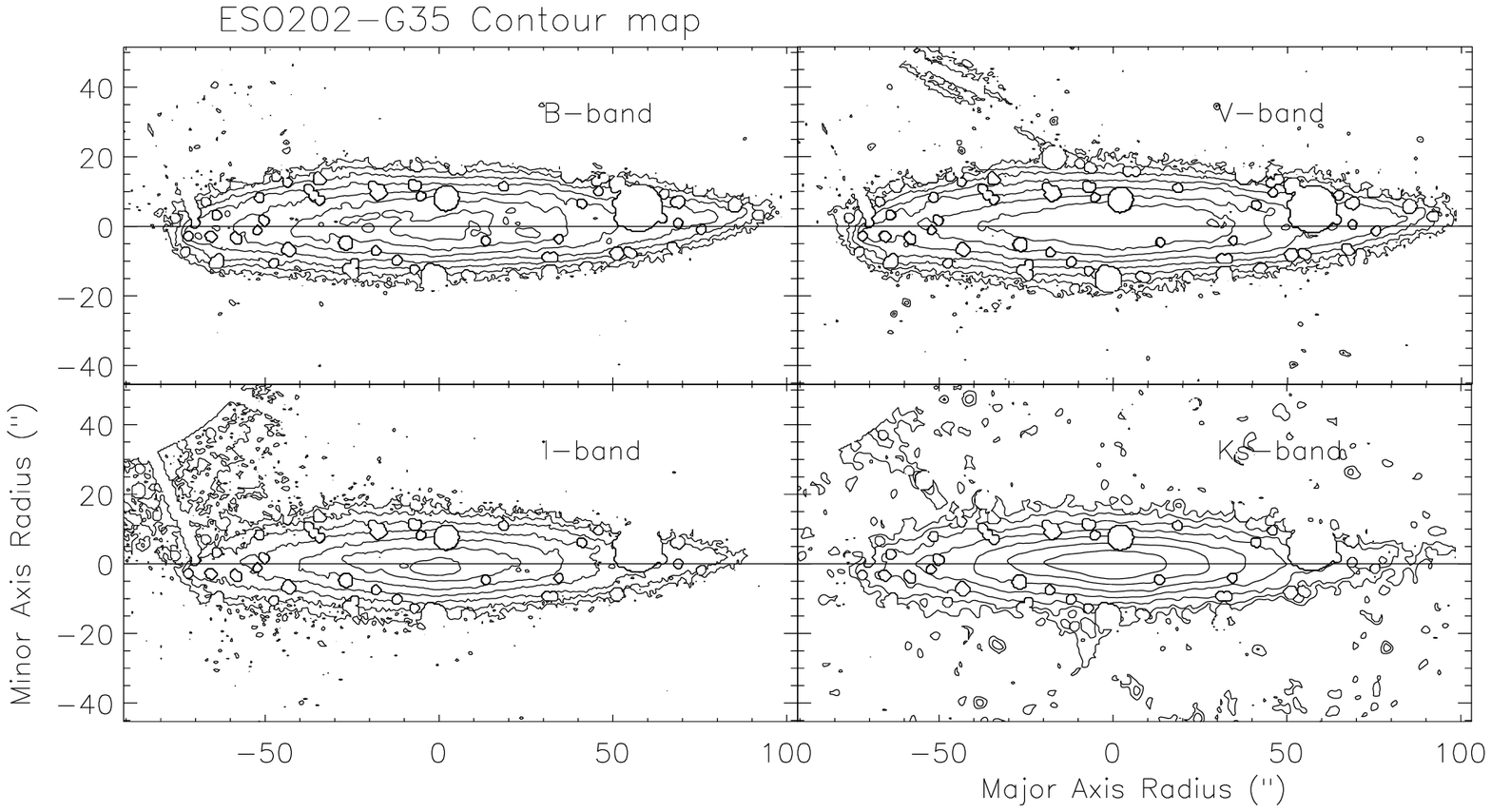}}
\resizebox{\hsize}{!}{\includegraphics{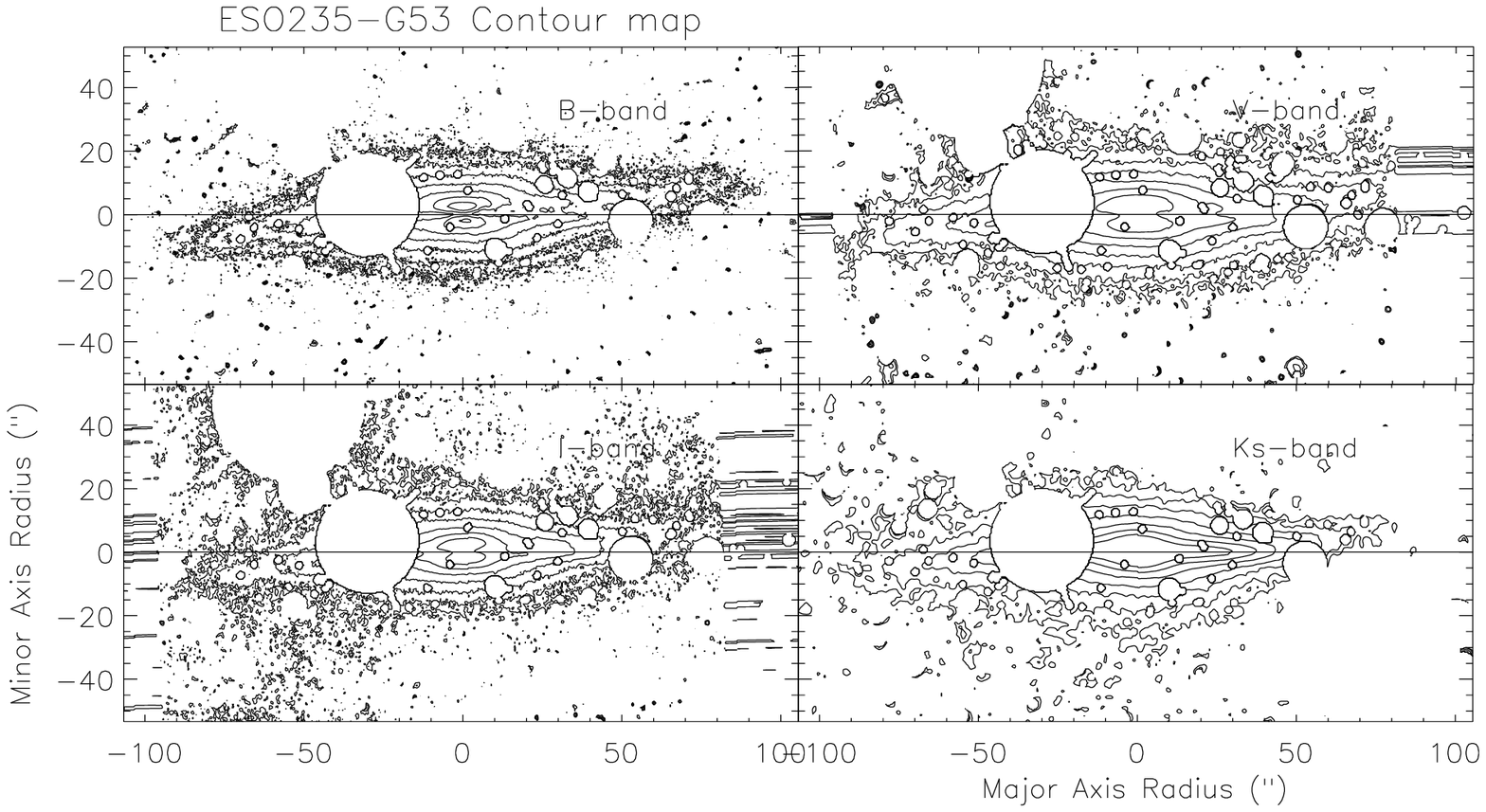}\includegraphics{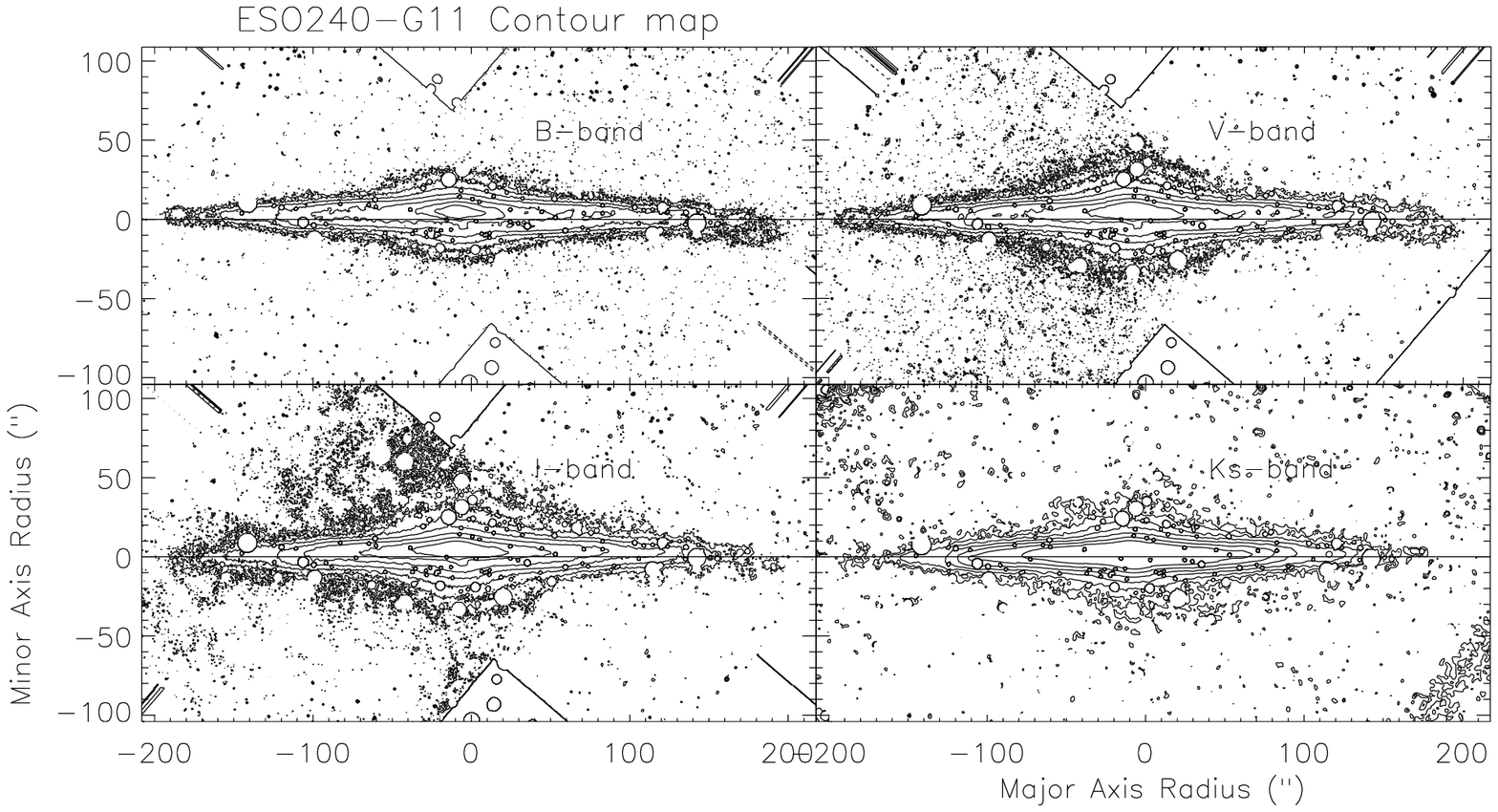}}
\resizebox{\hsize}{!}{\includegraphics{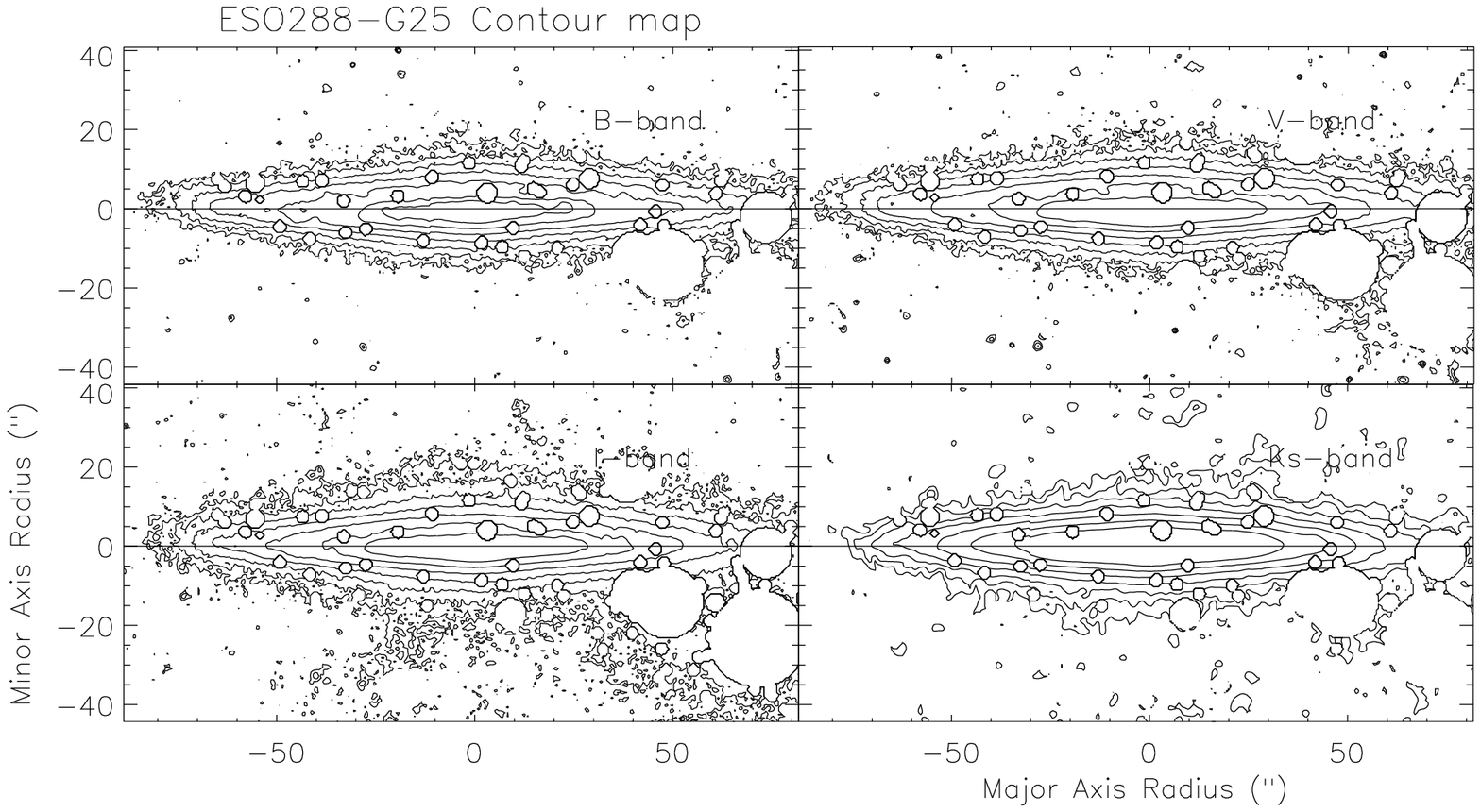}\includegraphics{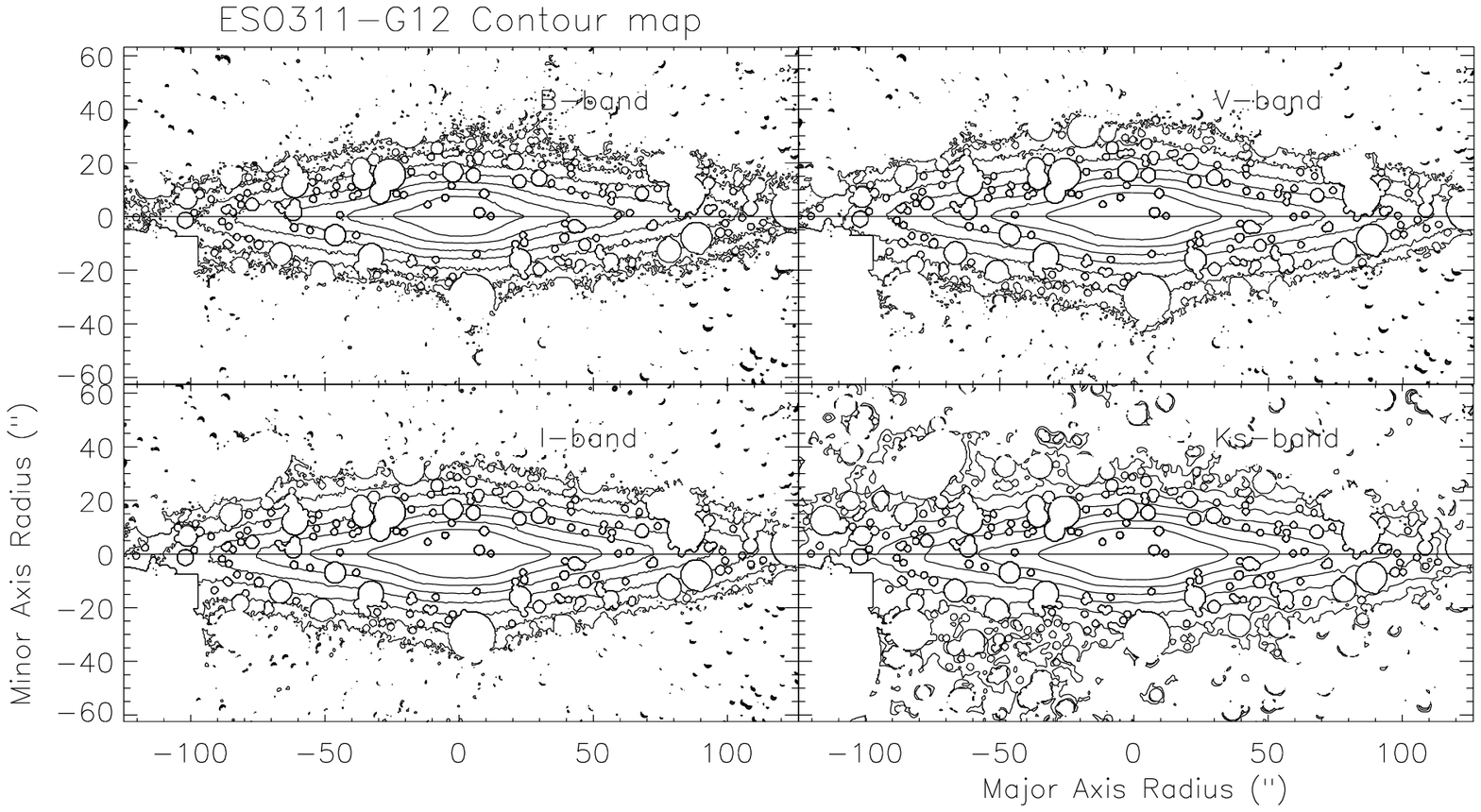}}
\caption{Contour maps of our 20 galaxies in the four filters
  analysed. The isophotes are equidistant (in units of $n \times
  3\sigma$ with $n = 1,2,3, \cdots$, equivalent to steps of 0.75 mag
  arcsec$^{-2}$), starting at $\sim 3\sigma$ above the sky
  background. The galaxies have been rotated, but N is close to the
  top, E to the left.}
\label{isofotas}
\end{figure*}

\begin{figure*}
\setcounter{figure}{1}
\resizebox{\hsize}{!}{\includegraphics{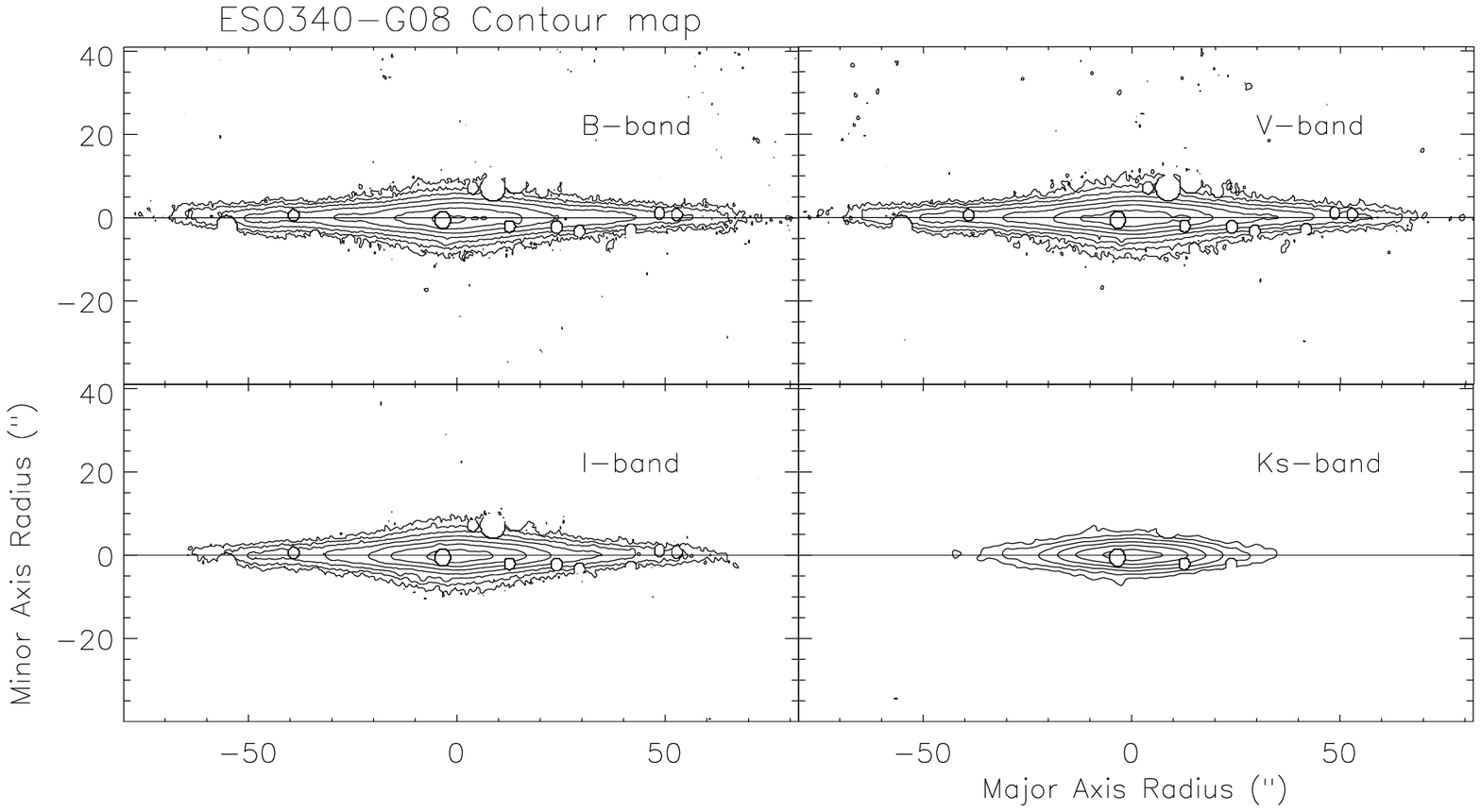}\includegraphics{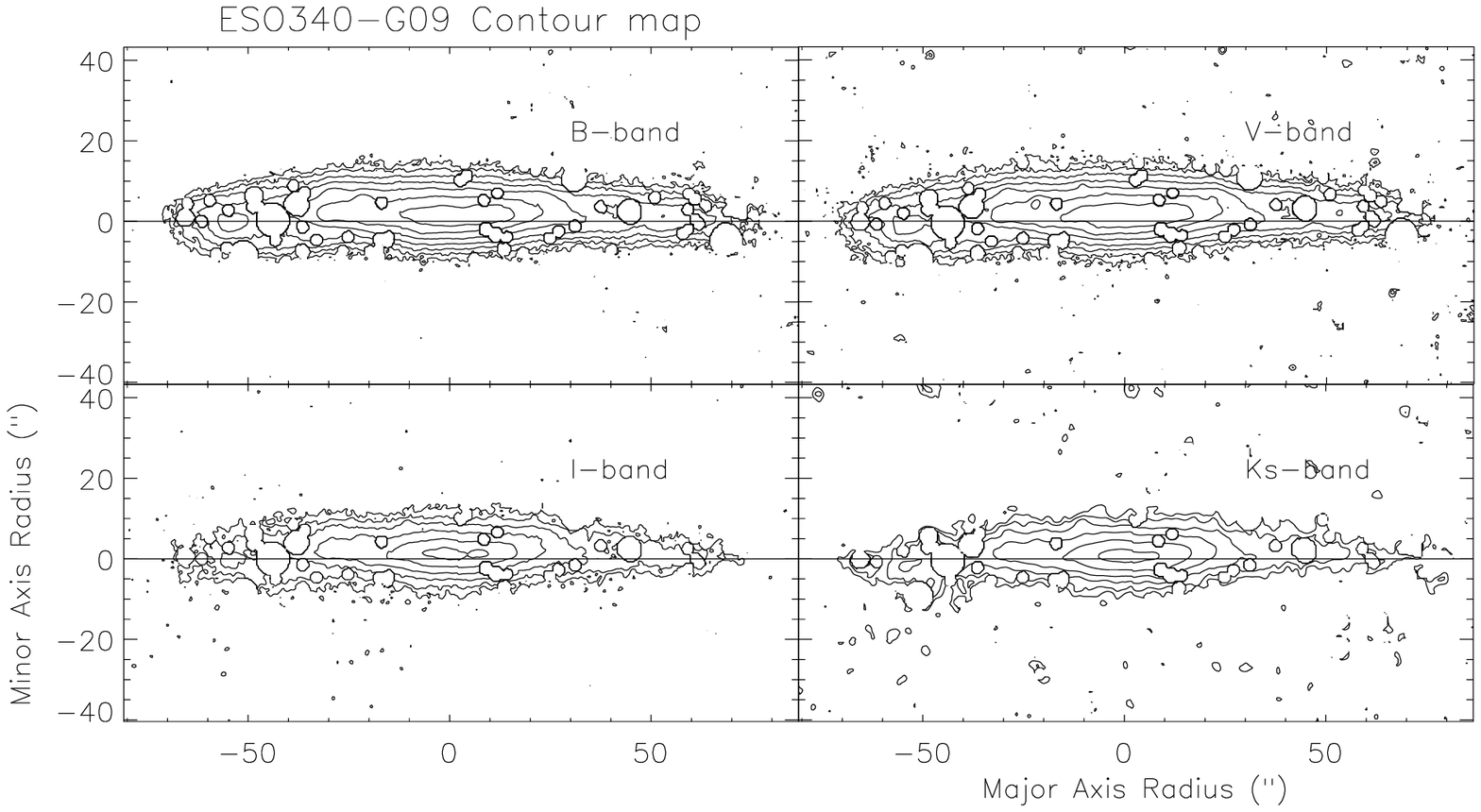}}
\resizebox{\hsize}{!}{\includegraphics{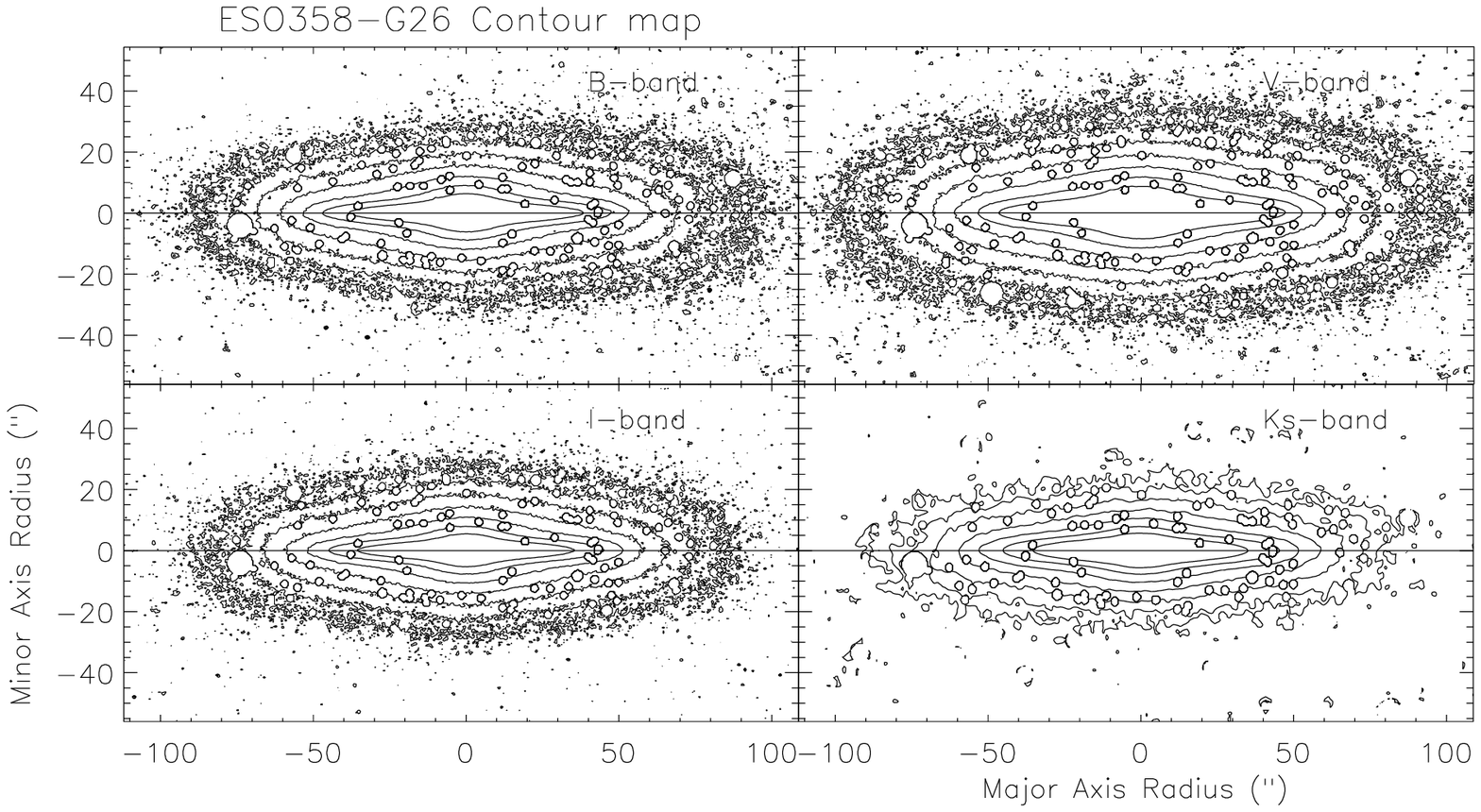}\includegraphics{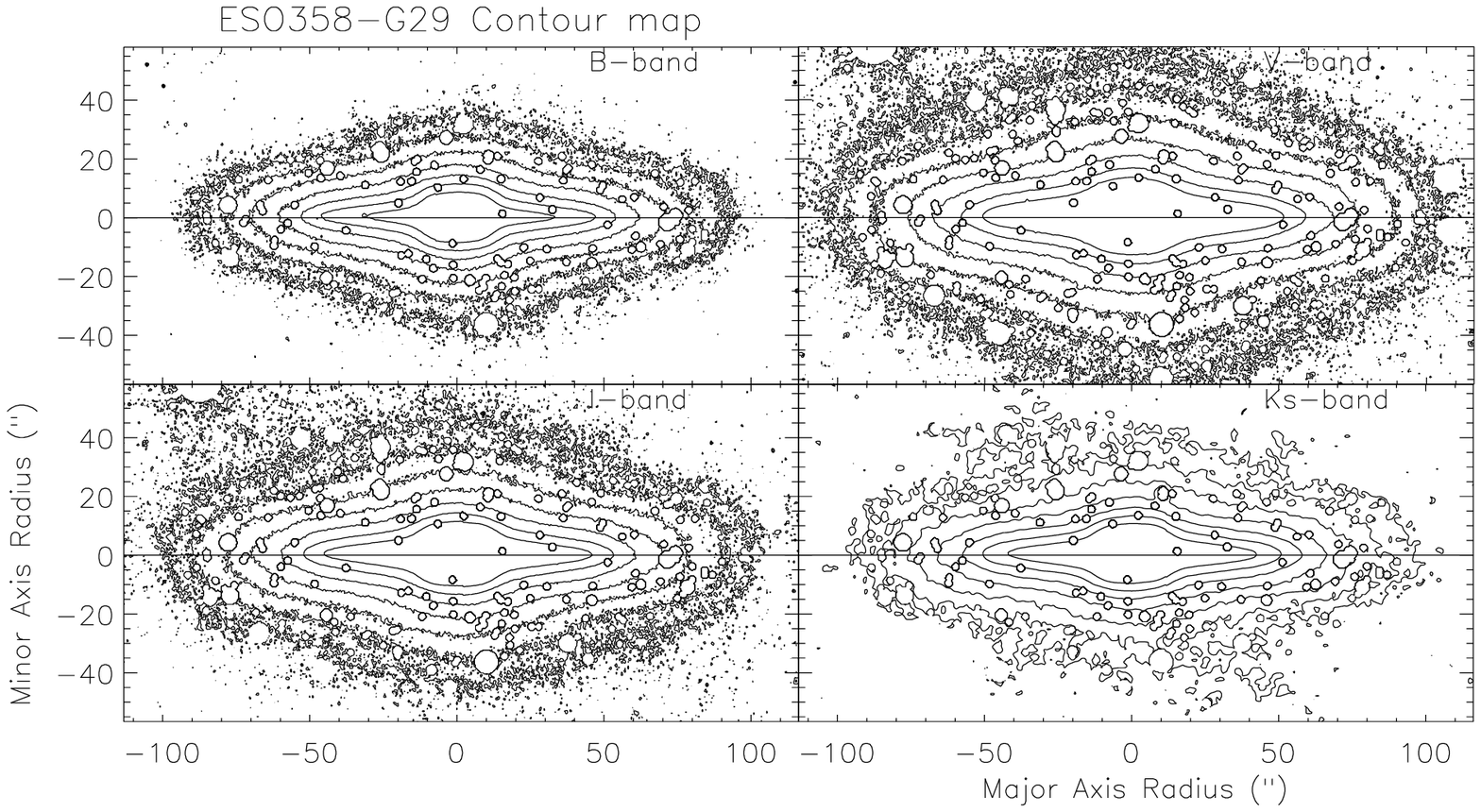}}
\resizebox{\hsize}{!}{\includegraphics{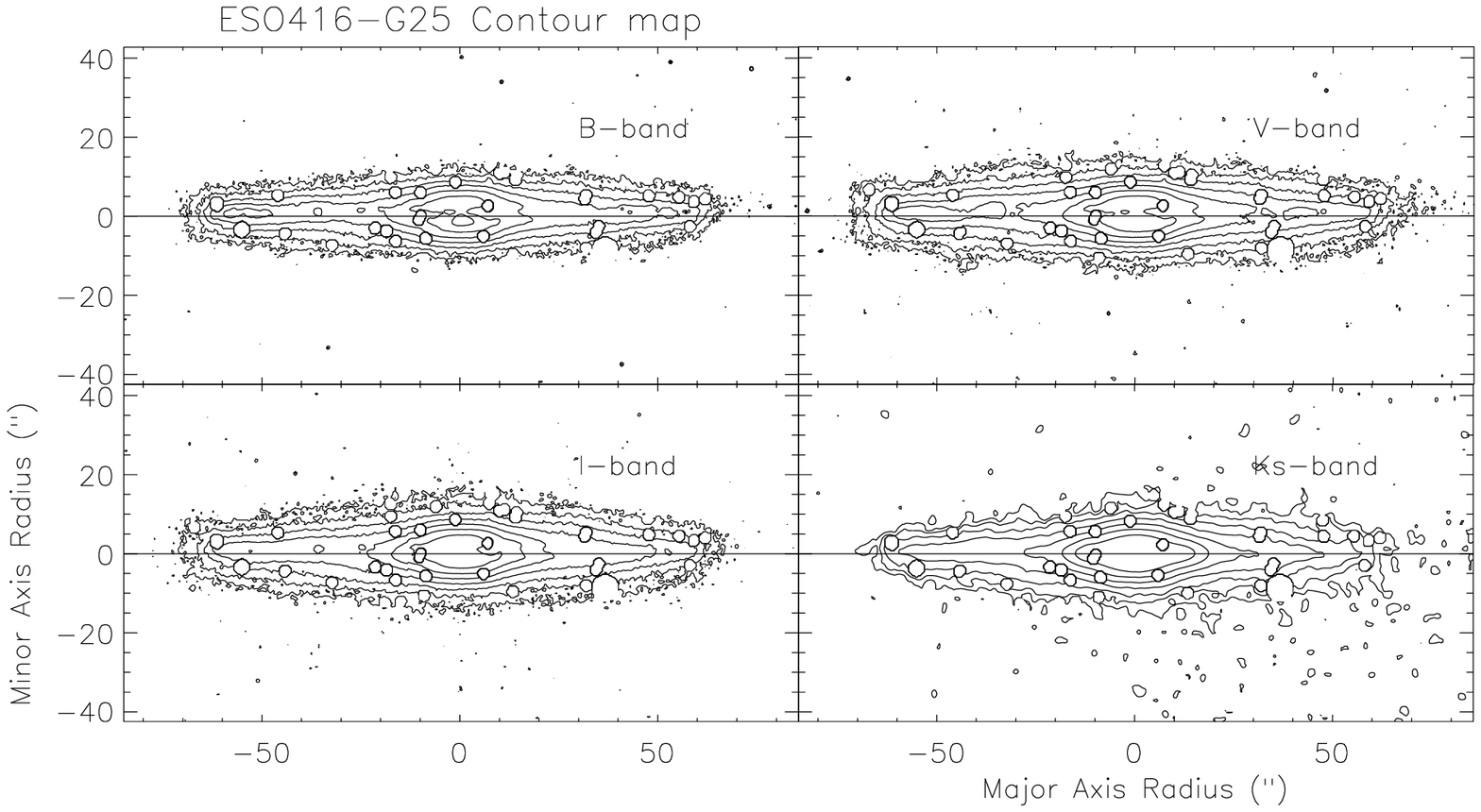}\includegraphics{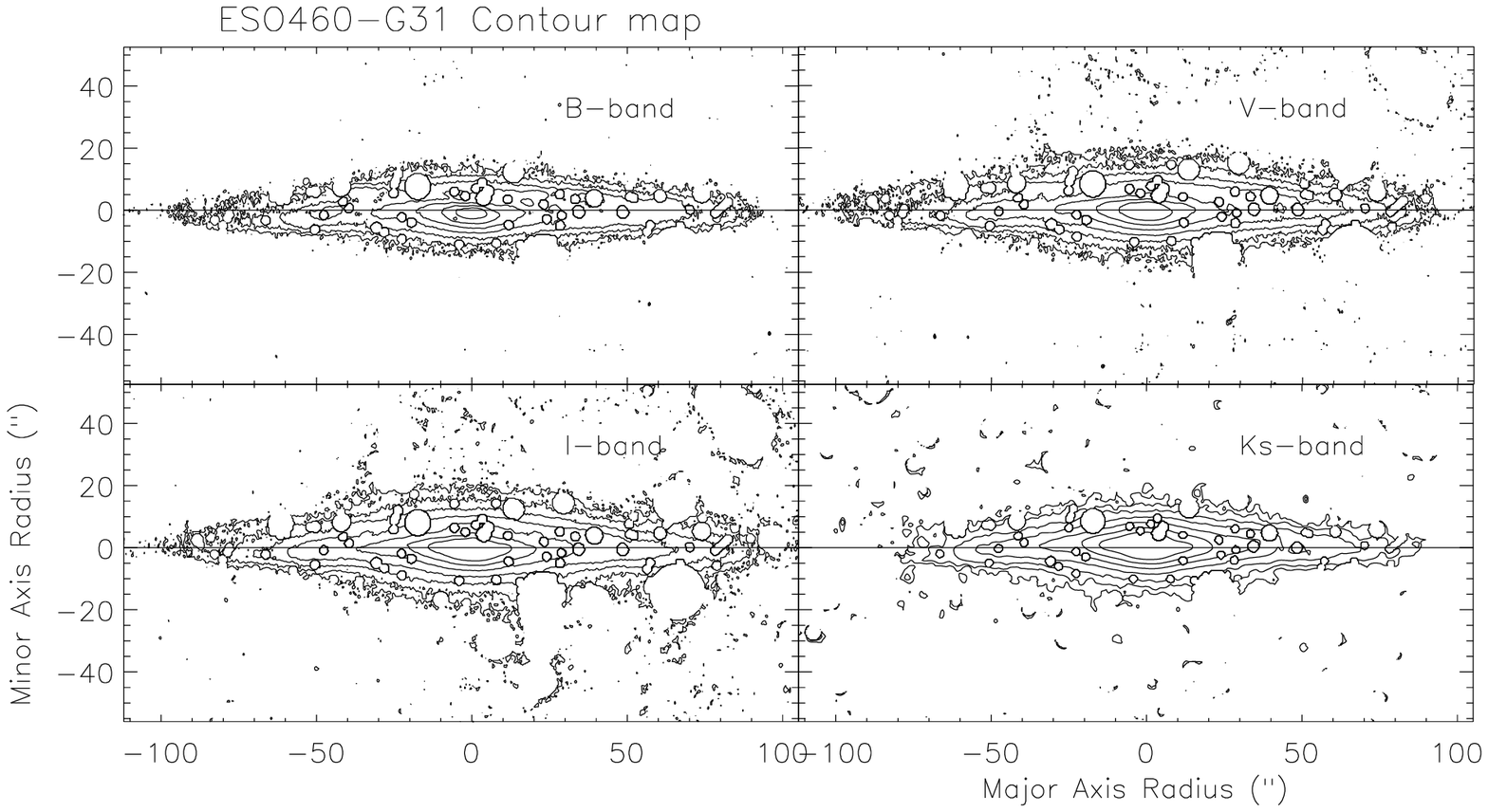}}
\resizebox{\hsize}{!}{\includegraphics{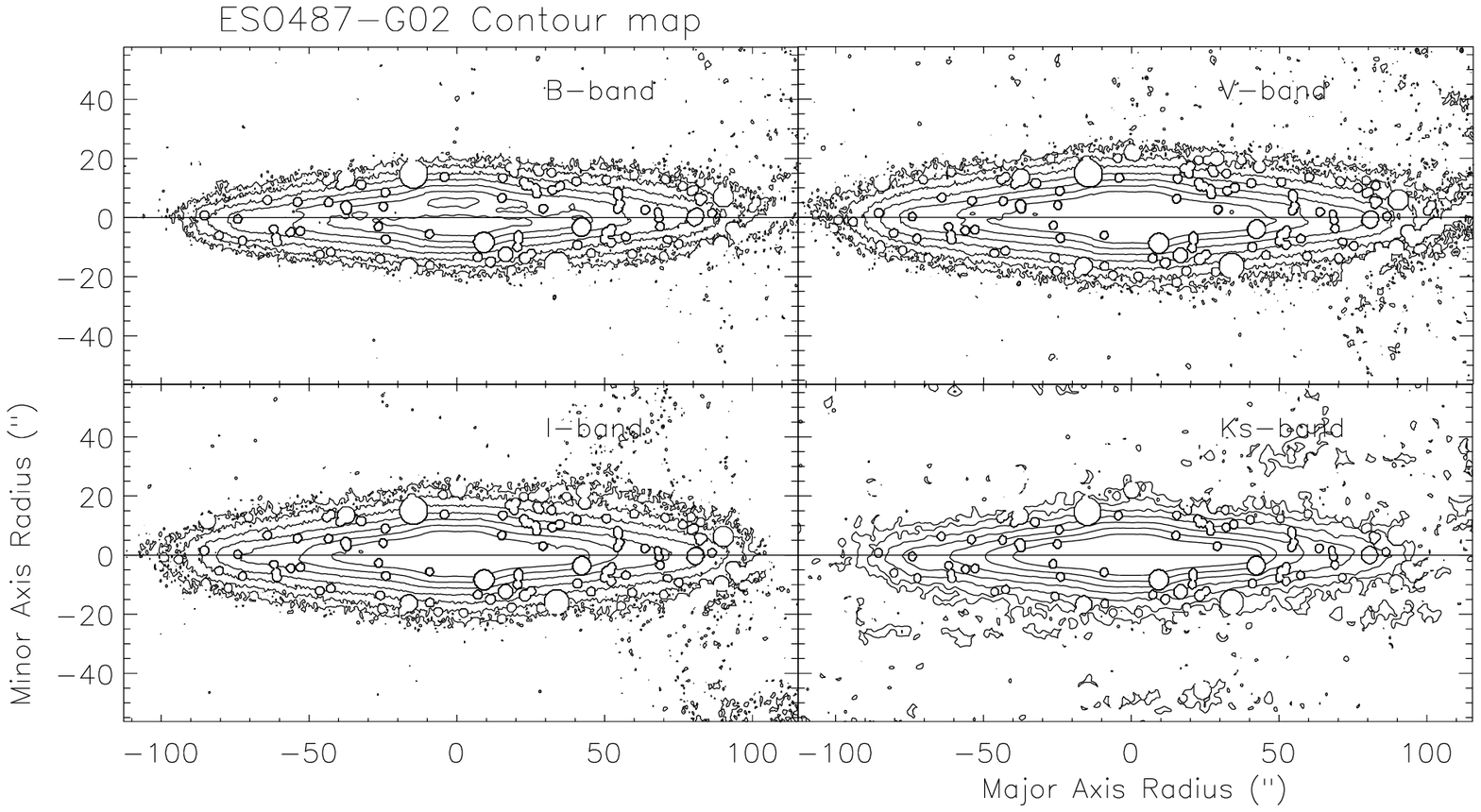}\includegraphics{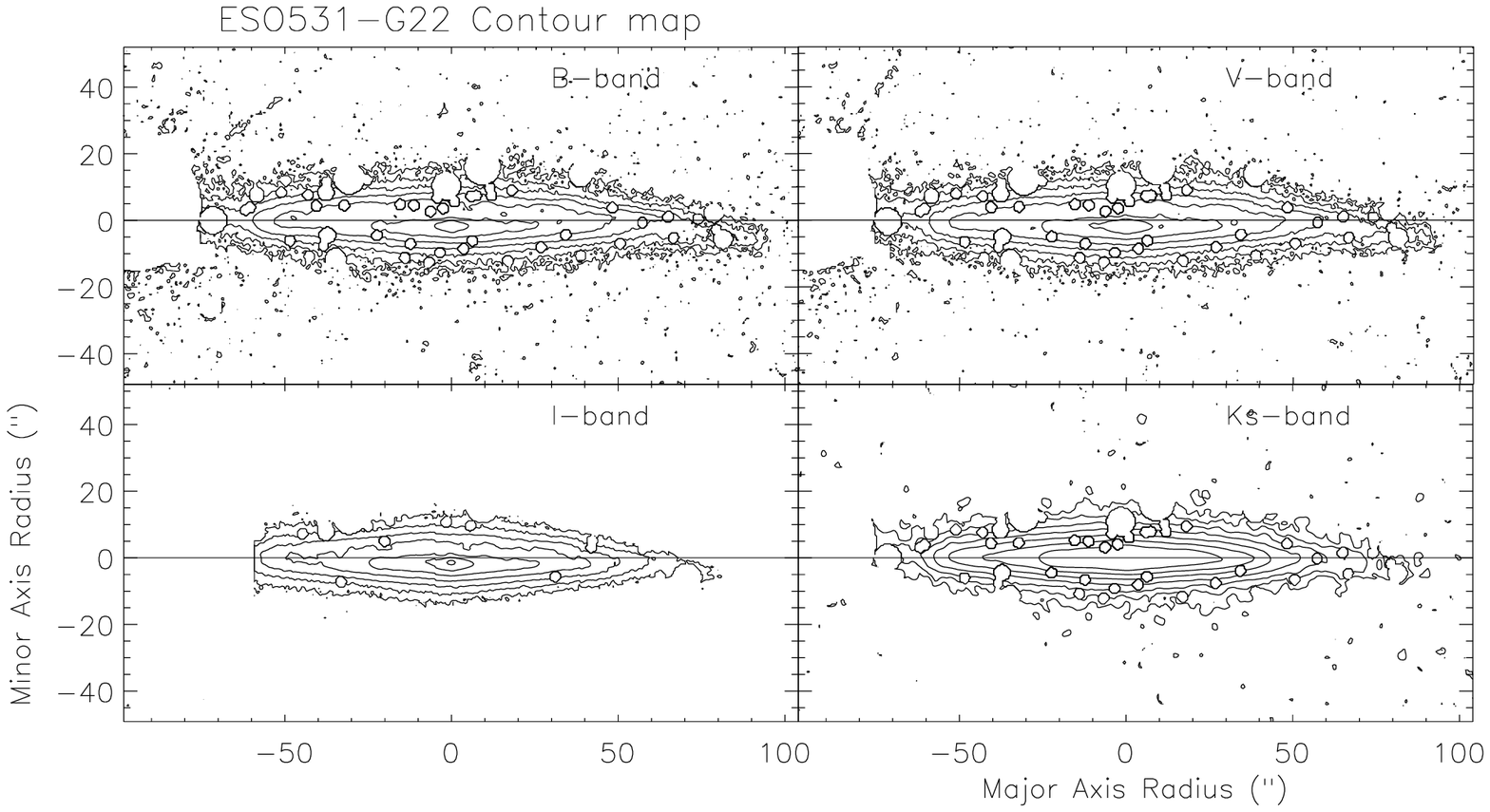}}
\resizebox{\hsize}{!}{\includegraphics{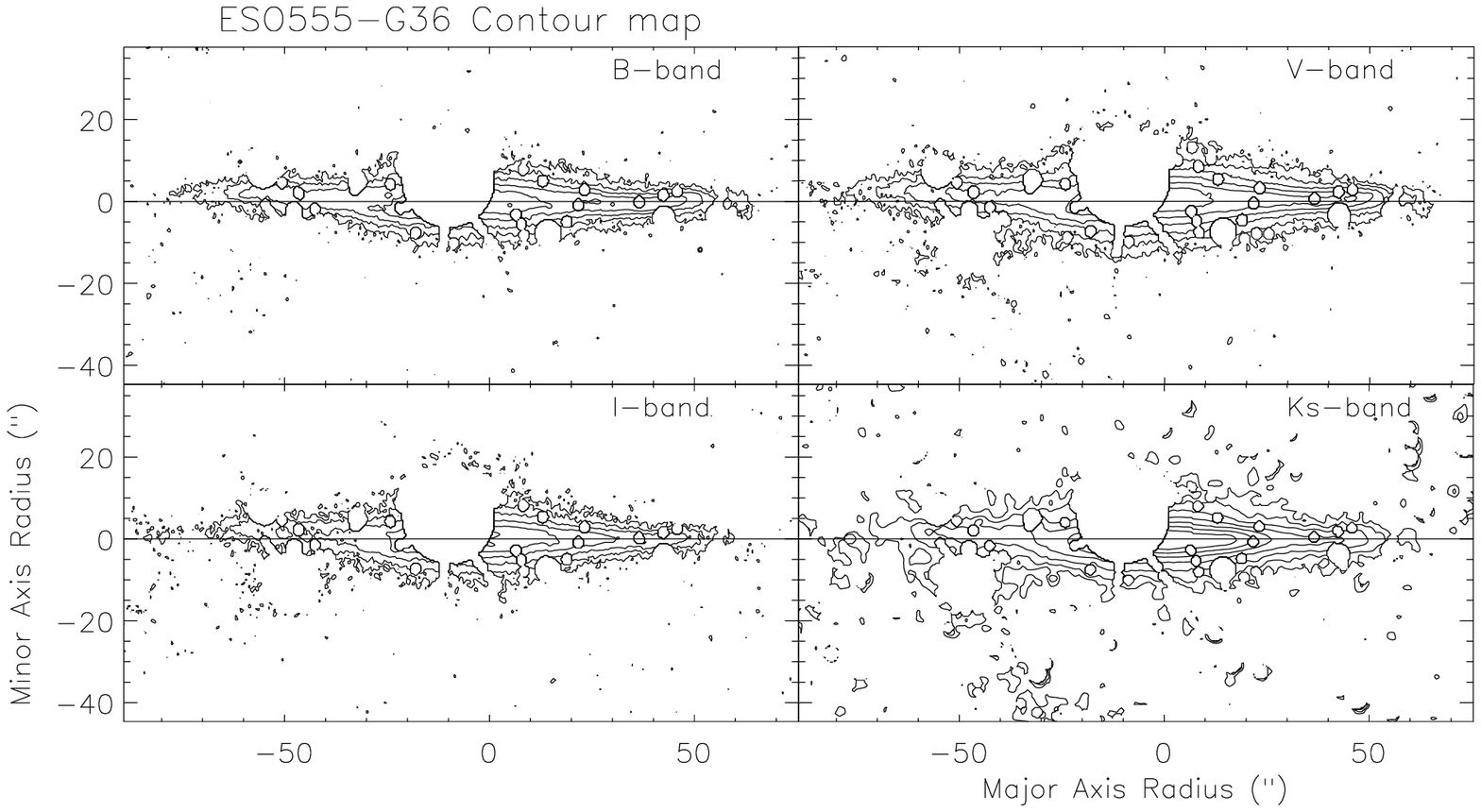}\includegraphics{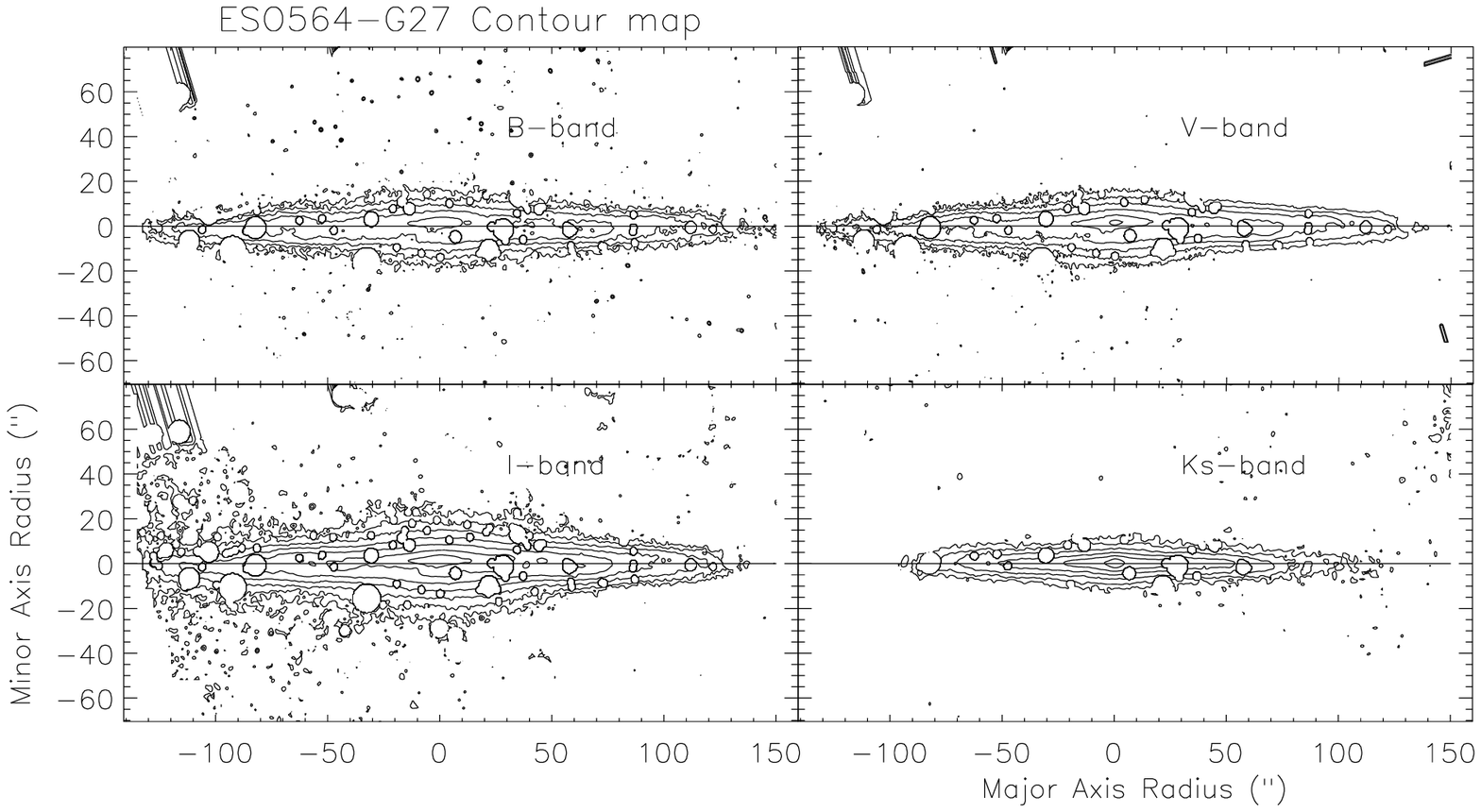}}
\caption{Continued.}
\end{figure*}

To calculate the warp curves, we fitted Gaussians to the vertical
profiles of the galaxies in each filter. We only considered data with
a signal-to-noise ratio greater than 3 ($\sigma$ =
$\sqrt{\sigma^{2}_{\rm{sys}} + \sigma^{2}_{\rm{std}}}$, where
$\sigma_{\rm{sys}}$ is the systematic error associated with the random
background noise and $\sigma_{\rm{std}}$ the standard deviation),
and with the FWHM of the peak smaller than 100 pixels. The results of
our analysis are presented in Fig.~\ref{alabeos}, where the warp
curves are shown only for those values with error bars smaller than
0.5$\arcsec$. This estimated error bar can be computed by scaling the
standard deviation ($1\sigma$ error) by the measured $\chi^2$ value.

\subsection{Warp parameters}
\label{warppar}

It is convenient to use a geometric definition of a warp, e.g., for
statistical studies. For an edge-on galaxy, the suitable coordinates correspond to the two directions contained in the plane of the sky: the direction defined by the major axis of the galaxy, $x$, and the direction of the rotation axis of the galaxy, $y$. The centre of the galaxy is the origin of coordinates. A warp curve is defined as the locus of points
($x_i, y_i$) tracing the distribution of the highest intensities,
where $y_i$ is the position of the centre of the best-fitting Gaussian
at a given $x_i$.

We derived the parameter $\omega$ (defined in Jim\'enez-Vicente et
al. 1997),
\begin{equation}
\omega=\frac{1}{L^3} \int_{-L/2}^{+L/2} x y {\rm d} x,
\end{equation}
where $L$ is the full size of the edge-on galaxy (the diameter at
3$\sigma$ above the sky background). The absolute value of $\omega$ is
a measurement of the strength of a warp, while its sign distinguishes
between N-like ($\omega > 0$) and S-like ($\omega < 0$) warps. These types are defined from the similarity of the warp shape and the type letters (see Fig.~\ref{dibu} for definition of types of warps). This
definition is nondimensional and, therefore, independent of a galaxy's
distance and size. It is also independent of the units used and does
not strongly depend on the angular resolution of the observations
(although a higher resolution would enable a more precise
evaluation). For numerical purposes, we have
\begin{equation}
\omega=\frac{1}{L^3} \sum_{i=-L/2}^{L/2} x_i y_i.
\end{equation}

We can also define this parameter for both sides of our galaxies:
$\omega_{\rm R}$ for the right-hand side $(^1 \! /_{L^3}
\sum_{i=0}^{L/2} x_i y_i)$ and $\omega_{\rm L}$ for the left-hand side
$(^1 \! /_{L^3} \sum_{i=-L/2}^{0} x_i y_i)$. Because of the flattening
of the radial profile in the inner regions of our sample galaxies, it
is not critical to have a precise estimate of the position of the
galaxy centre. Warps can appear in three flavours
(cf. L\'opez-Corredoira et al. 2008). The N- or S-like warps have
$\omega_{\rm R}$ and $\omega_{\rm L}$ of the same sign, the signs of
$\omega_{\rm R}$ and $\omega_{\rm L}$ for U-like warps differ, and
L-like warps apply to galaxies in which only one side is warped. The
difference between N- and S-like warps is unimportant from a physical
point of view. However, we have retained this information because it
is needed, for instance, when considering the orientation of warps in
a cluster of galaxies (see, e.g., Battaner et al. 1991).

With this parameter we can, for example, objectively compare the
appearance of a galaxy at several wavelengths to determine whether a colour gradient exists within the warp. We can also detect warps
that would otherwise not have been found. Warps are not always
completely symmetric, in which case this parameter would hide the
information of the clearly warped side of the galaxy. To avoid this
problem, we calculated the warp parameters on either side of our
galaxies independently, along with $\alpha_{\rm s}$, the degree of
asymmetry (S\'anchez-Saavedra et al. 2003),
\begin{equation}
\alpha_{\rm s} = \frac{| \omega_{\rm R} - \omega_{\rm L}
|}{\omega_{\rm R} + \omega_{\rm L}}.
\end{equation}

One of the most important errors in estimating $\omega$ arises from imperfectly determining the PA of the major axis, resulting in an
error $\theta$ in the PA. Since $\theta \neq 0$, we introduce the
error
\begin{equation}
\Delta \omega =\frac{1}{L^3} \int_{-\infty}^{\infty} x (x \tan \theta)
{\rm d}x \approx \frac{1}{3L^3} {L^3} \theta \approx \frac{\theta}{3}.
\end{equation}
As $\theta$ is, in general, approximately 0.05$^\circ$
($9\times10^{-4}$ radians), we estimate $\Delta \omega$ to be of the
order of $3\times10^{-4}$ rad. A large warp can have a value of
$10\times10^{-4}$ and a barely perceptible warp $5\times10^{-4}$
rad. In the remainder of this paper, we consider a galaxy as
warped if its calculated $\omega$ is clearly greater than the
associated error. (See Table 2 for our results.)

For those galaxies showing an appreciably warped disc, we fitted a
number of additional parameters to the warp curves. We tried to use the function proposed by Jim\'enez-Vicente et al. (1997), but it had too many fitting parameters so the fits were resticted to the simple description:

\begin{equation}
y=\left\{
\begin{array}{ll}
0       & |x|<|A| \\
C (x-A) & |x|\geq|A|
\end{array}   .
\right.
\end{equation}

This function reproduces the warp's shape as a first approximation; i.e., it is flat up to a
point and then deviates from the symmetry plane until it
asymptotically reaches a new direction. The interpretation of the
parameters $A$ and $C$ is as follows.

\begin{itemize}
\item $A$ is the starting point of the warp. It has dimensions of
length.
\item $C$ is the (nondimensional) value of the asymptotic slope.
\end{itemize} 

Throughout the rest of the paper, we refer to parameters A and C as those obtained by fitting the experimental points to Eq. 5. These values are shown in Table 3 for the different bands.
In addition, we use the angles $\alpha$ and $\beta$, which are most
widely used to represent the warp amplitude (S\'anchez-Saavedra et
al. 2003) Here, $\alpha$ is the angle between the line connecting the
galaxy's centre and the outermost detected point, and the disc's major
axis, while $\beta$ is the angle between the disc's major axis and the
line from the outermost detected point to the point where the warp
starts ($A$): see Fig.~\ref{dibu}.

\begin{figure}
\setcounter{figure}{3}
\resizebox{\hsize}{!}{\includegraphics{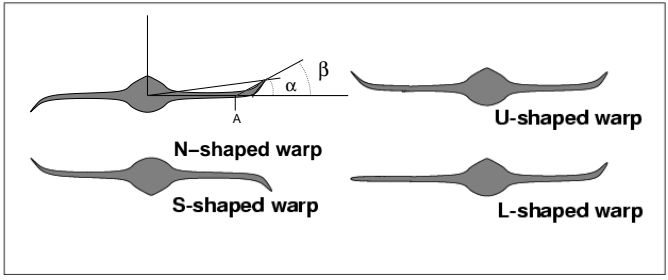}}
\caption{Definitions of angles, parameters, and types of warps.}
\label{dibu}
\end{figure}

We obtained the same result for the warp parameter $\omega$, whether it
was calculated starting from the galaxy centre or from the starting
point of the warp, $A$. This confirms that it is not critical to have
a precise estimate of the position of the galaxy centre because of the
radial flattening in the inner regions of our galaxies. In some
galaxies, such as ESO142-G24 or ESO157-G18, the warp sets off in one
direction, turns back to the mean plane, and ends in the opposite
hemisphere. The Milky Way's warp in the Southern Hemisphere is an
example of this type of behaviour (Porcel et al. 1997). Florido et
al. (1991) also report such `elbow'-type warps. This effect may come from the presence of a more strongly warped dust lane (than
the stellar disc) or be an intrinsic effect associated with warps. For
such galaxies, a four-parameter fit would have provided a better
description. However, since an excessive number of fit parameters
renders any interpretation of simple warps unclear, we prefer to use
three-parameter fits.

\begin{table*}
\centering
\begin{flushleft}
\caption{Warp parameters for our four filters ($\times 10^4$) and
degree of asymmetry for each warped galaxy, with warp values listed in
boldface.}  
{\scriptsize
\begin{tabular}{|l|c|c|c||c||c|c|c||c||c|c|c||c||c|c|c||c||} \hline
Galaxy & \multicolumn{4}{c|}{$B$ filter} & \multicolumn{4}{c|}{$V$ filter}
& \multicolumn{4}{c|}{$I$ filter} & \multicolumn{4}{c|}{$K_{\rm s}$
filter} \\ \cline{2-17} & $\omega$ & $\omega_{\rm R}$ & $\omega_{\rm L}$ &
$\alpha_{\rm s}$ & $\omega$ & $\omega_{\rm R}$ & $\omega_{\rm L}$ & $\alpha_{\rm s}$ &
$\omega$ & $\omega_{\rm R}$ & $\omega_{\rm L}$ & $\alpha_{\rm s}$ & $\omega$ &
$\omega_{\rm R}$ & $\omega_{\rm L}$ & $\alpha_{\rm s}$ \\ \hline \hline

  ESO026-G06 & 1$\pm$3 & 1$\pm$3 & $-$0$\pm$3 &  & 0$\pm$2 & 1$\pm$2 & $-$1$\pm$2 &  & 2$\pm$3 & 2$\pm$3 & 1$\pm$3 &  & $-$3$\pm$3 & $-$2$\pm$3 & $-$1$\pm$3 & \\
  ESO033-G22 & $-$1$\pm$4 & 0$\pm$4 & $-$1$\pm$4 &  & $-$3$\pm$6 & $-$1$\pm$6 & $-$2$\pm$6 &  & $-$1$\pm$4 & $-$1$\pm$4 & 0$\pm$4 &  & 3$\pm$4 & 1$\pm$4 & 1$\pm$4 & \\
  ESO142-G24 & \bf{3$\pm$2} & \bf{4$\pm$2} & $-$1$\pm$2 & 1.0 & \bf{3$\pm$2} & \bf{4$\pm$2} & 0$\pm$2 & 1.0 & \bf{4$\pm$2} & \bf{4$\pm$2} & 0$\pm$2 & 1.0 & 2$\pm$2 & 1$\pm$2 & 0$\pm$2 & -\\
  ESO157-G18 & \bf{$-$4$\pm$3} & $-$2$\pm$3 & $-$1$\pm$3 & - & \bf{$-$4$\pm$2} & \bf{$-$3$\pm$2} & $-$1$\pm$2 & $-$1.0 & 0$\pm$3 & $-$1$\pm$3 & 1$\pm$3 & - & 1$\pm$3 & 1$\pm$3 & 0$\pm$3 & -\\
  ESO201-G22 & \bf{4$\pm$2} & \bf{3$\pm$2} & 1$\pm$2 & 1.0 & 2$\pm$2 & \bf{3$\pm$2} & $-$1$\pm$2 & 1.0 & 2$\pm$2 & \bf{3$\pm$2} & $-$1$\pm$2 & 1.0 & 2$\pm$3 & 2$\pm$3 & 1$\pm$3 & -\\
  ESO202-G35 & \bf{17$\pm$6} & \bf{12$\pm$6} & 5$\pm$6 & 1.0 & \bf{16$\pm$3} & \bf{12$\pm$3} & \bf{5$\pm$3} & 0.4 & \bf{11$\pm$6} & \bf{9$\pm$6} & 3$\pm$6 & 1.0 & 4$\pm$4 & 4$\pm$4 & 2$\pm$4 & -\\
  ESO235-G53 & \bf{29$\pm$6} & \bf{10$\pm$6} & \bf{19$\pm$6} & 0.3 & \bf{39$\pm$7} & \bf{16$\pm$7} & \bf{22$\pm$7} & 0.2 & \bf{28$\pm$3} & \bf{11$\pm$3} & \bf{17$\pm$3} & 0.2 & \bf{25$\pm$2} & \bf{10$\pm$2} & \bf{15$\pm$2} & 0.2\\
  ESO240-G11 & \bf{$-$4$\pm$3} & \bf{$-$4$\pm$3} & 0$\pm$3 & $-$1.0 & \bf{$-$2$\pm$1} & \bf{$-$3$\pm$1} & 1$\pm$1 & $-$1.0 & 0$\pm$2 & $-$1$\pm$2 & 0$\pm$2 & - & 0$\pm$2 & 0$\pm$2 & 0$\pm$2 & -\\
  ESO288-G25 & \bf{$-$6$\pm$5} & 1$\pm$5 & \bf{$-$8$\pm$5} & $-$1.0 & \bf{$-$7$\pm$2} & $-$2$\pm$2 & \bf{$-$5$\pm$2} & $-$0.4 & \bf{$-$6$\pm$3} & $-$1$\pm$3 & \bf{$-$5$\pm$3} & $-$1.0 & \bf{$-$5$\pm$1} & \bf{$-$3$\pm$1} & \bf{$-$2$\pm$1} & $-$0.2\\
  ESO311-G12 & \bf{5$\pm$1} & \bf{3$\pm$1} & 1$\pm$1 & 0.5 & \bf{5$\pm$1} & \bf{3$\pm$1} & \bf{2$\pm$1} & 0.2 & \bf{5$\pm$1} & \bf{3$\pm$1} & \bf{2$\pm$1} & 0.2 & \bf{4$\pm$1} & \bf{2$\pm$1} & 1$\pm$1 & 0.3\\
  ESO340-G08 & $-$1$\pm$1 & 1$\pm$1 & \bf{$-$2$\pm$1} & $-$1.0 & $-$1$\pm$1 & 1$\pm$1 & \bf{$-$2$\pm$1} & $-$1.0 & $-$1$\pm$2 & 1$\pm$2 & $-$2$\pm$2 & - & $-$1$\pm$1 & $-$1$\pm$1 & $-$1$\pm$1 & 0.0\\
  ESO340-G09 & $-$2$\pm$7 & $-$6$\pm$7 & 4$\pm$7 &  & 4$\pm$5 & $-$1$\pm$5 & 5$\pm$5 &  & 1$\pm$8 & $-$6$\pm$8 & 7$\pm$8 &  & 9$\pm$9 & 2$\pm$9 & 8$\pm$9 & \\
  ESO358-G26 & $-$1$\pm$1 & $-$1$\pm$1 & $-$1$\pm$1 &  & 0$\pm$1 & 0$\pm$1 & 0$\pm$1 &  & 0$\pm$0 & 0$\pm$0 & 0$\pm$0 &  & 2$\pm$2 & 1$\pm$2 & 2$\pm$2 & \\
  ESO358-G29 & 0$\pm$1 & 0$\pm$1 & 0$\pm$1 &  & 1$\pm$1 & 1$\pm$1 & 1$\pm$1 &  & $-$1$\pm$2 & 0$\pm$2 & 0$\pm$2 &  & $-$2$\pm$2 & $-$1$\pm$2 & $-$1$\pm$2 & \\
  ESO416-G25 & 0$\pm$1 & \bf{2$\pm$1} & $-$1$\pm$1 & 1.0 & 0$\pm$1 & \bf{2$\pm$1} & $-$1$\pm$1 & 1.0 & $-$2$\pm$2 & 1$\pm$2 & \bf{$-$4$\pm$2} & $-$1.0 & \bf{$-$4$\pm$2} & $-$1$\pm$2 & \bf{$-$3$\pm$2} & $-$1.0\\
  ESO460-G31 & 8$\pm$9 & 1$\pm$9 & 7$\pm$9 &  & 4$\pm$4 & 2$\pm$4 & 3$\pm$4 &  & 7$\pm$8 & 1$\pm$8 & 6$\pm$8 &  & 0$\pm$3 & $-$3$\pm$3 & 3$\pm$3 & \\
  ESO487-G02 & $-$1$\pm$3 & 0$\pm$3 & 0$\pm$3 &  & $-$1$\pm$2 & 0$\pm$2 & $-$1$\pm$2 &  & 1$\pm$2 & 0$\pm$2 & 1$\pm$2 &  & 4$\pm$4 & 0$\pm$4 & 4$\pm$4 & \\
  ESO531-G22 & \bf{$-$20$\pm$6} & \bf{$-$14$\pm$6} & $-$6$\pm$6 & $-$0.4 & \bf{$-$16$\pm$4} & \bf{$-$11$\pm$4} & \bf{$-$5$\pm$4} & $-$0.4 & \bf{$-$11$\pm$4} & $-$4$\pm$4 & \bf{$-$7$\pm$4} & $-$0.3 & \bf{$-$6$\pm$2} & \bf{$-$2$\pm$2} & \bf{$-$4$\pm$2} & $-$0.3\\
  ESO555-G36 & \bf{$-$10$\pm$4} & 0$\pm$4 & \bf{$-$11$\pm$4} & $-$1.0 & \bf{$-$12$\pm$8} & $-$1$\pm$8 & \bf{$-$11$\pm$8} & $-$1.0 & \bf{$-$12$\pm$5} & $-$1$\pm$5 & \bf{$-$11$\pm$5} & $-$1.0 & \bf{$-$10$\pm$5} & $-$2$\pm$5 & \bf{$-$9$\pm$5} & $-$1.0\\
  ESO564-G27 & 3$\pm$3 & 0$\pm$3 & 2$\pm$3 & - & \bf{4$\pm$3} & 2$\pm$3 & 3$\pm$3 & 1.0 & \bf{3$\pm$2} & 1$\pm$2 & 2$\pm$2 & 1.0 & \bf{$-$5$\pm$1} & \bf{$-$3$\pm$1} & \bf{$-$2$\pm$1} & $-$0.2\\
\hline
\end{tabular}}
\end{flushleft}
\end{table*}

\section{Analysis}

We consider a galaxy as warped if it is warped (according to the
definition in Sect. \ref{warppar}) in any of the four observed
bands. Based on this assumption, we find that 13 of our 20 sample
galaxies (65\%) are warped. This fraction supports previous results
based on larger galaxy samples with similar optical characteristics to
our sample (S\'anchez-Saavedra et al. 1990, 2003; Reshetnikov \&
Combes 1999). Given the difficulties of observing warps with their
line of nodes perpendicular to the line of sight, this high fraction
of warped discs indicates that essentially all spirals are
warped. Optical warps are less perceptible than HI warps, but at both
wavelengths the frequency of warps is the same, i.e., close to 100\%
(Sancisi 1976; Garc\'{\i}a-Ruiz 2001; Garc\'{\i}a-Ruiz et al. 2002b;
van der Kruit 2007).

As expected, the N- and S-like warp frequencies are similar, since
these characteristics depend on the observer rather than the
galaxy. We observe six galaxies with N-like and four galaxies with
S-like warps. The frequency of asymmetric warps is lower: ESO416-G25
exhibits a U-like warp, while ESO240-G11 and ESO555-G36 show L-like
warps. Therefore, the most frequent warp morphologies are N- and
S-like warps (see also Ann \& Park 2006).

\begin{figure}
\resizebox{\hsize}{!}{\includegraphics{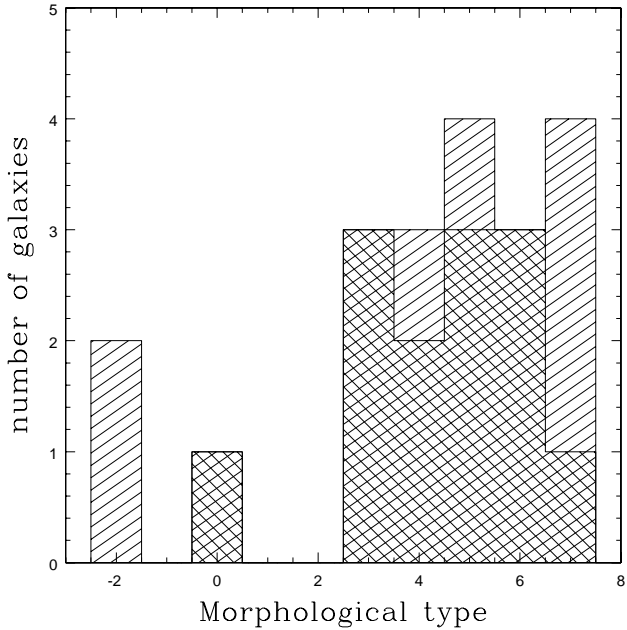}}
\caption{Morphological-type distribution of our warped
galaxies. Hatched area: All galaxies. Cross-hatched area: Warped
galaxies.}
\label{morfo}
\end{figure}

The distribution of warped galaxies as a function of morphological
type is shown in Fig.~\ref{morfo}, where the hatched area shows all
galaxies and the cross-hatched area covers warped galaxies. The warp
frequency does not seem to depend significantly on galaxy morphology,
except for S0 and very-late-type galaxies, which both yield lower
frequencies.

A noticeable result from S\'anchez-Saavedra et al. (2003) was that S0
galaxies are generally not warped. This places strong constraints
on any hypothesis proposed to explain warps. Among our sample
galaxies, ESO311-G12 has a small warp in both the optical and the NIR,
but both of the other S0 sample galaxies, ESO358-G26 and ESO358-G29, do not
exhibit any detectable warps. Although our sample only contains three
S0 galaxies, so that we cannot draw any statistical conclusions on
that basis, our results are not in contradiction to previously
published results.

Since we have data in both optical and NIR bands, we can compare both,
enabling us to either detect or reject colour dependences in the
warps. This comparison is difficult, however, because of inherent
uncertainties and because the differences as a function of wavelength
are not very large. Five of our galaxies (ESO235-G53, ESO288-G25,
ESO311-G12, ESO416-G25, and ESO555-G36) show similar warp curves in
all four observed bands, one galaxy (ESO531-G22) is warped in all four
filters but exhibits a clearly weaker IR warp, one galaxy (ESO564-G27)
is warped in the $V, I$, and $K_{\rm s}$ filters, and six galaxies
(ESO142-G24, ESO157-G18, ESO201-G22, ESO202-G35, ESO240-G11, and
ESO340-G08) are warped in the optical bands but not in the IR.

\begin{table*}
\begin{flushleft}
\caption{Warp parameters as a function of passband. }  
{\scriptsize
\begin{tabular}{|l|c|c|cccc|cccc|cccc|} \hline
\underline{Galaxy name}  & \multicolumn{2}{c|}{} & \multicolumn{4}{c|}{$y$ ($\pm$0.5)} & \multicolumn{4}{c|}{$A$ ($\pm$1)} & \multicolumn{4}{c|}{$C$ ($\pm$0.01)} \\
\cline{4-7}\cline{8-11}\cline{12-15}
& Side & $r_{\rm last}$ $\pm$1 (\arcsec) & $B$ & $V$ & $I$ & $K_{\rm s}$ & $B$ & $V$ & $I$ & $K_{\rm s}$ & $B$ & $V$ & $I$ & $K_{\rm s}$ \\
 &  & (3) & (4) & (5) & (6) & (7) & (8) & (9) & (10) & (11) & (12) & (13) & (14) & (15) \\
\hline
\hline
  ESO142-G24 & Right & 110 & $-$1 & $-$1.5 & 1 & 0 & 108 & 106 & 103 & - & 0.17 & 0.12 & 0.27 & - \\
             & Left  & 103 & 0 & 0.5 & 0 & 0 & $-$105 & $-$103 & $-$104 & - & 0.10 & 0.04 & 0.08 & - \\
  ESO157-G18 & Right & 90 & $-$1.5 & $-$2 & $-$1 & 0 & 70 & 68 & - & - & $-$0.06 & $-$0.07 & - & - \\
             & Left  & 92 & 0.5 & 0.5 & 0 & 1 & $-$81 & $-$115 & - & - & $-$0.05 & 0.04 & - & - \\   
  ESO201-G22 & Right & 71 & 0 & 0.5 & 0.5 & 0 & 46 & 63 & 29 & - & 0.07 & 0.10 & - & - \\
             & Left  & 72 & 1 & 1 & 0.5 & 0 & $-$66 & $-$86 & $-$63 & - & 0.05 & 0.21 & 0.02 & - \\
  ESO202-G35 & Right & 82 & 2 & 2 & 2 & 0 & 30 & 33 & 43 & - & 0.90 & 0.07 & 0.05 & - \\
             & Left  & 72 & $-$2 & $-$2 & $-$2 & $-$1.5 & $-$41 & $-$44 & $-$45 & - & 0.07 & 0.07 & 0.09 & - \\
  ESO235-G53 & Right & 70 & 7 & 6 & 5 & 6 & 27 & 36 & 30 & 30 & 1.52 & 0.16 & 0.15 & 0.39 \\
             & Left  & 70 & $-$7 & $-$5 & $-$6 & $-$4 & $-$11 & $-$35 & $-$35 & $-$39 & 0.11 & 0.16 & 0.11 & 0.10 \\
  ESO240-G11 & Right & 180 & $-$5.5 & $-$5.5 & $-$3 & 0 & 101 & 132 & - & - & - & - & - & - \\
             & Left  & 165 & $-$0.5 & $-$1.5 & $-$1 & 0 & - & $-$103 & - & - & 0.00 & 0.02 & - & - \\
  ESO288-G25 & Right & 67 & $-$0.5 & $-$1 & $-$3.5 & $-$2.5 & 65 & 60 & 60 & 43 & $-$0.08 & $-$0.09 & $-$0.49 & $-$0.04 \\
             & Left  & 73 & 2.5 & 2 & 1 & 0 & $-$26 & $-$28 & $-$29 & - & $-$0.03 & $-$0.05 & $-$0.02 & - \\
  ESO311-G12 & Right & 118 & 1.5 & 1 & 1 & 0 & 46 & 32 & 47 & 32 & 0.03 & 0.02 & 0.02 & 0.02 \\
             & Left  & 102 & - & - & - & 0 & $-$49 & $-$75 & $-$63 & $-$60 & 0.01 & 0.06 & 0.03 & 0.02 \\
  ESO340-G08 & Right & 37 & 0 & 0 & 0 & 0 & 64 & 59 & - & - & $-$0.08 & $-$0.12 & - & - \\
             & Left  & 39 & 0 & 0 & 0 & 0 & $-$50 & $-$50 & - & - & $-$0.05 & $-$0.05 & - & - \\
  ESO416-G25 & Right & 63 & 1.5 & 1.5 & 1.5 & $-$3 & 51 & 51 & 51 & 59 & 1.00 & 0.68 & 0.61 & $-$0.68 \\
             & Left  & 69 & 0 & 0 & - & 1 & - & - & $-$46 & $-$54 & - & - & $-$0.04 & $-$0.06 \\
  ESO531-G22 & Right & 88 & $-$4 & $-$3.5 & - & $-$2.5 & 58 & 44 & 58 & 44 & $-$0.14 & $-$0.12 & $-$0.11 & - \\
             & Left  & 69 & 2 & 1.5 & - & 2 & $-$39 & - 48 &  $-$12 & $-$57 & $-$0.05 & $-$0.04 & $-$0.04 & $-$0.14 \\
  ESO555-G36 & Right & 52 & 0 & 0 & 0 & 0 & - & 53 & 50 & - & - & - & $-$0.07 & - \\
             & Left  & 69 & 2.5 & 2 & - & 1.5 & $-$28 & $-$25 & $-$28 & $-$21 & $-$0.04 & $-$0.02 & $-$0.04 & $-$0.02 \\ 
  ESO564-G27 & Right & 117 & 0 & 0.5 & 0.5 & $-$1 & - & - & - & 48 & - & - & $-$0.04 & $-$0.04 \\
             & Left  & 83 & 0 & $-$1 & $-$0.5 & $-$0.5 & - & $-$25 & $-$70 & $-$37 & - & 0.01 & 0.02 & $-$0.02 \\
\hline
\end{tabular}}\\
(3) Radius of the last measured point in $K_{\rm s}$ (arcsec). 
(4), (5), (6) and (7) $y$ at the same point (arcsec). \\
(8), (9), (10) and (11) Warp starting points (arcsec). 
(12), (13), (14) and (15) Asymptotic slope.
\end{flushleft}
\end{table*}
 
\begin{figure}
\resizebox{\hsize}{!}{\includegraphics{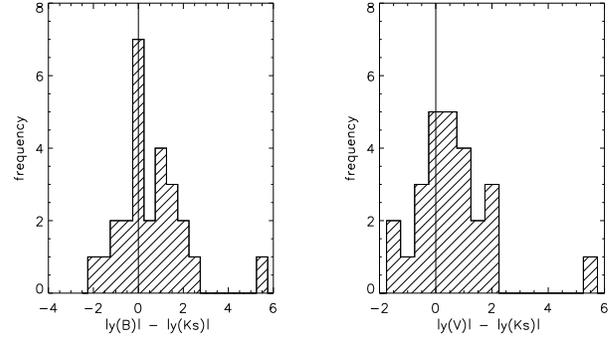}}
\caption{Histogram of $y(B) - y(K_{\rm s})$ and $y(V) - y(K_{\rm s})$
at the last measured point in $K_{\rm s}$.}
\label{histo}
\end{figure}

Since NIR profiles reach, in general, smaller radii than their optical
counterparts, it is not straightforward to compare the warp distortion
at different wavelengths. To do so, we compared the value of the fitted warp curve's $y$
at the same radial distance in each band, for which we
adopted the last measured point in $K_{\rm s}$ (i.e., the $h$
parameter of Ann \& Park 2006): see Table 3, column (3). Based on this
table, we appreciate that the optical $B$ and $V$ bands do not exhibit
significant differences. However, we find a difference between $V$ and
$K_{\rm s}$: $y(V)$ is usually greater than $y(K_{\rm s}$) (in 14 of
22 cases; see Fig.~\ref{histo}). On the other hand, $y(K_{\rm s}$) is
greater than $y(V)$ for only six of the 22 cases. This difference can
be quite large. The most extreme case is ESO240-G11 (right-hand side),
where the $y(V)-z(K_{\rm s}$) = 5.5 $\arcsec$. We find a marginal
tendency for NIR warps to have lower absolute values of the fit
parameter $C$. Our attempts at statistically detecting a colour
dependence of the observed warps are, therefore, inconclusive. A
better understanding would require more studies of this type. If the
colour dependence were eventually confirmed, mechanisms acting
directly on the gas as the perturbing agent should be favoured. A
colour difference has already been found by Florido et al. (1991). They
also noticed that the dust lane in NGC 4013 is much more strongly
warped than its optical disc. Recently, a study of 2MASS star counts
in the Milky Way (Reyl\'e et al. 2009) confirmed that the Galactic
warp is less obvious in stars than in the gas and that the disc's
shape is different at negative and positive longitudes. However, other
authors have not been able to confirm such a difference (e.g., Momany
et al. 2006 for the Milky Way).

Figure~\ref{beta} shows the behaviour of the warp angle $\beta$ as a
function of passband. This angle represents the strength of the warp, i.e. 
how much the outer disc deviates from the plane defined by the inner (unwarped) disc.. Its
highest value is obtained for galaxies with revised Hubble type ${\rm
T}=3$, corresponding to Sb. For these galaxy types, a tendency exists
toward lower values of $\beta$ in the NIR filter than in the $B$ band.
The starting point of the warp, $A$, does not appear to be a function
of wavelength.

\begin{figure}
\resizebox{\hsize}{!}{\includegraphics{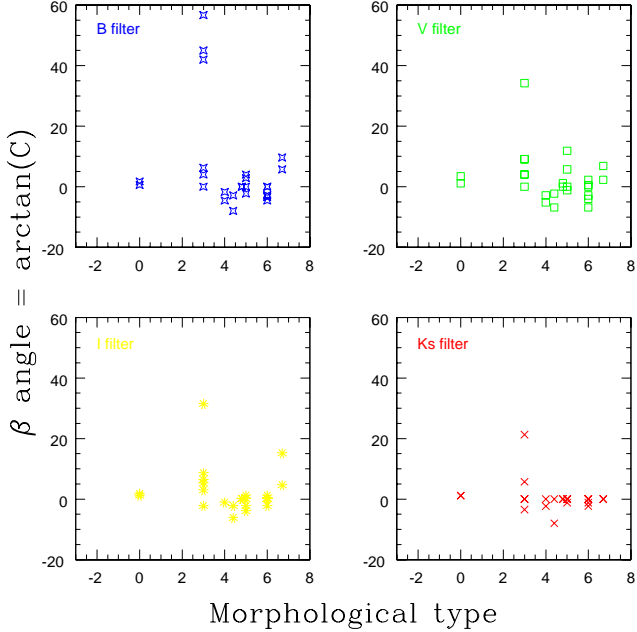}}
\caption{Distribution of warp angle as a function of revised Hubble
type.}
\label{beta}
\end{figure}

Kregel et al. (2002) analysed and used the same optical observations
discussed here to study the flattening and truncation of the stellar
discs. We take their measurements of the truncation radius (see van der Kruit 2007 for definition),
$R_{\rm max}$, in the $I$ band for 11 of our warped galaxies. It would be very useful to
compare this truncation radius with the starting radius of the warp ($A$),
because the relation between warps and disc truncations in edge-on
galaxies (van der Kruit 2007), if any, is unclear. Obviously we obtain
from this comparison that the warp radius, $A$, is approximately equal to 0.8 $R_{\rm max}$. Van der Kruit
(2007) concludes that when an HI warp is present, it starts at
approximately $1.1 R_{\rm max}$. Therefore it appears that optical
warps always start earlier than HI warps (HI warps start at
approximately 1.4 $A$), although we note that the low resolution of
the HI data makes it difficult to detect low-amplitude warps.

Kregel et al. (2004) also presented HI synthesis observations of 15
edge-on galaxies. They have seven galaxies in common with our sample,
of which ESO487-G02 was observed with insufficient sensitivity, and
five galaxies are found to have a warped HI disc. We now analyse
the objects in common with Kregel et al. (2004) for which warps have
been detected either by them, by us, or by both. They find that the
neutral hydrogen distribution in ESO142-G24 appears warped on both
sides of the galaxy, following a similar pattern to the
stellar disc. In our optical observations, the warp starts at 106
$\arcsec$ in the south and 104 $\arcsec$ in the north, but on the
basis of our NIR observations, we could not detect the same
behaviour. Instead, we observe a corrugated disc.

Kregel et al. (2004) find that ESO157-G18 is lopsided, which may be
related to the small companion, APMBGC157+052+052,
although this galaxy had no previously determined redshift. They did
not detect any clear HI warp. We also detected some differences
between both sides of this galaxy. Our warp curves exhibit a similar
behaviour in all optical bands, while in the NIR the warp is not clear.

ESO201-G22 shows a similar warp curve in HI and in all optical bands,
but it is uncertain in the NIR, where the data are not as deep as in
the optical. The $K_{\rm s}$-band warp curve finishes at a smaller
radius than in the optical bands. The bright star on the eastern
(left-hand) side of the galaxy hinders our analysis of this region.

ESO240-G11 is very interesting because a region in the galaxy's northwest
(right) is clearly warped in the $B$ and $V$ bands, but the warp is not as
marked in the NIR. Kregel et al. (2004) detected warping
of the HI layer and they point out that the HI warp can be traced by faint emission in the
$I$-band image. In the optical bands, the dust
lane is not completely coincident with the galactic plane, which means
that this galaxy is either not perfectly edge-on or that it has a
warped or asymmetric bulge, which also results in a rather large error
bar. On the southeastern (left-hand) side, the warp radius is
$103\pm2$ $\arcsec$ with $\beta = -2.3^\circ$, and on the
northwestern side this radius is $132.3\pm0.6$ $\arcsec$ and $\beta
-63.7^\circ$, so this galaxy exhibits an L-like warp. The `elbow'
where the warp begins in a given direction turns back to the mean
plane and ends in the opposite hemisphere should show the beginning
of the spiral arms. A more detailed study of corrugations in the
galaxy's stellar and dust discs would also be very interesting.

The U-like type galaxy ESO416-G25 shows a similar warp curve in all
bands, including in HI and the NIR.

The final galaxy also studied by Kregel et al. (2004) is
ESO564-G27. This is a clearly corrugated Sc galaxy with some bright
foreground stars projected onto the major axis, so it is very
difficult to obtain robust conclusions about the stellar warp. Kregel
et al. (2004) find that the neutral hydrogen is rather symmetric and
extends further than the stellar disc. It is warped on both sides,
with the warp apparently starting beyond the stellar disc. We cannot
compare this with our results because of the large uncertainties.

Then we can conclude from this comparative study that 5 out of 6 galaxies have an HI warp (only ESO157-G18 does not show any clear HI warp), and these 5 galaxies have an optical warp, while in the NIR the presence of a warp is unclear or absent in all of them. This would indicate that the old star population is less warped than the gas and the more recently born stars.

We calculated $\alpha_{\rm s}$ -- see Eq. (3) -- for all warped
galaxies, and the results are shown in the Table 2. The asymmetry in the warp angles seems related to the warp's
intrinsic properties. Garc\'{\i}a-Ruiz et al. (2002b) show that the
largest asymmetries are observed in galaxies that display obvious
tidal features. We can only show, based on our sample, that the three
galaxies that exhibit asymmetries, ESO235-G53, ESO311-G12, and
ESO531-G22, all show a large warp amplitude, which supports the
positive correlation found by Garc\'{\i}a-Ruiz et al. (2002b). However,
none of the galaxies in our sample are involved in strong tidal
interactions.

\section{Discussion and conclusions}

Are all warps produced by the same mechanism? For example,
gravitationally driven warps should exhibit no differences between
gaseous and stellar morphologies, as opposed to mechanisms acting on
the gas (e.g., magnetically or accretion-driven warps). Are the warp
curves the same for different colours; i.e., is there any difference
in shape and magnitude at different wavelengths?

The present study suggests that, in some of the observed galaxies, the
warp is smaller at NIR wavelengths than in the optical. This -- combined
with the important result of S\'anchez-Saavedra et al. (2003) that
lenticulars do not exhibit warps (which is also compatible with the
results in this paper) -- suggests that gas seems to be a necessary
ingredient in the development of warps. We have not been able to
robustly confirm the absence of warps in S0-type galaxies because our
sample only contains three S0s. Therefore, we cannot draw any
statistical conclusions on that basis, but our results (at least) do 
not contradict previously published results that most (if
not all) lenticulars are not warped. This must be explained by some theoretical model. Models based on gravity alone might have
significant difficulties to explain these observations. A recent study
using 2MASS star counts in the Milky Way (Reyl\'e et al. 2009) shows
that the warp is less pronounced in stars than in the gas, and it also
shows different disc shapes at negative and positive longitudes. The
magnetic model of Battaner et al. (1990) is consistent with all of
these observations.

Our results show that NIR warps exist with a frequency that is nearly
the same as that of optical warps. This suggests that warps are
long-lived structures that were formed in the early stages of the
evolution of most galaxies. Since warps can be dissipated on a
time scale close to the galaxy's rotation period, this
indicates that the responsible mechanism, whatever it may be, seems to be
acting permanently. The intergalactic magnetic-field model remains an
interesting possibility. It is well-known that a warp will disappear
on a time scale of approximately 2 Gyr (Binney 1992). However, in the
magnetic model of warps, the field configuration is not an initial
condition but extragalactic fields act permanently. The extragalactic
magnetic-field lines could be frozen in in the intergalactic medium,
therefore corotating with galaxy clusters and presenting a relatively
stable configuration of the field with respect to the warped galaxy.

The magnetic model was designed to explain N- and S-like warps ($m=1$
modes), the most frequently occurring type (51\% out of 73\% of warped discs in
the large sample of 325 edge-on galaxies of Ann \& Park 2006). This model does not manage to explain $m=0$ warps (U-shaped
profiles). They should be interpreted as coming from gradients in
the extragalactic field, with characteristic lengths close to a
galaxy's size. Generations of asymmetric warps through other mechanisms
have been considered by Saha \& Jog (2006) and L\'opez-Corredoira et
al. (2002). Other mechanisms have been proposed to induce warping. In
fact, the variety of warp morphologies could come from different
causes that may act differently from galaxy to galaxy. However,
observations of NIR warps and direct comparison with their optical
counterparts could provide some constraints.

The frequency of warped discs in our sample is very high (65\%).
Given the difficulty of detecting warps with their lines of nodes in
the plane of the sky, this suggests a fraction of nearly 100\% for the
frequency of warps in spirals. We have also been able to confirm the
finding by Reshetnikov \& Combes (1999) that warps are equally present
in all types of spirals, ${\rm T}>0$.

The abrupt break between the undisturbed inner disc and the warped
region (with a roughly constant slope) suggests that these two regions
have different formation and evolutionary histories (e.g., van der
Kruit 2007). However, van der Kruit concludes that HI warps start at
$1.1 R_{\rm{max}}$. If this were the case, optical and NIR warps could
not be observed. That they are indeed observed allows us to conclude
that $R_{\rm{max}} > A$. Nevertheless, we cannot discard a possible
relationship between truncations and warps. It appears that optical
warps always start closer in than HI warps, although we note that the
low resolution of the HI data makes it difficult to detect
low-amplitude warps.

\begin{acknowledgements}
We are grateful for the helpful cooperation of the Ursuline Mothers
during the period of this work. This paper has been supported by the
`Plan Andaluz de Investigaci\'on' (FQM-108) and by the
`Secretar\'{\i}a de Estado de Pol\'{\i}tica Cient\'{\i}fica y
Tecnol\'ogica' (AYA2000-1574).
\end{acknowledgements}

\begin{figure*}
\setcounter{figure}{2}
\resizebox{\hsize}{!}{\includegraphics{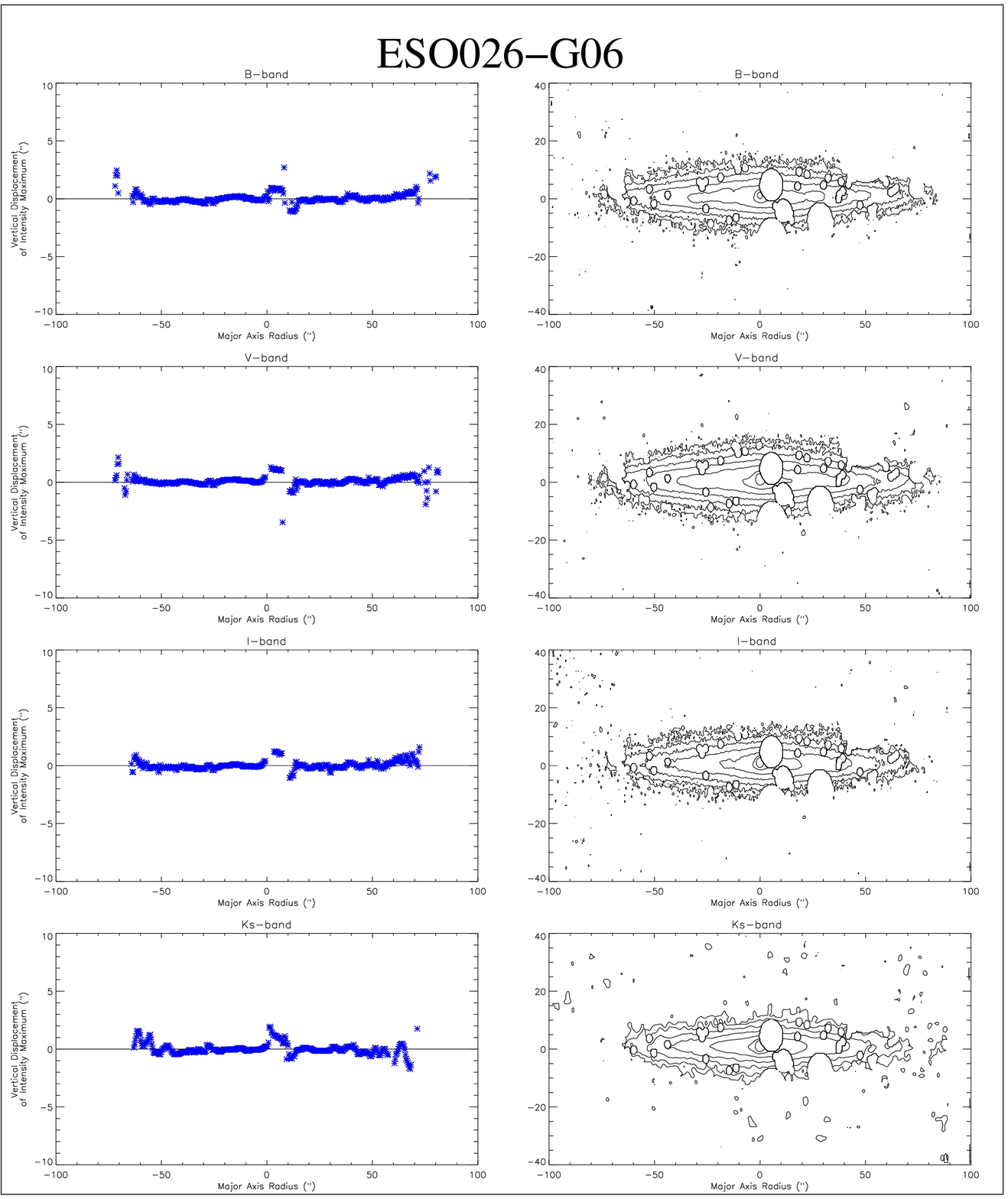}\includegraphics{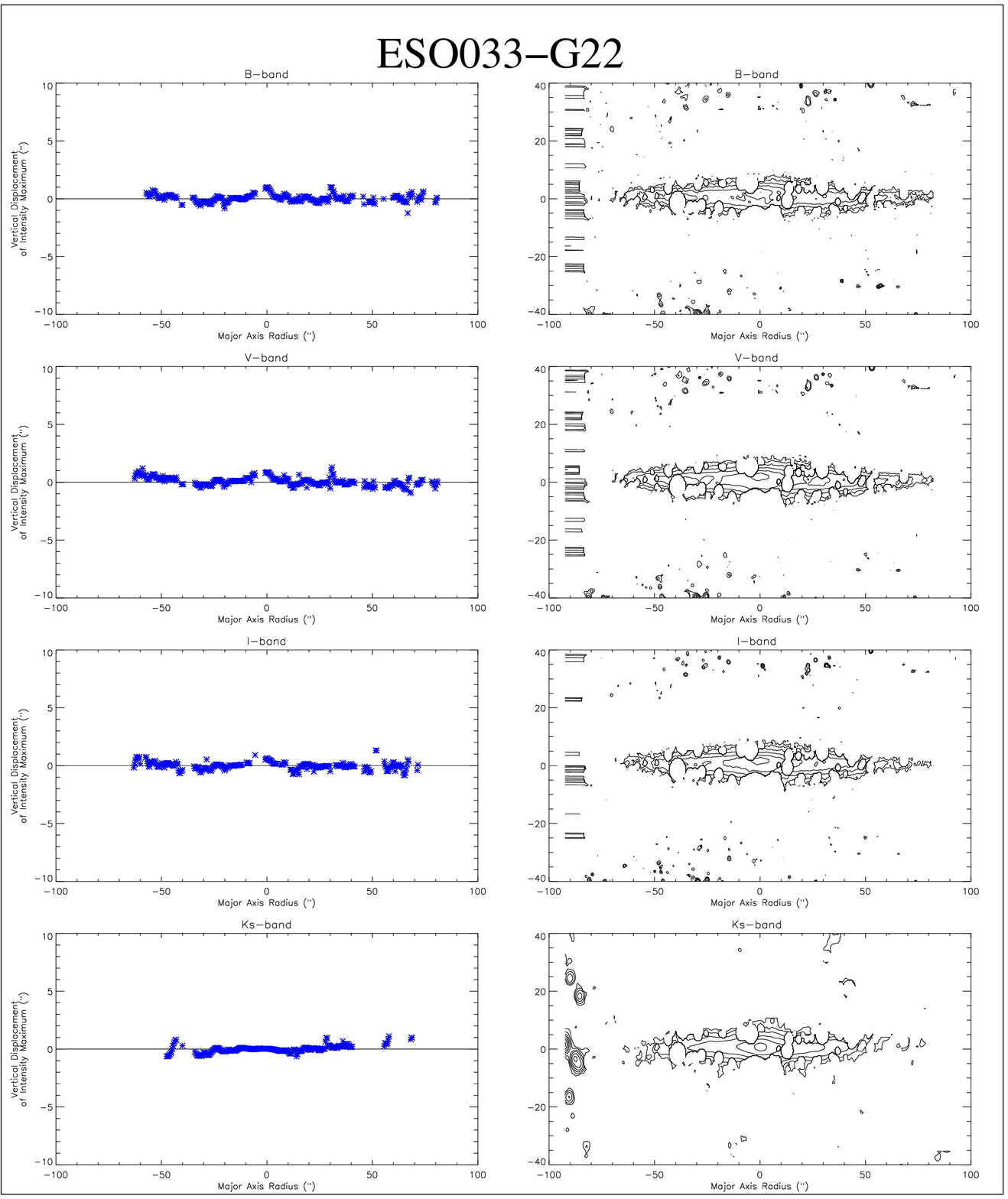}}
\resizebox{\hsize}{!}{\includegraphics{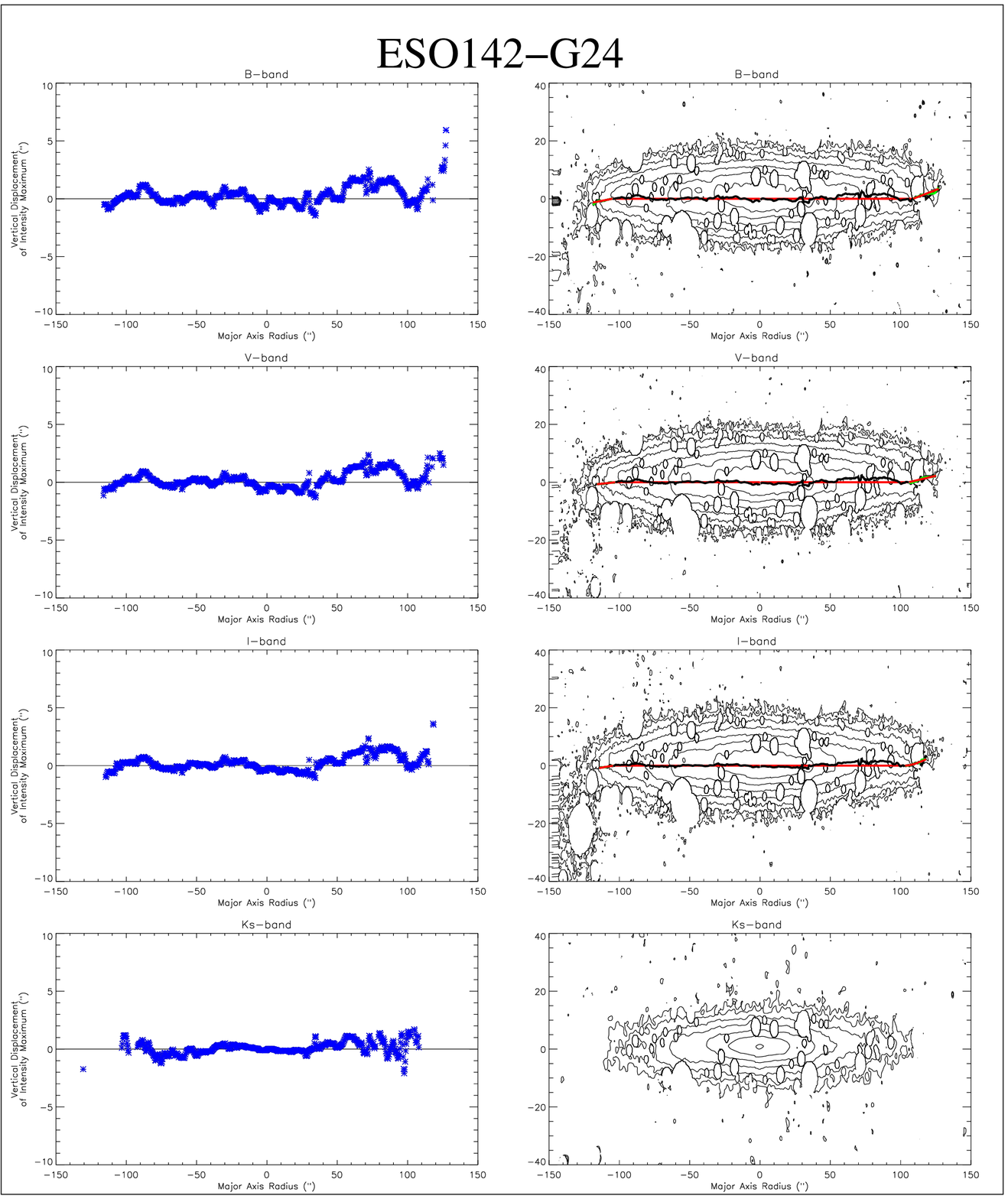}\includegraphics{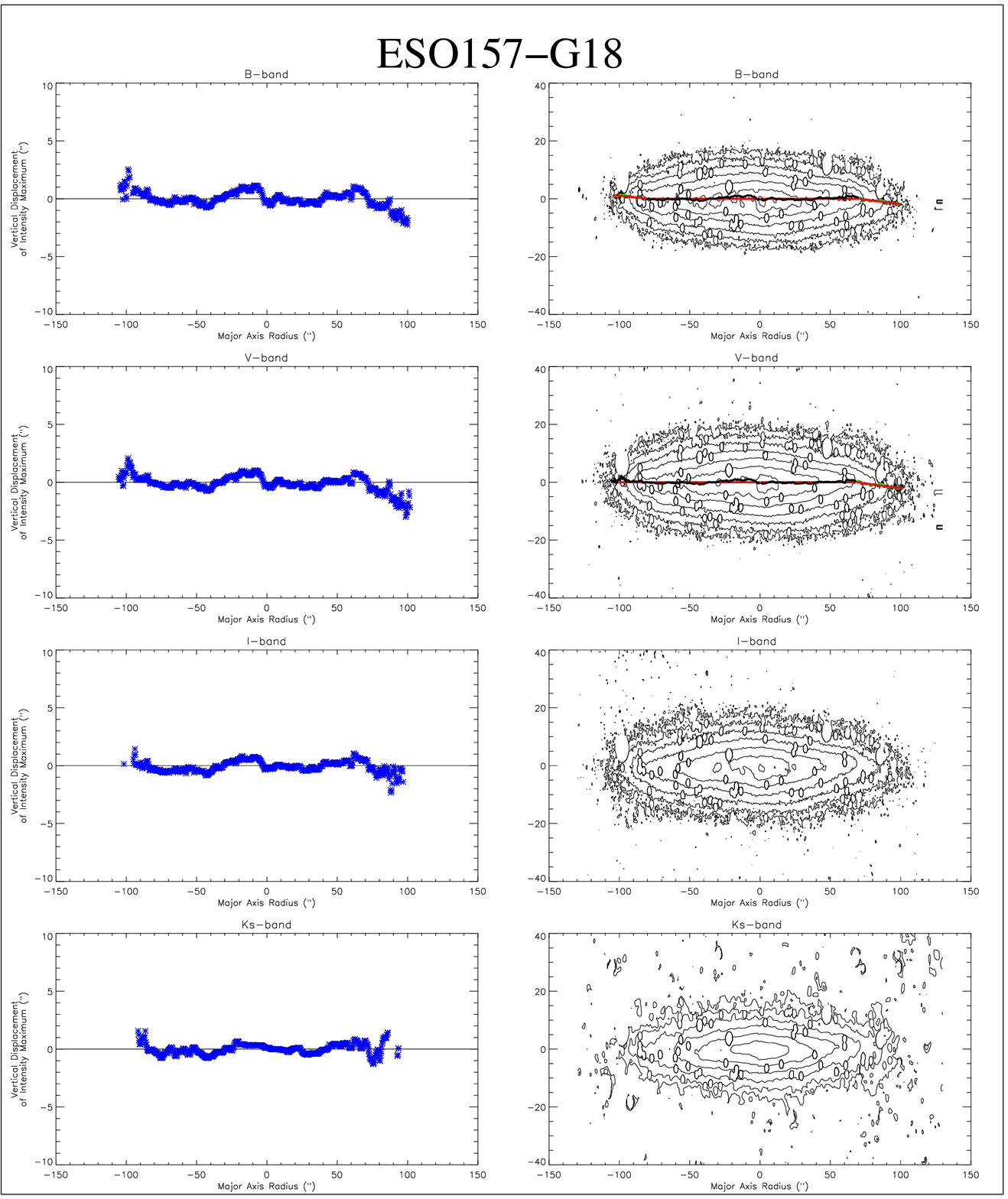}}
\caption{Left: Warp curves and right: Contour maps, for the different filters
  of our 20 galaxies. The warp curve (green line) is put on top only for those
  values with an error bar less than 0.5$\arcsec$. The red line shows the
  fitted warp curve. The isophotes are equidistant (in units of n*3$\sigma$
  equiv. to a step of +0.75 mag/arcsec$^2$) starting at a level of about
  3$\sigma$ above the sky background. The galaxy is rotated, but N is closer
  to the top and E to the left. The scale on the y-axis is enlarged to show a better curve.}
\label{alabeos}
\end{figure*}
\begin{figure*}
\setcounter{figure}{2}
\resizebox{\hsize}{!}{\includegraphics{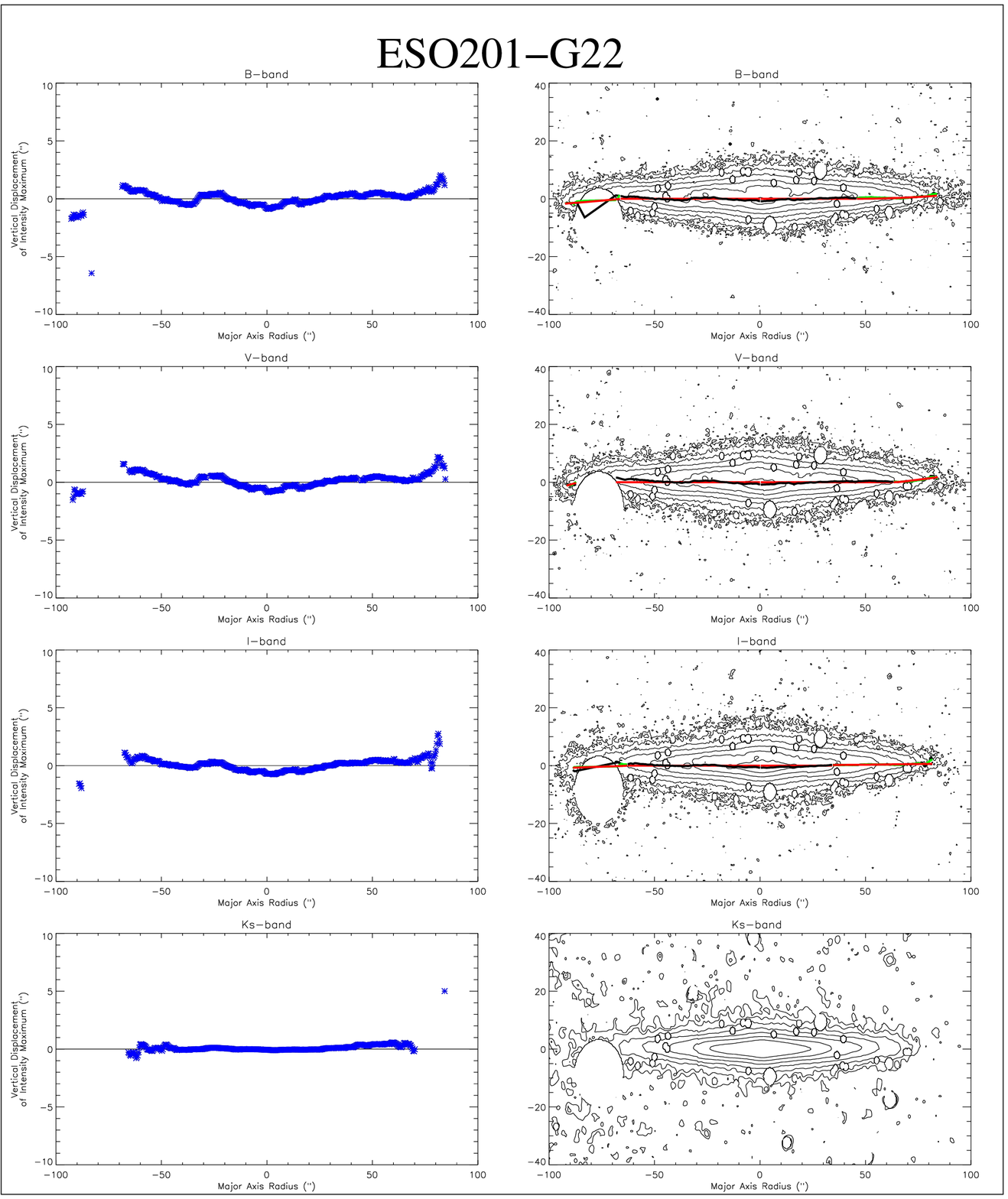}\includegraphics{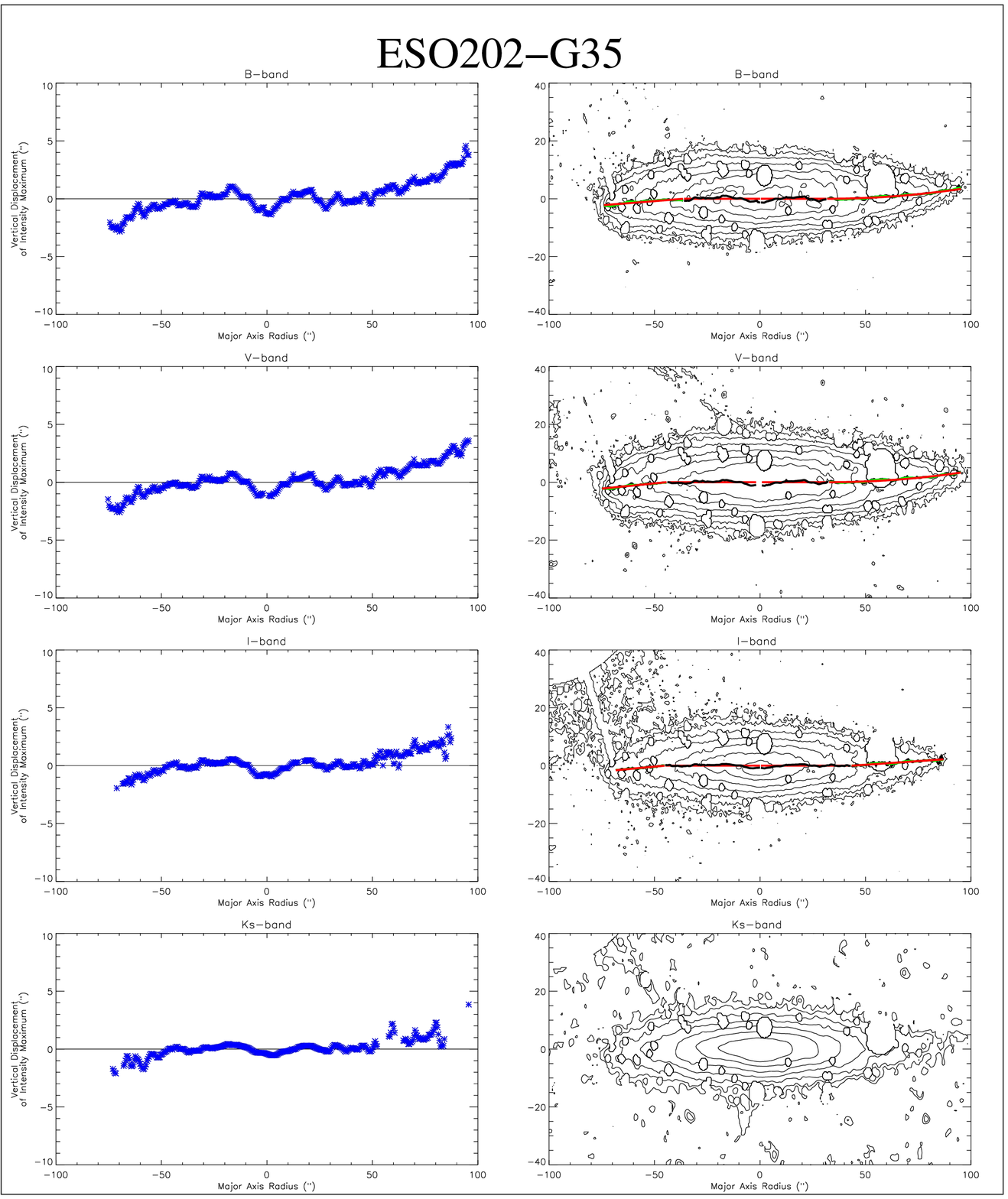}}
\resizebox{\hsize}{!}{\includegraphics{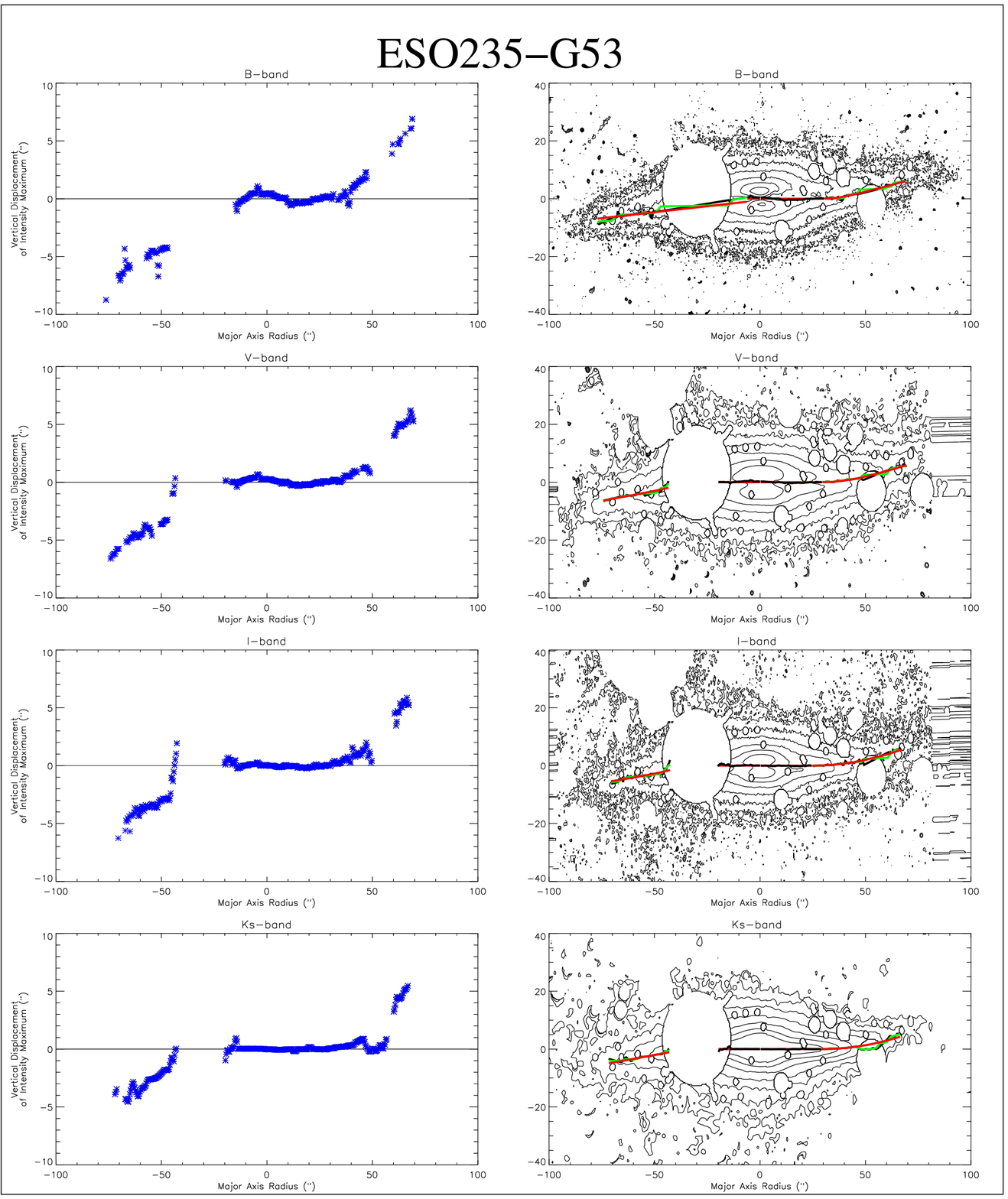}\includegraphics{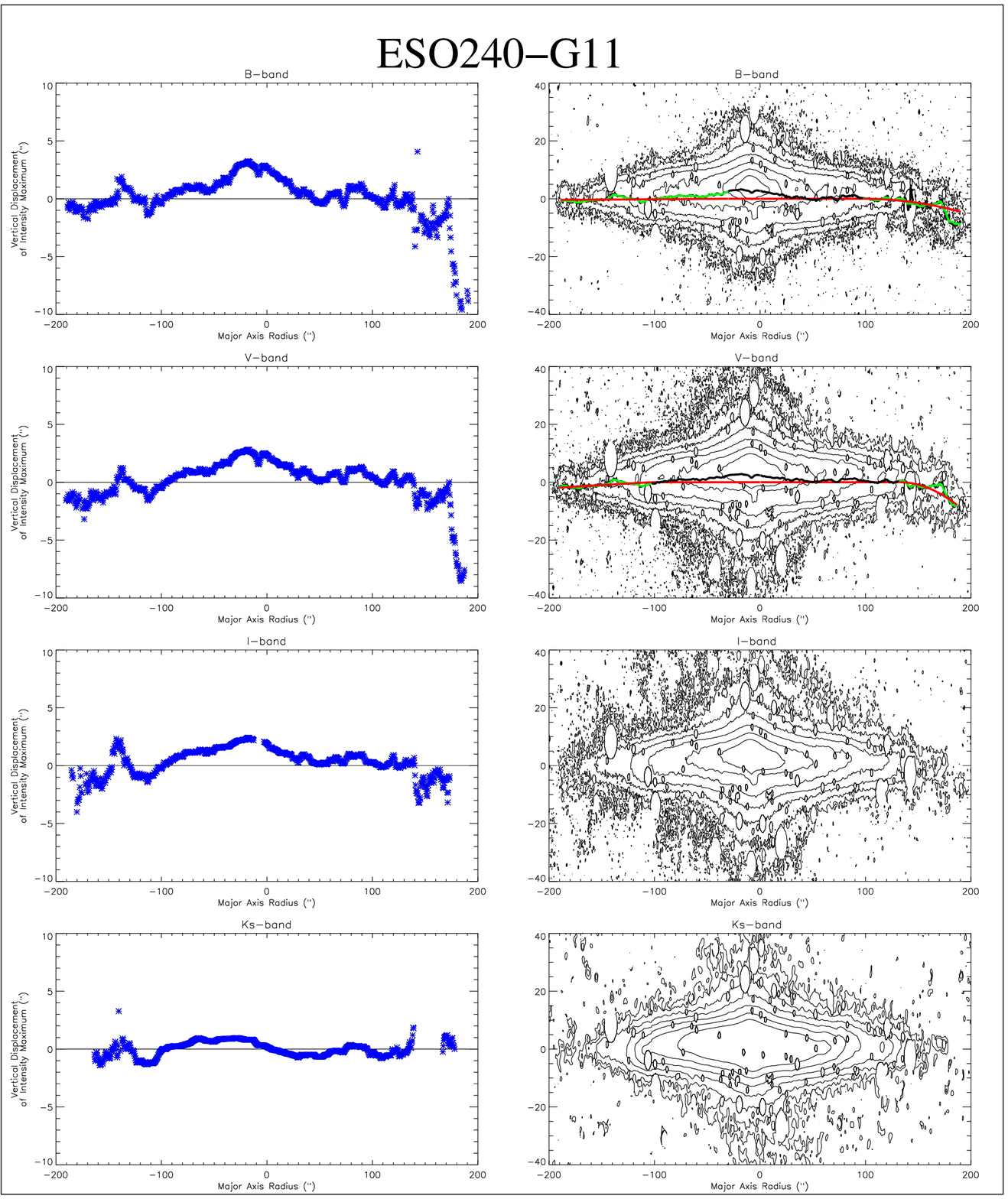}}
\caption{continued}
\end{figure*}
\begin{figure*}
\setcounter{figure}{2}
\resizebox{\hsize}{!}{\includegraphics{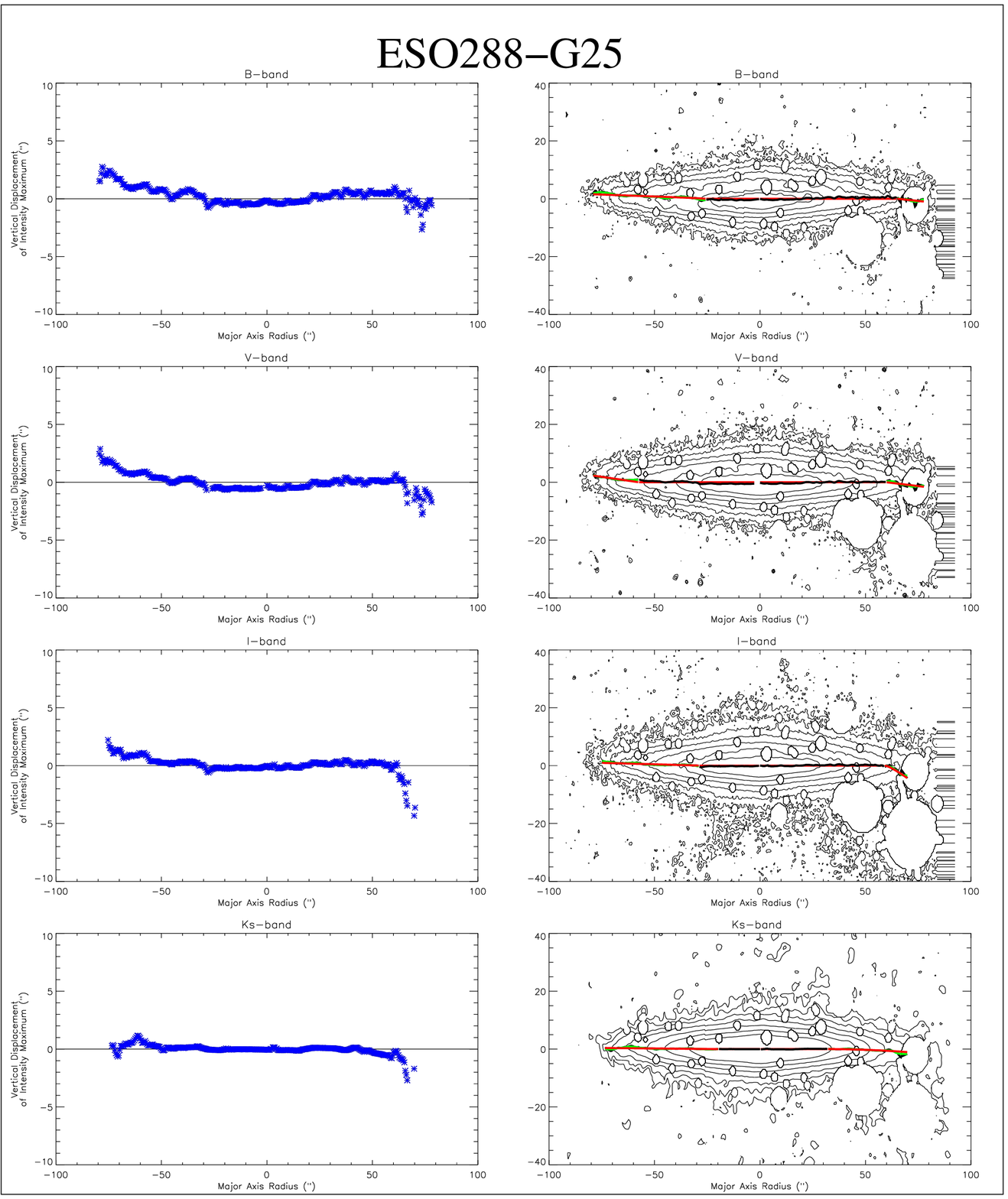}\includegraphics{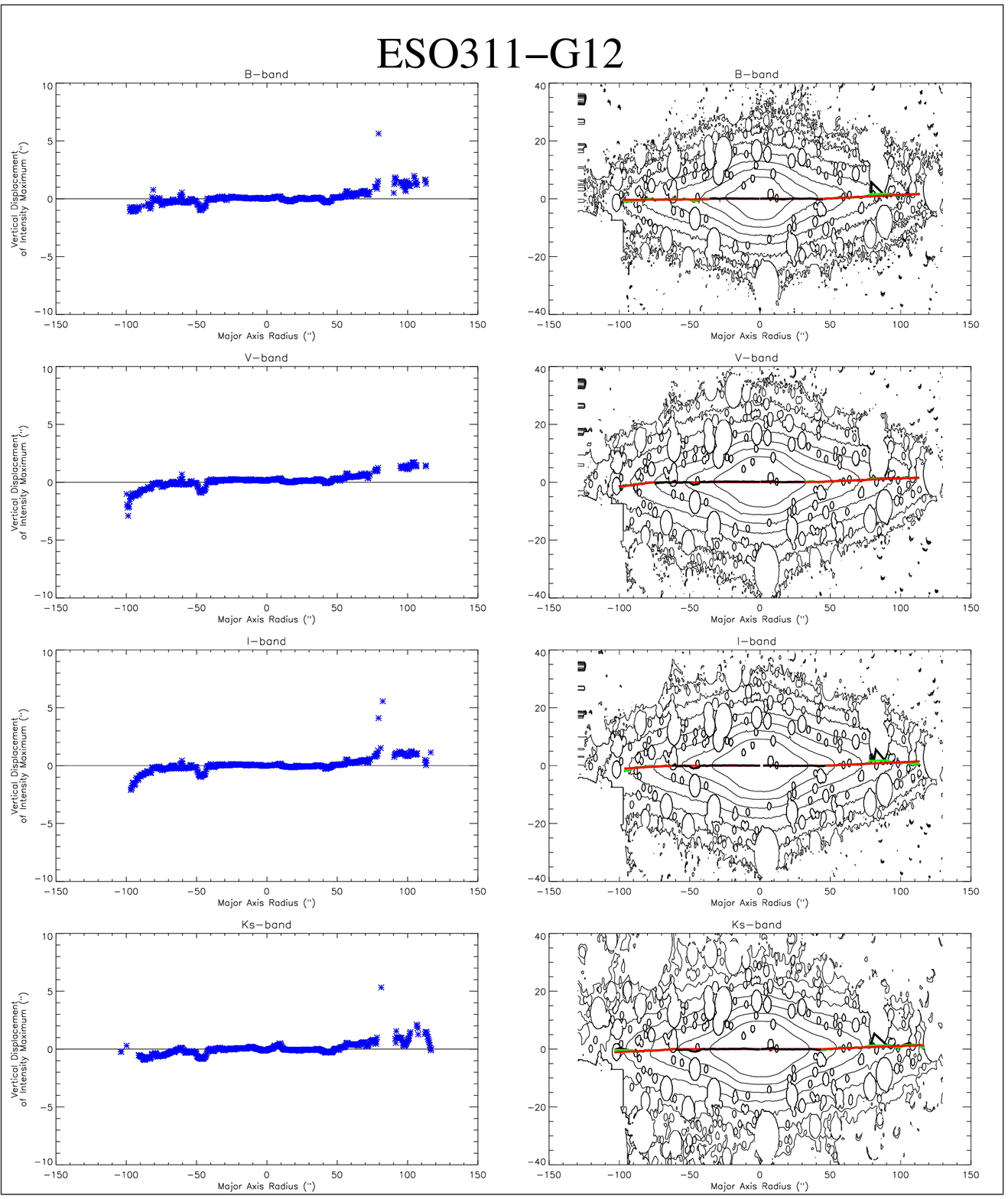}}
\resizebox{\hsize}{!}{\includegraphics{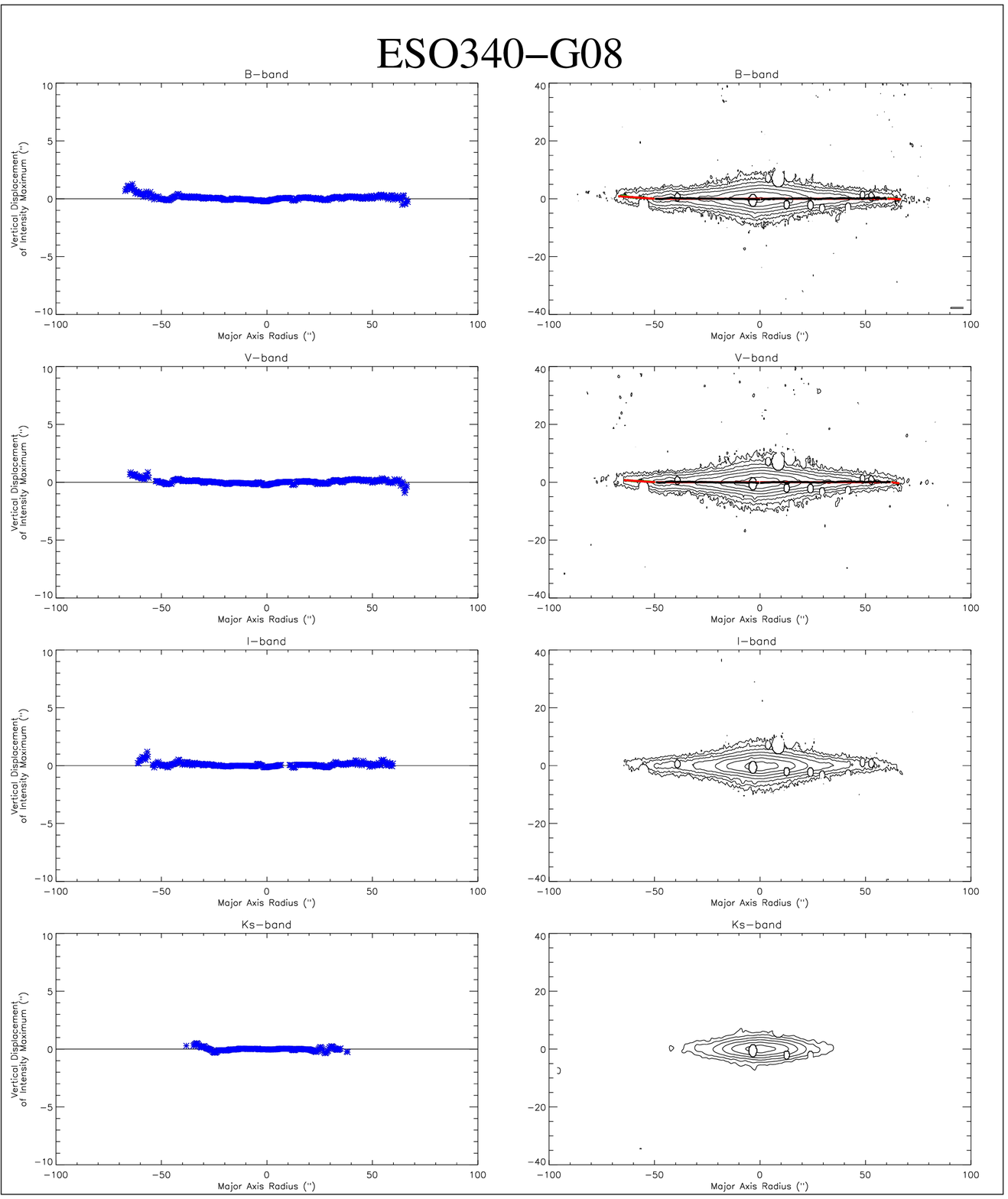}\includegraphics{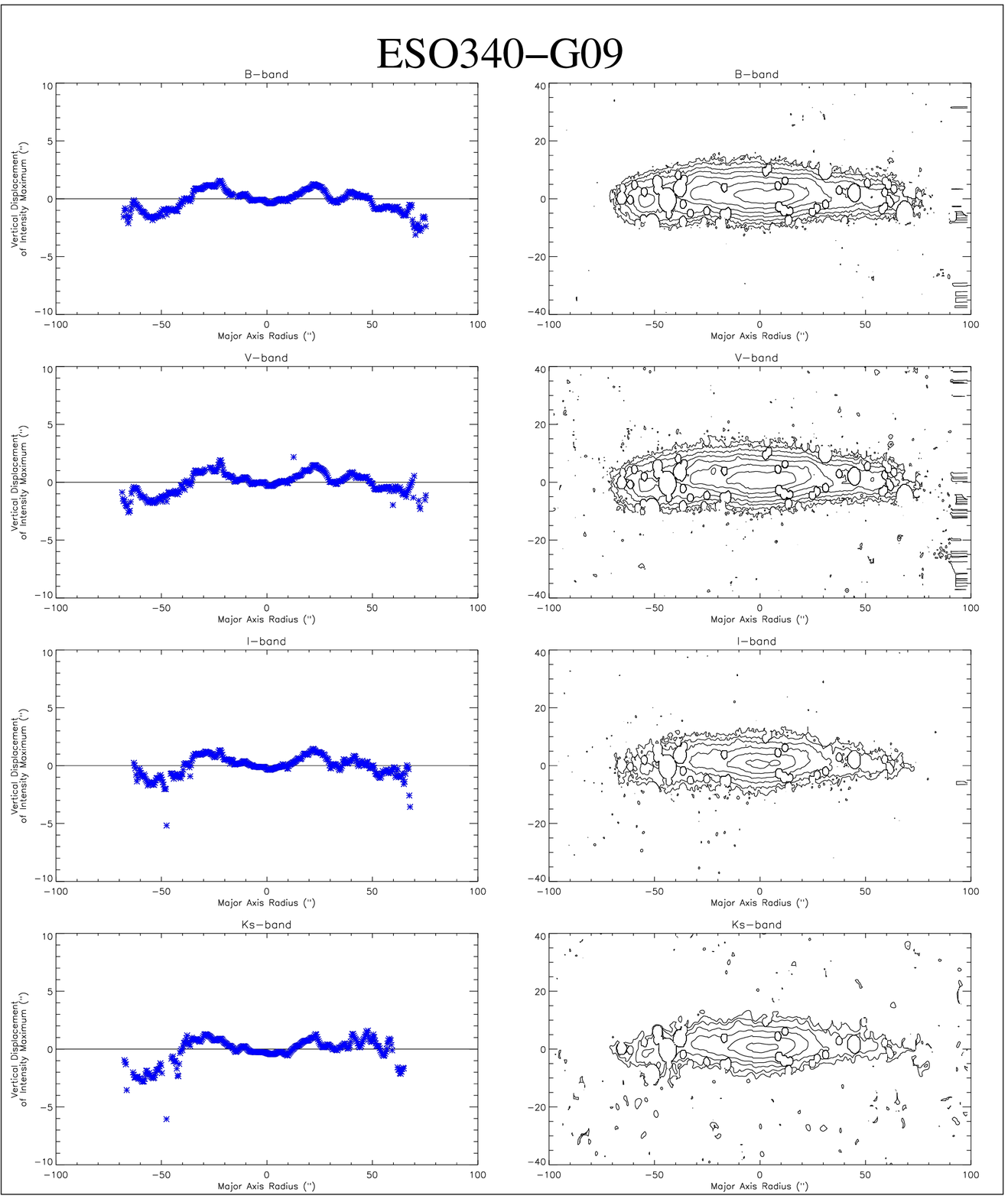}}
\caption{continued}
\end{figure*}
\begin{figure*}
\setcounter{figure}{2}
\resizebox{\hsize}{!}{\includegraphics{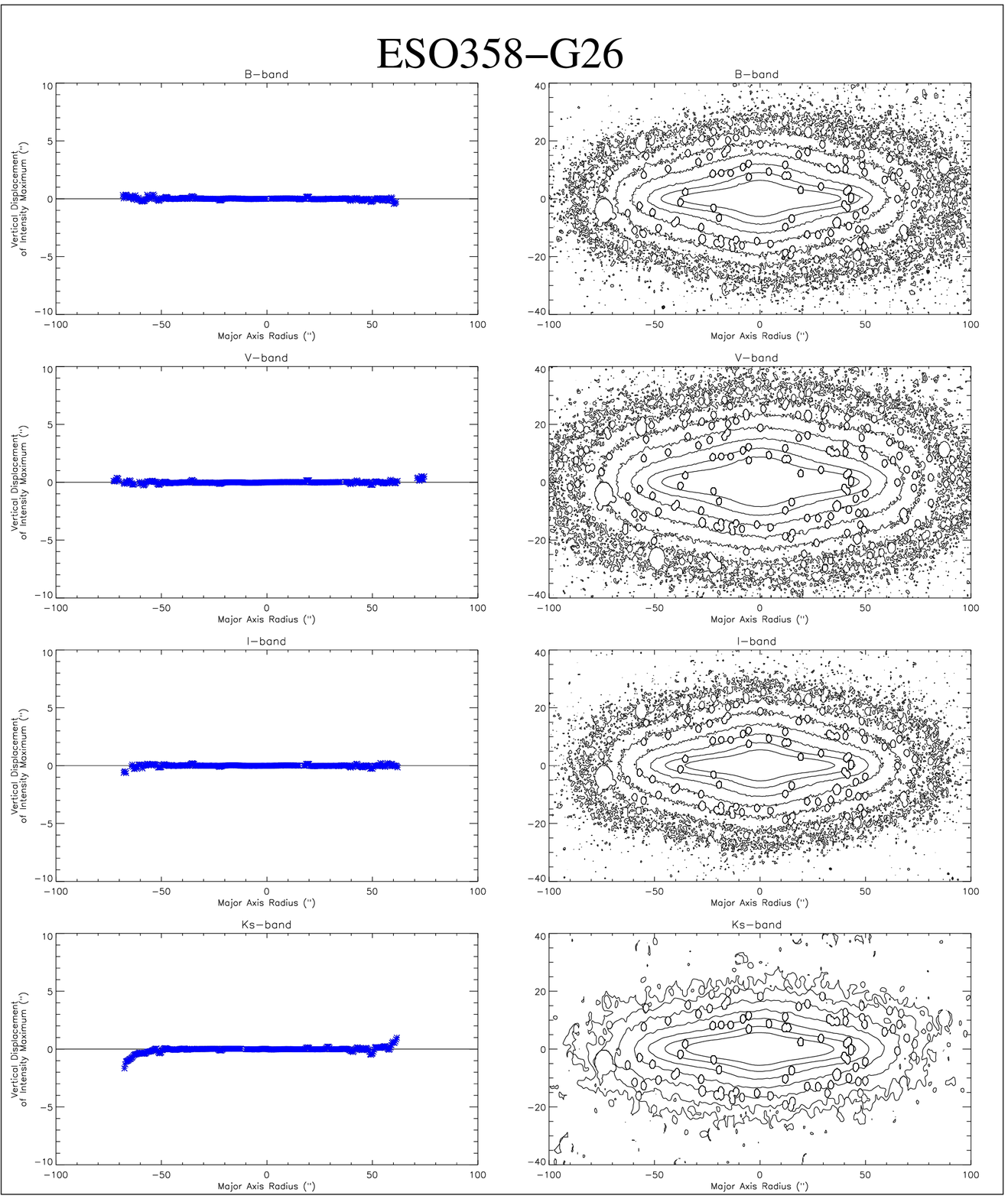}\includegraphics{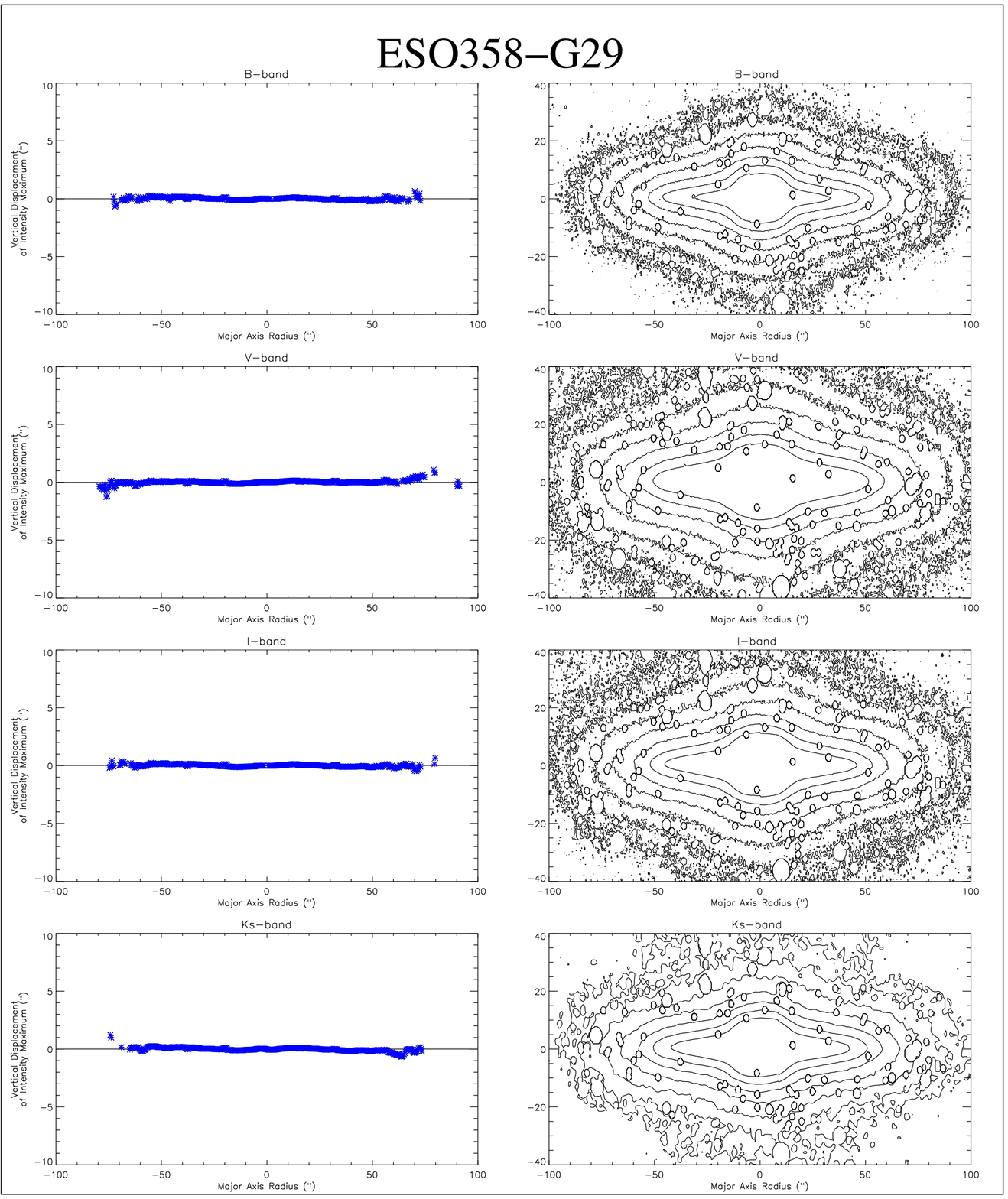}}
\resizebox{\hsize}{!}{\includegraphics{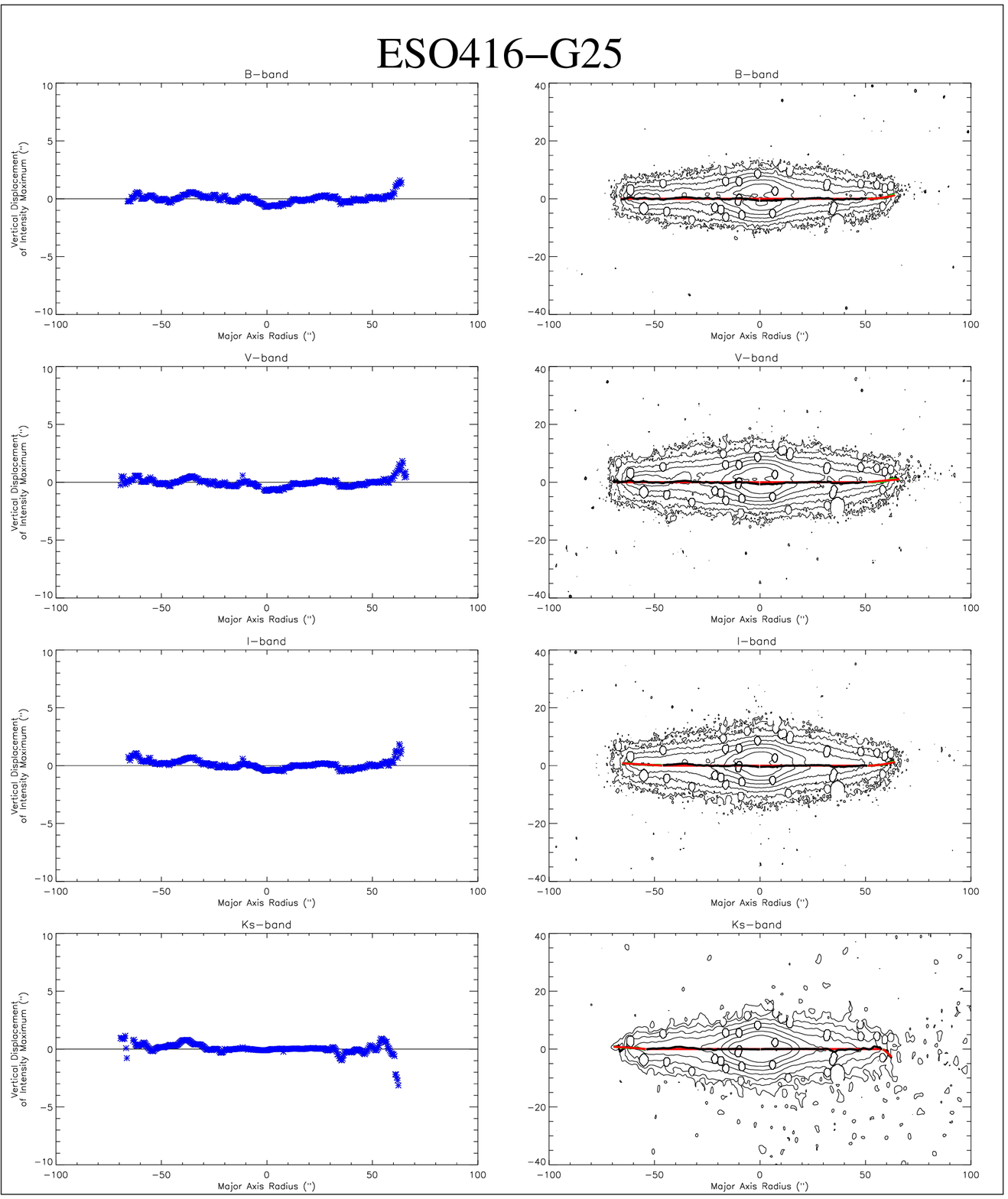}\includegraphics{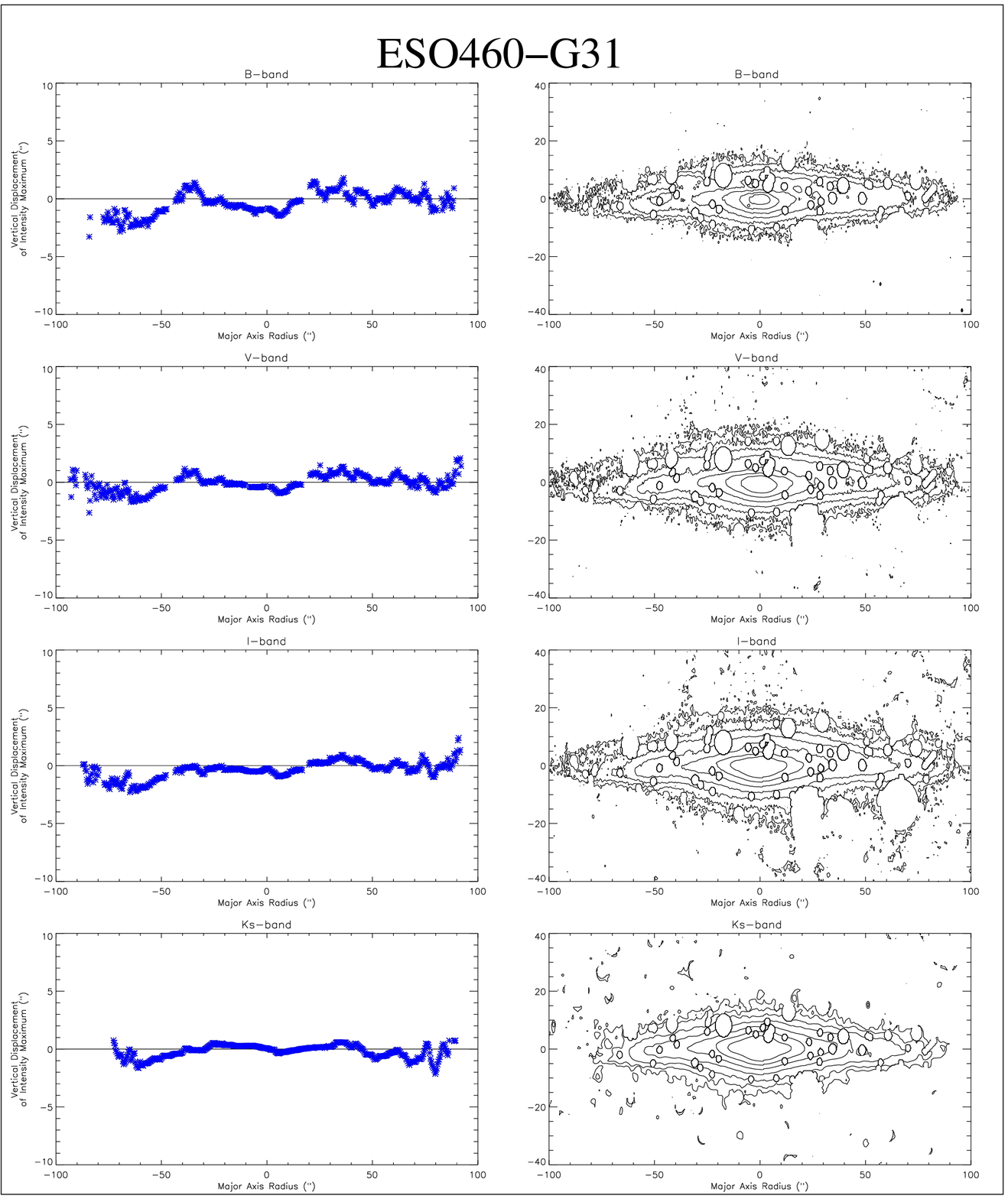}}
\caption{continued}
\end{figure*}
\begin{figure*}
\setcounter{figure}{2}
\resizebox{\hsize}{!}{\includegraphics{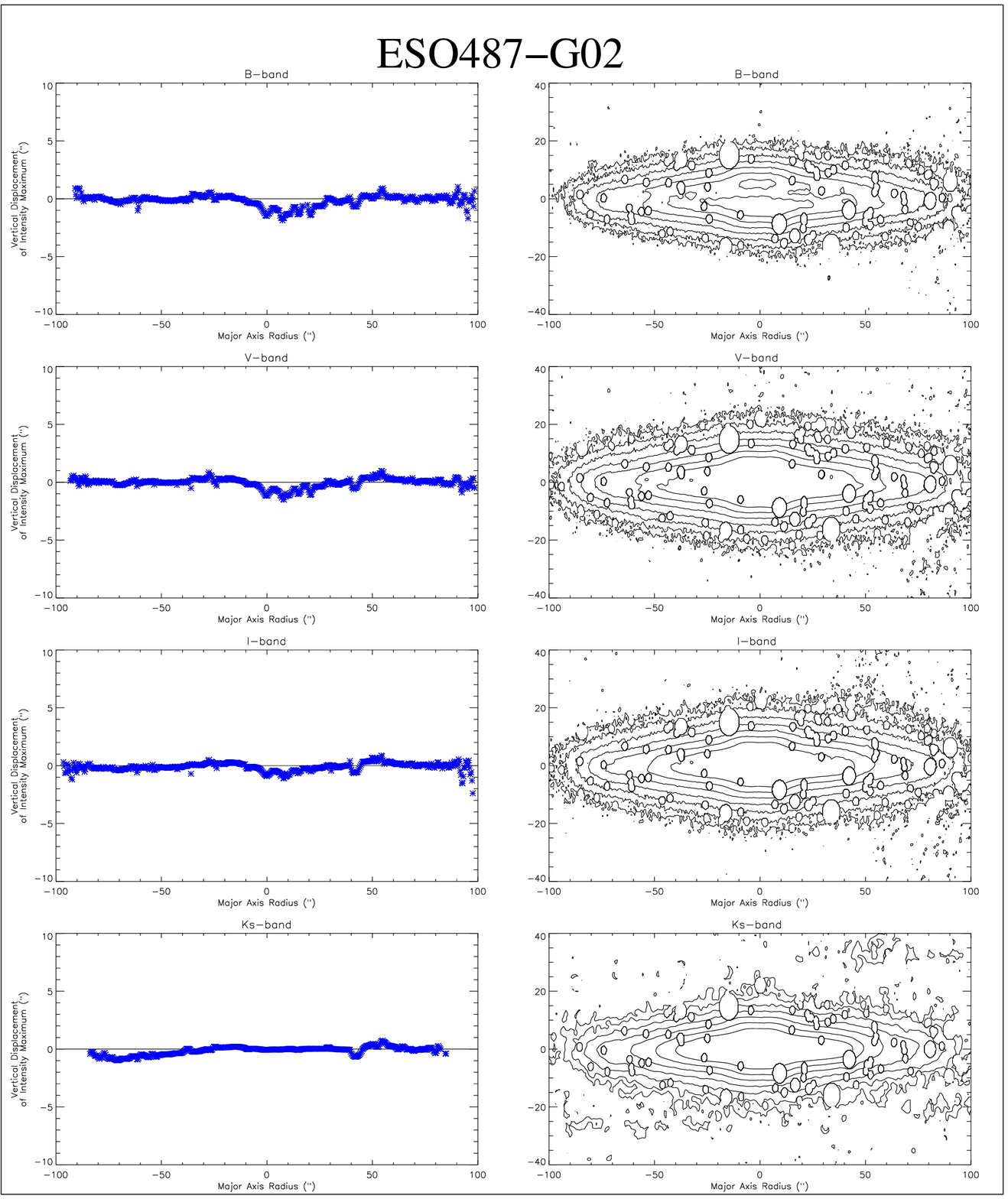}\includegraphics{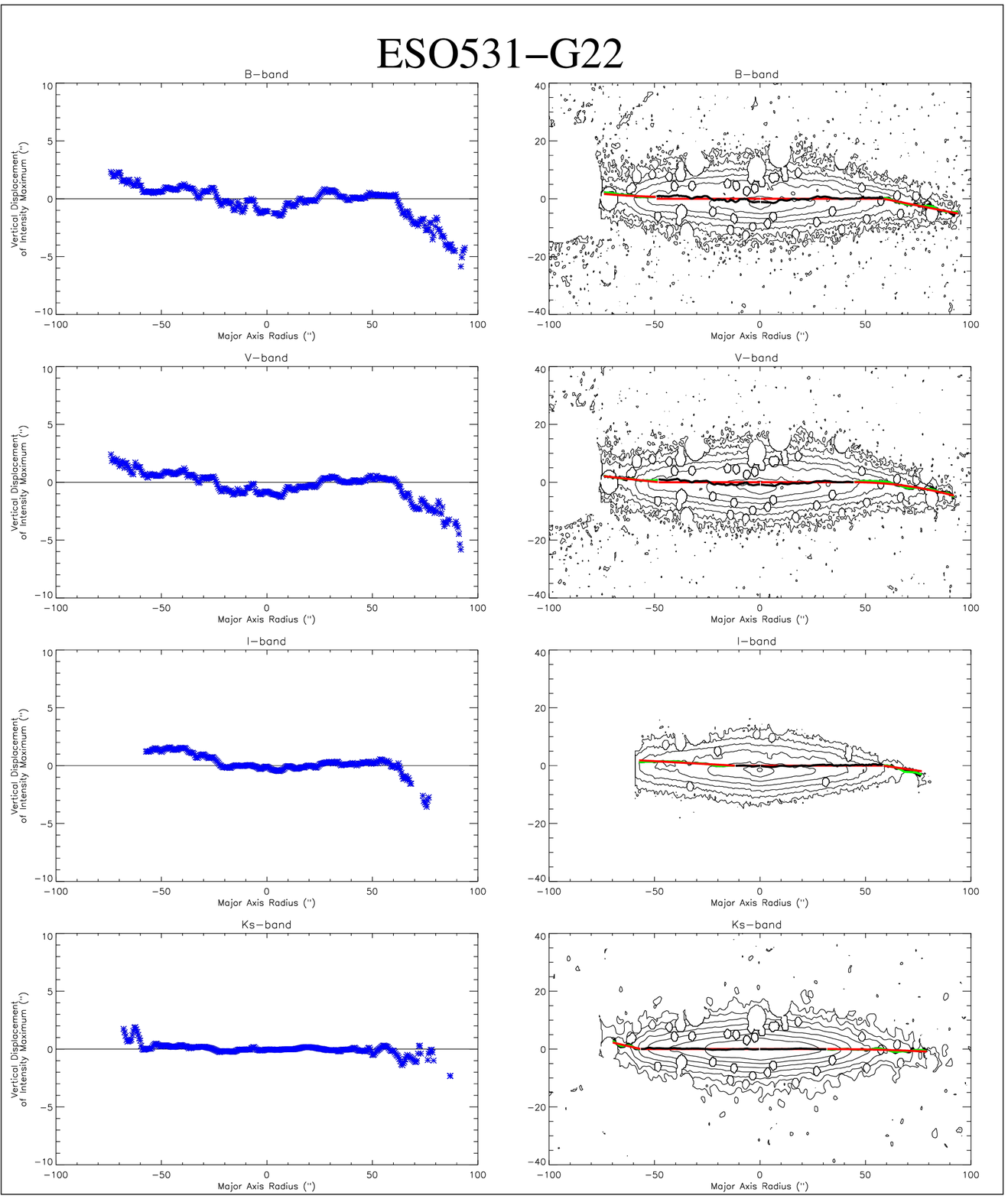}}
\resizebox{\hsize}{!}{\includegraphics{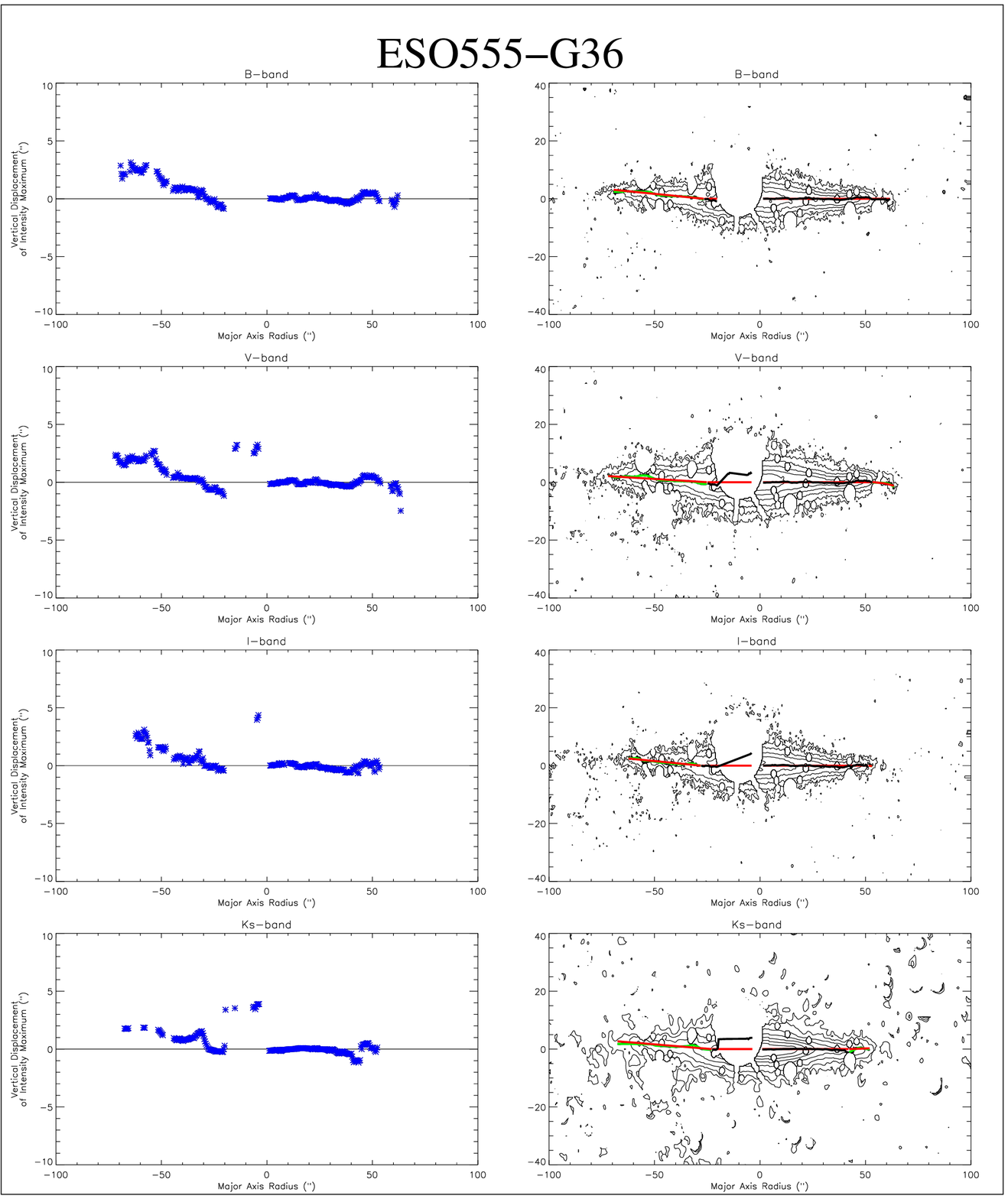}\includegraphics{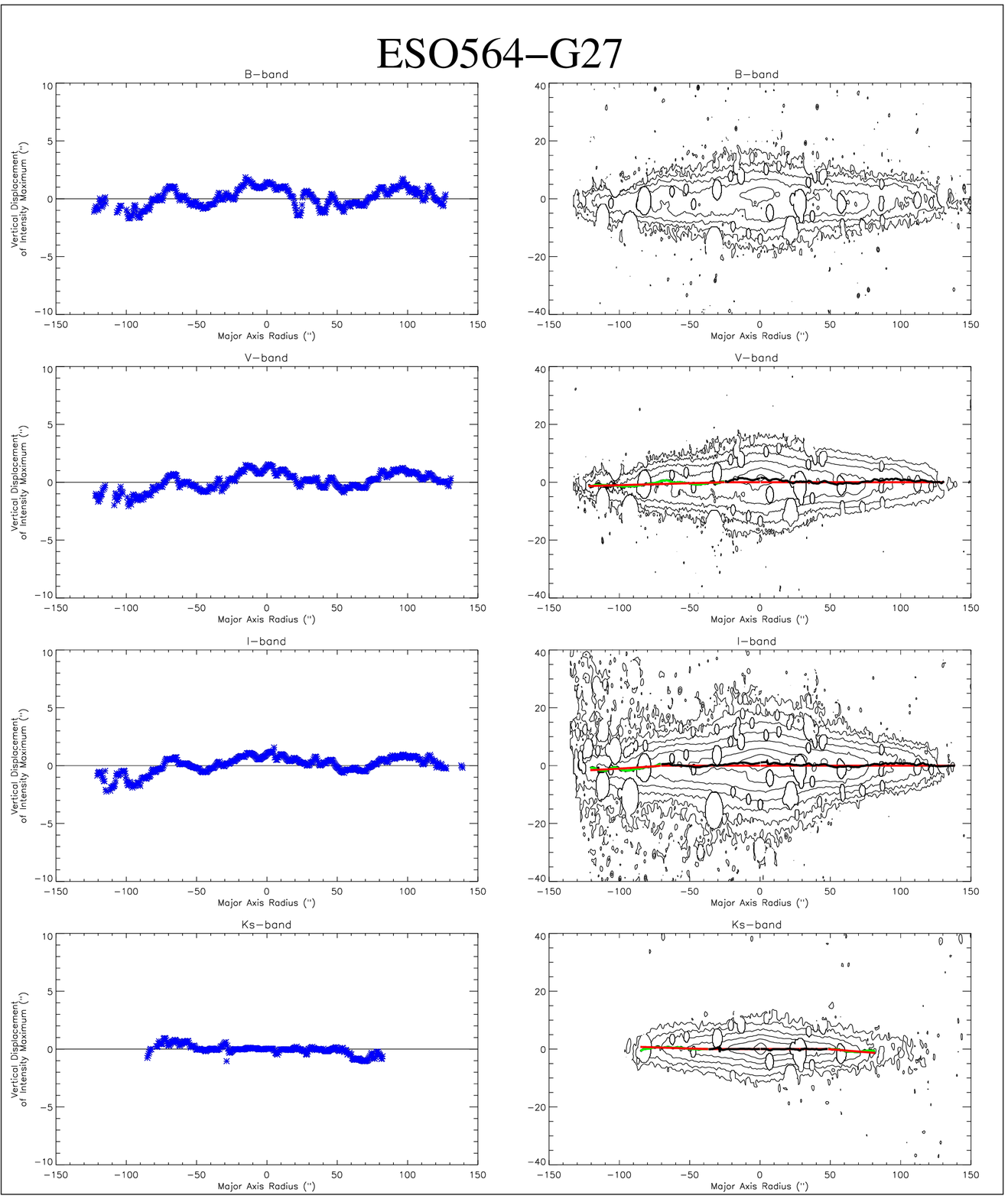}}
\caption{continued}
\end{figure*}

\end{document}